\documentclass[12pt,tikz]{article}

\usepackage[paper=letterpaper,margin=.85in]{geometry}

\usepackage{authblk}

\usepackage{amsfonts}
\usepackage{amsmath}
\usepackage{amsthm}
\usepackage{amssymb}
\usepackage{graphicx}
\usepackage{xcolor}

\usepackage{braket}
\usepackage{esint}
\usepackage{simpler-wick}

\usepackage{bm}
\usepackage{upgreek}
\usepackage{blkarray}

\usepackage[numbers,sort&compress]{natbib}
\usepackage{physics}
\usepackage{hyperref}
\usepackage{cleveref}
\crefname{equation}{Eq.}{Eqs.}
\usepackage{subfigure}

\usetikzlibrary{arrows.meta,decorations.pathmorphing,hobby}
\tikzset{
	eye/.pic={
		\draw[thick, fill=white]
		(-0.30,0) .. controls (-0.14,0.24) and (0.14,0.24) .. (0.30,0)
		.. controls (0.14,-0.24) and (-0.14,-0.24) .. cycle;
		\fill (0,0) circle (0.095);
		\fill[white] (0.05,0.055) circle (0.031);
	}
}

\usetikzlibrary{decorations.text}

\renewcommand{\Tr}{\text{Tr}}

\renewcommand{\dd}{\mathrm{d}}
\newcommand{\be}{\begin{equation}}
	\newcommand{\ee}{\end{equation}}
\newcommand{\ba}{\begin{aligned}}
	\newcommand{\ea}{\end{aligned}}

\newcommand{\dissipationpanel}[3]{%
	\begin{tikzpicture}
		\node[anchor=south west,inner sep=0pt] (img)
		{\includegraphics[width=\linewidth]{#3}};
		\node[anchor=south,font=\small]
		at ([xshift=20pt,yshift=-4pt]img.north) {#2};
		\node[anchor=north west,font=\small,inner sep=0pt]
		at ([xshift=8pt,yshift=-2.5pt]img.north west) {#1};
	\end{tikzpicture}%
}

\newcommand{\twopanelfig}[3]{%
	\begin{tikzpicture}
		\node[anchor=south west,inner sep=0pt] (img)
		{\includegraphics[width=\linewidth]{#1}};
		\begin{scope}[x={(img.south east)},y={(img.north west)}]
			\fill[white] (0.045,0.90) rectangle (0.085,1.0);
			\fill[white] (0.550,0.90) rectangle (0.590,1.0);
			\node[anchor=north west,inner sep=0pt,font=\small] at (0.050,0.985) {#2};
			\node[anchor=north west,inner sep=0pt,font=\small] at (0.555,0.985) {#3};
		\end{scope}
	\end{tikzpicture}%
}

\title{Closed- and open-system characterization of the Yukawa-SYK model in real time}

\author[1,2]{Alex Windey\thanks{\href{mailto:alex.windey@unitn.it}{alex.windey@unitn.it}}}
\author[1]{Christian H. Johansen}
\author[1,2]{Philipp Hauke\thanks{\href{mailto:philipp.hauke@unitn.it}{philipp.hauke@unitn.it}}}
\author[1,2]{Andrea Legramandi\thanks{\href{mailto:andrea.legramandi@unitn.it}{andrea.legramandi@unitn.it}}}

\affil[1]{Pitaevskii BEC Center, CNR-INO and Department of Physics, University of Trento, Via Sommarive 14, Trento, I-38123, Italy}
\affil[2]{INFN-TIFPA, Trento Institute for Fundamental Physics and Applications, Via Sommarive 14, Trento, I-38123, Italy}

\date{\small \today}

\begin{document}
	
	\maketitle

	\vspace{-1cm}

	\begin{abstract}
		
		Strongly correlated matter without quasiparticles is central to many-body physics, quantum chaos, and quantum gravity, motivating a considerable effort toward its realization in tabletop experiments. Two challenges remain: laboratory platforms are open systems, so conformal behavior must survive dissipation, and it must be identified through measurable quantities. We address both challenges in the Yukawa--Sachdev--Ye--Kitaev (YSYK) model, in which bosons and fermions interact as in, e.g., light--matter systems, but through random all-to-all couplings that drive the boson to criticality without fine tuning. Solving the model in real time using the Schwinger--Keldysh formalism, we treat closed and open systems within a unified framework and, in both cases, consider a thermal steady state to map their phase diagrams. In addition to the known conformal and impurity (or SYK$_2$) regimes, we identify novel high-temperature regimes in which the fermions are driven by an essentially bare boson and show that their dynamics is captured by a quadratic model with time-dependent random hopping. Across all regimes, the fermionic softening of the boson is bounded by a ``criticality budget'' determined only by the Yukawa coupling and the temperature. Coupling the boson with a thermal environment induces a mass shift that deforms the boundaries between different regimes and exhausts the criticality budget, thus imposing a bound on temperature and dissipation rate compatible with self-tuned criticality. We further introduce a set of markers built from the two-point functions that separate all identified regimes, are accessible in linear response, and require no fit of the conformal exponents. We complement them with the quantum Fisher information, deriving upper bounds with and without dissipation. These results may help characterize holographic quantum matter beyond the closed setting.
		
	\end{abstract}
	
	\clearpage
	
	\tableofcontents
	
	\clearpage
	
	\section{Introduction}
	\label{sec:intro}

	Strongly correlated fermionic systems that fail to organize into quasiparticles represent some of the most challenging and intriguing problems in many-body physics, as they lie beyond the reach of standard theoretical techniques. The Sachdev--Ye--Kitaev (SYK) model~\cite{SY93, Kitaev2015, Maldacena:2016hyu}  is one of the rare examples in which such a strongly correlated state can be described analytically. Its all-to-all random interactions render the disorder-averaged theory solvable in the large-$N$ limit, where the resulting saddle point develops an emergent reparameterization symmetry at low energies and realizes a maximally chaotic~\cite{Maldacena_2016} non-Fermi liquid (NFL)~\cite{Chowdhury2022}. Remarkably, the system can also be used to describe, through the holographic duality~\cite{Maldacena:1997re}, the physics of a near-extremal black hole~\cite{Sachdev:2015efa, Maldacena:2016upp,Kitaev:2017awl}. The SYK model therefore brings together three questions that are usually addressed separately: how to describe non-Fermi liquids and strange metals; how to characterize quantum chaos in strongly interacting systems; and how quantum gravity behaves in a non-perturbative regime. These connections have motivated efforts to realize holographic quantum matter in table-top experiments~\cite{Danshita:2016xbo, Pikulin:2017mhj, Chew_2017, Chen2018, cavity_proposal, Brzezinska2023, Baumgartner:2024ysk,Baumgartner:2025uue,Granet:2025yys}. Two obstacles stand in the way: the SYK interaction must be engineered, as it does not occur spontaneously in nature; and any realistic laboratory platform is an open system, so that the strongly coupled critical state has to survive the effect of dissipation. Even if these challenges are overcome and an experimental realization is achieved, a further difficulty remains: one must still determine, from experimentally accessible observables, whether the system has actually entered the critical regime.
	
	The Yukawa--SYK (YSYK) model introduced in Refs.~\cite{Esterlis_2019,wang_solvable_2020} and further studied in, e.g., Refs.~\cite{wang_quantum_2020,wang_phase_2021,classen_superconductivity_2021,inkof_quantum_2022,valentinis_correlation_2023,valentinis_bcs_2023,patel_universal_2023, Grunwald:2023plc, davis_quantum_2023,Qiu:2025ynz,Solis_2026,Cao:2026uxg}, in which $M$ bosons are coupled to $N$ fermions through random Yukawa couplings, addresses the first of these obstacles. Its vertex is the ordinary boson--fermion vertex of quantum electrodynamics (QED) or of the electron--phonon problem. However, what makes SYK interesting is nevertheless retained: the interaction dynamically cancels the bare boson mass, a mechanism known as self-tuned quantum criticality \cite{pan_yukawa-syk_2021}, and the system enters a conformal NFL regime. The model exhibits a considerably richer phase diagram than the SYK model, schematically displayed in Fig.~\ref{fig:phase-diagram}. The conformal regime occupies the low-temperature part of the plane spanned by the Yukawa coupling $g$ and the temperature $T$, and is separated by smooth crossovers from regimes in which the physics is effectively free or quadratic. 
	However, some of these regimes share characteristic features with the conformal one despite having a different underlying low-energy behavior. 
	In particular, the bosonic gap can close even outside the conformal regime,~\footnote{A behavior which is observed in the $\mathrm{SYK}_2$ regime of Fig.~\ref{fig:phase-diagram}.} complicating the precise identification of this NFL phase. A natural fingerprint of the conformal behavior is the power-law scaling of the two-point functions, but this can be very hard to measure in practice, since it is visible only in a low-frequency window that is limited from below by the finite temperature and from above by the effect of dissipation. It is therefore crucial to identify experimentally accessible observables that can discriminate the conformal phase from the other regimes without fitting the conformal exponents, to determine how the different phases are affected by coupling to an external bath, and to establish how much dissipation the self-critical state can tolerate.

	%
	%
	
	\definecolor{cFree}  {HTML}{C87ECC}   
	\definecolor{csidebands}{HTML}{4C72B0}
	
	\definecolor{cDriven}{HTML}{C44E52}   
	\definecolor{cImp}   {HTML}{55A868}   
	\definecolor{cConf}  {HTML}{DD8452}   
	\newcommand{\pdTint}{26}              
	
	\colorlet{fillFree}  {cFree!\pdTint!white}
	\colorlet{fillDriven}{cDriven!\pdTint!white}
	\colorlet{fillImp}   {cImp!\pdTint!white}
	\colorlet{fillConf}  {cConf!\pdTint!white}
	\colorlet{fillSide}  {csidebands!\pdTint!white}
	
	\newcommand{\pdW}{17.0cm}    
	\newcommand{\pdH}{17.0cm}    
	\newcommand{\pdImgW}{6.7cm} 
	\newcommand{\pdFade}{14mm}   
	\newcommand{\pdNs}{160}      
	
	\def\Mx{0.9}
	\def\xRz{0.9}
	\def\xUz{0.23}
	\def\xLz{-0.02}
	
	\def\xSz{-0.22}   
	
	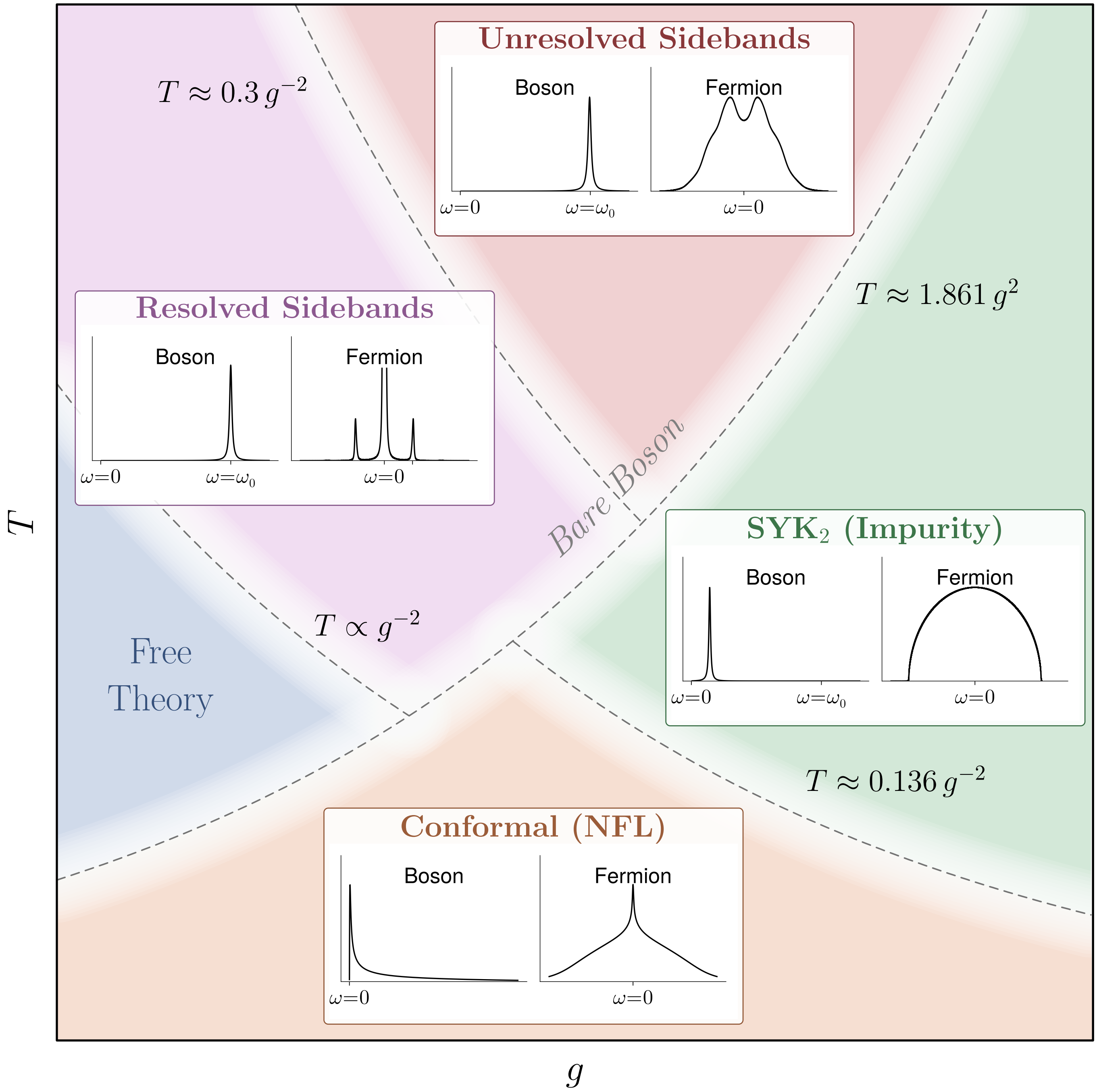
\begin{figure*}[p]
		\centering
		\begin{tikzpicture}[
			x=\pdW, y=\pdH, line join=round, line cap=round,
			declare function={
				yR(\x) = pow(10,  \Mx*(\x-\xRz));
				yU(\x) = pow(10, -\Mx*(\x-\xUz));
				yL(\x) = pow(10, -\Mx*(\x-\xLz));
				yS(\x) = pow(10, -\Mx*(\x-\xSz));
			},
			phasecard/.style={          
				draw=#1!65!black, fill=white, fill opacity=0.88, text opacity=1,
				line width=0.5pt, rounded corners=2pt, inner sep=2.5pt,
				align=center, font=\small},
			crossoverline/.style={black!55, line width=0.7pt,
				dash pattern=on 4pt off 3pt},   
			eqlabel/.style={inner sep=1pt}
			]
			
			\pgfmathsetmacro{\xA}{0.5*(\xRz+\xUz)}
			\pgfmathsetmacro{\xB}{0.5*(\xRz+\xLz)}
			
			\pgfmathsetmacro{\xC}{0.5*(\xRz+\xSz)}   
			\def\pS#1#2{plot[domain=#1:#2,samples=\pdNs] (\x,{yS(\x)})}
			
			\def\pR#1#2{plot[domain=#1:#2,samples=\pdNs] (\x,{yR(\x)})}
			\def\pU#1#2{plot[domain=#1:#2,samples=\pdNs] (\x,{yU(\x)})}
			\def\pL#1#2{plot[domain=#1:#2,samples=\pdNs] (\x,{yL(\x)})}
			
			\newcommand{\halo}[1]{%
				\foreach \k in {1,...,12}{%
					\draw[white,opacity=0.15,line width={(13-\k)/12*\pdFade}] #1;}}
			
			\begin{scope}
				\clip (0,0) rectangle (1,1);
				
				\fill[fillFree]   (0,{yR(0)})                                  
				-- plot[domain=0:\xA    ,samples=\pdNs] (\x,{yR(\x)})
				-- plot[domain=\xA:\xUz ,samples=\pdNs] (\x,{yU(\x)})
				-- (0,1) -- cycle;
				\fill[fillDriven] (\xUz,1)                                     
				-- plot[domain=\xUz:\xA ,samples=\pdNs] (\x,{yU(\x)})
				-- plot[domain=\xA:\xRz ,samples=\pdNs] (\x,{yR(\x)})
				-- cycle;
				\fill[fillImp]    (\xRz,1)                                     
				-- plot[domain=\xRz:\xB ,samples=\pdNs] (\x,{yR(\x)})
				-- plot[domain=\xB:1    ,samples=\pdNs] (\x,{yL(\x)})
				-- (1,1) -- cycle;
				\fill[fillConf]   (0,{yR(0)})                                  
				-- plot[domain=0:\xB    ,samples=\pdNs] (\x,{yR(\x)})
				-- plot[domain=\xB:1    ,samples=\pdNs] (\x,{yL(\x)})
				-- (1,0) -- (0,0) -- cycle;
				
				\fill[fillSide] (0,{yR(0)})                                    
				-- plot[domain=0:\xC   ,samples=\pdNs] (\x,{yR(\x)})
				-- plot[domain=\xC:0   ,samples=\pdNs] (\x,{yS(\x)})
				-- cycle;
				
				\halo{\pR{0}{\xRz}}
				\halo{\pU{\xUz}{\xA}}
				\halo{\pL{\xB}{1}}
				\halo{\pS{0}{\xC}}               
				
				\draw[crossoverline] \pR{0}{\xRz};
				\draw[crossoverline] \pU{\xUz}{\xA};
				\draw[crossoverline] \pL{\xB}{1};
				\draw[crossoverline] \pS{0}{\xC};  
			\end{scope}
			
			\draw[line width=1pt] (0,0) rectangle (1,1);
			\node[below=7pt] at (0.5,0) {\Large $g$};
			\node[left=7pt]  at (0,0.5) {\rotatebox{90}{\Large $T$}};
			
			\node[eqlabel] at (0.85,0.72) {\large $T \approx 1.861 \, g^{2}$};
			\node[eqlabel] at (0.17,0.915) {\large $T \approx 0.3\,g^{-2}$};
			\node[eqlabel] at (0.3,0.4) {\large $T \propto g^{-2}$};
			\node[eqlabel] at (0.81,0.25) {\large $T \approx 0.136\,g^{-2}$};
			
			\node[eqlabel, align=center, text=csidebands!70!black, font=\Large] at (0.1,0.35) {Free\\Theory};
			
			\path[decorate, decoration={text along path, text align=center, raise=15pt,
				text={|\Large\itshape\color{gray}| Bare Boson}}] \pR{0.41}{0.7};

			\newcommand{\phasepanel}[5]{%
				\node[phasecard=#1] at (#2,#3)
				{\makebox[\pdImgW][c]{\textcolor{#1!70!black}{\bfseries\strut #4}}\\[-0.6mm]
					\includegraphics[width=\pdImgW]{#5}};}
			
			\phasepanel{cFree}  {0.22}{0.62}{\large Resolved Sidebands}  {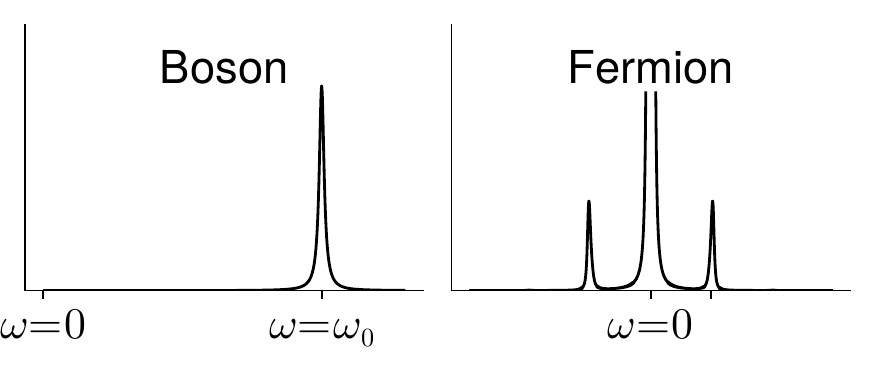}
			\phasepanel{cDriven}{0.567}{0.88}{\large Unresolved Sidebands}{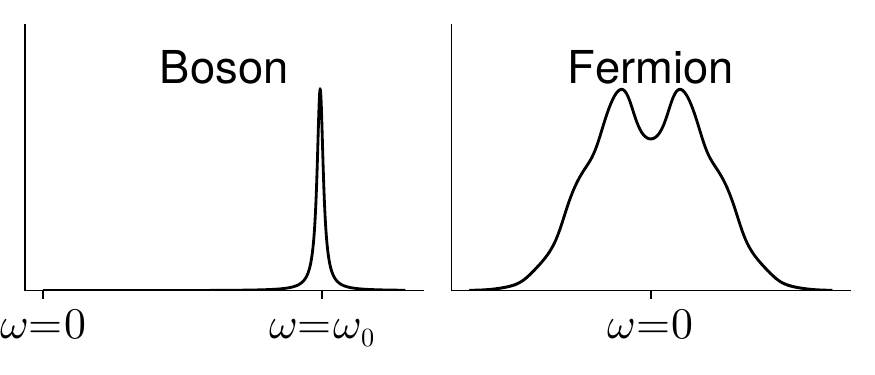}
			\phasepanel{cImp}   {0.79}{0.408}{\large SYK$_2$ (Impurity)}         {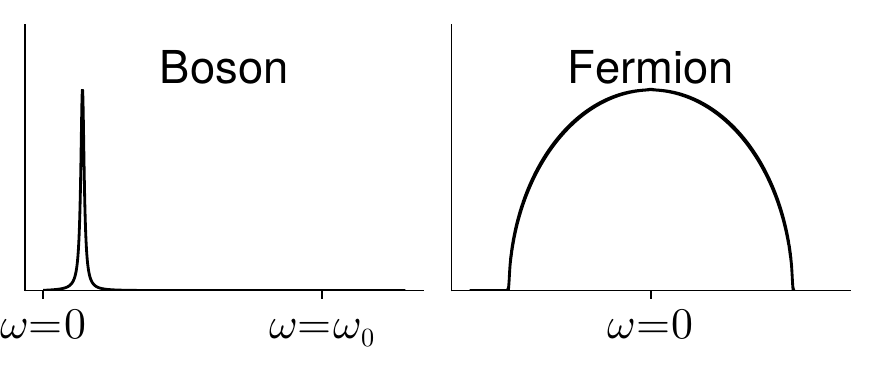}
			\phasepanel{cConf}  {0.46}{0.12}{\large Conformal (NFL)}                  {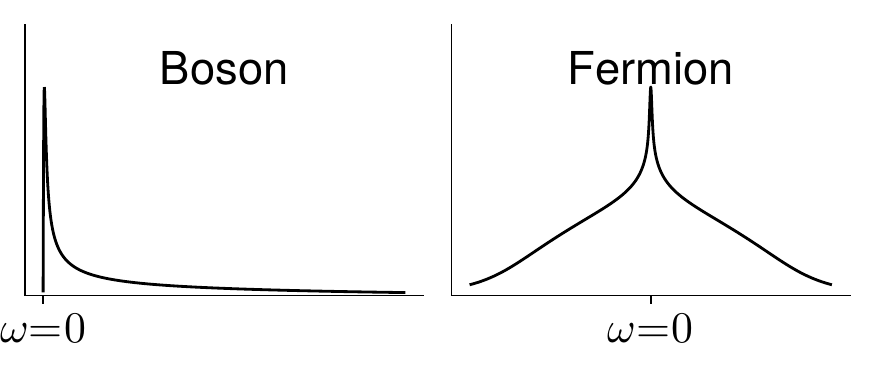}
			
		\end{tikzpicture}
		
		\caption{\label{fig:phase-diagram}%
			Schematic phase diagram of the Yukawa--SYK model in the plane spanned by the
			Yukawa coupling $g$ and the temperature $T$, in units of the bare boson mass $\omega_0$.
			Five regimes---labeled according to the salient behavior of the fermions---are separated by crossover lines that parametrically follow $T\propto g^{2}$ and $T\propto g^{-2}$. 
			Each regime is distinguished by a characteristic behavior of bosonic (left) and fermionic (right) spectral functions. 
			At low temperatures, there is the conformal NFL phase sustained by self-tuned criticality between fermion and boson. 
			When $g^{-2} \ll T \ll g^2$, the boson remains critical but the fermion becomes broad and is well described by an SYK$_2$ model. 
			The region $T\gg g^{2}$ is characterized by an essentially bare boson, which imprints itself in three qualitatively different ways on the fermion: a high-temperature strong-coupling regime with sidebands at multiples of $\omega_0$ but which overlap, one where the sidebands get neatly resolved, and a regime where the fermion is essentially free. 
		}
	\end{figure*}
	
	In this work, we perform an analytical and numerical study of the YSYK model in the thermodynamic limit using a Schwinger--Keldysh path-integral formulation. Formulating the large-$N$, $M$ saddle point on the Schwinger--Keldysh contour provides direct access to the spectral functions, without the need for analytic continuation, and allows us to treat the dynamics of both closed and open systems within a unified framework. Mapping out the $(g,T)$ phase diagram (see Fig.~\ref{fig:phase-diagram}), we analytically recover the conformal (NFL) and the $\mathrm{SYK}_2$ (impurity) regimes known from the imaginary-time analysis of Ref.~\cite{Esterlis_2019}, together with quantities that are less natural to extract in Matsubara space, such as the interaction-induced boson linewidth. 
	Both regimes rely on the interaction driving the boson to criticality. We show that this mechanism is bounded by a ``criticality budget'', a bound that follows solely from the thermal nature of the fermions and is independent of the details of the saddle-point solution.
	We also identify two regimes that were not described in previous studies, in which the temperature is the largest scale but the coupling is not small: in these regimes, the boson stays essentially bare while the fermion hybridizes with sidebands displaced by integer multiples of the bare boson mass $\omega_0$. A sharp crossover separates the regimes of resolved and unresolved sidebands. We derive a toy model valid at large temperatures that takes the form of a quadratic SYK$_2$ model with time-dependent random hopping and which reproduces this structure exactly. 
	
	Since the crossovers between the above regimes are smooth, we then construct a minimal set of observables that separates them, built entirely from the fermionic and bosonic two-point functions and therefore accessible in linear response. Importantly, these observables use only coarse features of the two-point functions and, in contrast to extraction of the conformal exponent, do not require any fitting. 
	These are: 
	(i) The inverse static susceptibility $\alpha$, which measures how far the boson sits from criticality. As the criticality budget becomes small, $\alpha$ approaches its free value $\omega_0$, which is reached only asymptotically. The crossover between the self-tuned (conformal and SYK$_2$) and bare-boson regimes is therefore defined by convention, as the point where the constant-$\alpha$ curves depart from their small-gap scaling and approach the saturated budget.
	(ii) The curvature $\xi$ of the inverse bosonic propagator at zero frequency distinguishes the conformal and SYK$_2$ regimes among the self-tuned regimes at small $\alpha$, being negative only in the conformal phase. 
	(iii) The fermionic weight between sidebands generated by coupling to an essentially free boson field, which permits to separate the resolved from the unresolved sideband regimes within the driven-fermion phase.   
	We complement these with the quantum Fisher information (QFI) \cite{Hauke_2016} of the real bosonic quadrature, which can be obtained simply from integrating the corresponding response function against a thermal kernel and which carries an operational meaning as a metrological sensitivity~\cite{Braunstein_Caves1994}. Using sum rules obeyed by the bosonic propagator, we derive an upper bound that is saturated in the SYK$_2$ regime, where the boson is a single sharp mode, but not in the conformal one, where its spectral weight spreads above the thermal scale. As a consequence, along lines of constant gap the QFI develops a maximum close to the crossover between the two regimes at a temperature set by the renormalized boson mass and with a growing height on approach to criticality. These signatures also permit us to present, rather than only overall scalings, quantitative parameter estimates for the crossover regions.

	In the second part of this work, we open the system by coupling the bosonic field to an environment, following Refs.~\cite{Hosseinabadi:2023mid, Cichutek_2024}. However, instead of studying quench dynamics, we fix the steady state to be thermal and focus on how each regime gets deformed by dissipation, and how much dissipation-induced spectral broadening is needed to destabilize the self-critical regime. Since, in the infrared, a generic bath enters only through its static shift of the boson mass and its low-frequency exponent, we model it by a single Markovian loss rate $\kappa$, which, among the baths that preserve the conformal phase, provides an upper bound on the effect of dissipation. The criticality budget,  which is fixed by the coupling strength and the temperature alone, limits the ability of the system to become self-critical, and once the bath stiffens the boson beyond what the criticality budget can compensate, no choice of parameters can restore criticality. 
	At weak coupling to the bath, the static shift acts in all regimes as a rescaling of the Yukawa coupling, which for a bath interaction that stiffens the boson, such as particle exchange, displaces the crossover lines of Fig.~\ref{fig:phase-diagram} toward larger $g$. The dynamical part of the bath, instead, narrows the conformal frequency window and steepens the conformal/SYK$_2$ boundary beyond this rescaling, so that the conformal regime shrinks in favor of the SYK$_2$ one, while in the driven-fermion regimes it turns the gaps between sidebands into pseudogaps, so that the resolved-sideband regime survives only at weak dissipation. At strong coupling to the bath, the fermions saturate the criticality budget by collapsing onto a sharp quasiparticle, while the boson is completely dominated by the bath interaction and reduces to a Drude-like response \cite{Leggett_1987,Hu_1992}. 
	
	More broadly, the open-system analysis presented here connects to ongoing efforts to
	characterize dissipative quantum chaos~\cite{Zhang2019,Sa:2020fpf,Kulkarni:2021gtt,
		Kawabata:2022cpr,Garcia-Garcia:2024tbd,Pelliconi2026, Liu2026}, the real-time dynamics of holographic
	quantum matter coupled to an environment~\cite{Chen:2017dbb,Almheiri:2019jqq,Chen:2020wiq}, and of its gravitational
	dual~\cite{Jana:2020vyx,Garcia-Garcia:2022adg,Wang:2023vkq,Vargas:2026ijf}. 
	
	Moreover, while our aim here is to understand the general thermodynamic behavior of YSYK and not so much to mimic a specific experimental setting, we expect the obtained results to provide useful guidance for developing platforms on which such dynamics may be probed
	experimentally~\cite{cavity_proposal,Peyruchat:2024akp,Baumgartner:2025uue,Wold:2025kcy,Ferrari:2026zkn,Pacchioni2026}.
	For example, as our discussions about the criticality budget suggest, 
	a cavity-QED implementation may best reach the conformal regime when the frequency detuning between cavity-transition and drive-laser is set to match the cavity decay rate (assuming this to be the dominant dissipation channel). 
	The developed probes also directly connect to observables accessible in experiments, in general through linear response and in the specific case of cavity QED through homodyne detection. 
	
	The rest of the paper is organized as follows. Section~\ref{sec:phase_diagram} sets up the YSYK model on the Schwinger--Keldysh contour, introduces the criticality budget and solves the saddle-point equations in each regime, both analytically and numerically, arriving at the phase diagram of Fig.~\ref{fig:phase-diagram}. Section~\ref{sec:observables} introduces the observables that discriminate between these regimes, uses them to locate the crossovers, and derives an upper bound on the QFI. Section~\ref{sec:dissipation} couples the boson to the environment in thermal equilibrium and quantifies its effect on each regime, deriving the saddle-point solutions in the presence of this specific type of dissipation and how the crossover lines and the QFI are modified. Section~\ref{sec:conclusions} collects our conclusions and outlines directions for future work. The appendices contain additional, more technical material: Appendix~\ref{app:path_integral} presents the path-integral derivation of the saddle-point equations for a generic stationary state and a generic bosonic bath, Appendix~\ref{app:regimes} the details of the analytic solutions in the different regimes, Appendix~\ref{app:measurement} the protocols by which the two-point functions can be extracted experimentally, including a balanced homodyne scheme for a cavity-QED realization, and Appendix~\ref{app:numerics} the numerical method, in particular the constant-$\alpha$ scheme used throughout, together with additional results with dissipation.

	\section{Model and phase diagram}
	\label{sec:phase_diagram}
	
	The Yukawa--SYK model provides a rich framework for studying strongly correlated fermion--boson systems with random interactions, exhibiting nontrivial quantum chaotic behavior in the infrared~\cite{davis_quantum_2023,Solis_2026}. Previous works have analyzed the equilibrium properties and phase diagram of this model using imaginary-time methods~\cite{Esterlis_2019}. However, several aspects remain less explored, including the real-frequency spectral properties of the thermal state at finite temperature, a more comprehensive characterization of the phase diagram, and a systematic investigation of line-width broadening across the different phases.
	
	In this section, we revisit the Yukawa--SYK model using the Schwinger--Keldysh formalism, which allows us to formulate the theory directly in real time. This approach has several advantages: it gives immediate access to observables such as the spectral density and two-point correlation functions without the need for analytic continuation, and provides a shared framework for studying both equilibrium and non-equilibrium phenomena. Using this formalism, we map out the phase diagram across a range of temperatures and interaction strengths. We further show that the fermions can soften the boson only up to a ``criticality budget'', fixed by the coupling and the temperature alone. This analysis not only sheds further light on the known phases, it also provides (a) evidence of two regimes with distinct characteristics not described in previous studies, and (b) quantitative parameter estimates at which the crossovers are expected, which will be considered in Sec.~\ref{sec:observables}. While in this paper we will mostly focus on thermal equilibrium, we also give a full derivation for arbitrary stationary states in Appendix~\ref{app:path_integral}.
	
	\subsection{Yukawa--SYK model in real time}
	
	We consider a version of the Yukawa--SYK (YSYK) model where $M$ bosons with bare frequency $\omega_0$ are coupled to $N$ spinless fermions~\cite{Solis_2026}, in contrast to the original spinful model introduced in Ref.~\cite{Esterlis_2019}. 
	The Hamiltonian of the system reads
	\begin{equation}
		\label{eq:YSYK_H}
		H = \omega_0 \sum_{m=1}^M a_m^\dag a_m - \mu_C \sum_{j=1}^N c_j^\dag c_j + \frac{1}{\sqrt{2MN}} \sum_{jkm} g^m_{jk} \, c_j^\dag c_k (a_m + a^\dag_m) \, ,
	\end{equation}
	where, for each boson mode $m$, the couplings $g^m_{ij}$ are sampled independently from the Gaussian unitary ensemble (GUE) with vanishing mean and variance $g^2$. 
	In contrast to  Ref.~\cite{Esterlis_2019}, this model does not possess any superconducting transition. 
	The boson frequency $\omega_0$ provides a natural scale for the problem, so in all the figures (as well as in some inequalities when there are no ambiguities) we express the dimensionful quantities in units of $\omega_0$.
	
	As discussed in Appendix \ref{app:path_integral}, the YSYK theory can be treated using the Schwinger--Keldysh path integral approach and becomes solvable in the thermodynamic limit after averaging over disorder. We take the thermodynamic limit by sending $M,N \to \infty$ while keeping the ratio $N/M$ fixed.
	In this limit, the theory closes self-consistently in terms of averaged fermionic and bosonic two-point functions and the associated self-energies.
	The real-time dynamics is characterized by the following fermionic and bosonic propagators:
	\begin{align}
		G^>(t,t') &= -\frac{i}{N} \sum_{j=1}^N \langle c_j(t) c_j^\dag(t')  \rangle \, , \hspace{2.5cm} G^<(t,t') = \frac{i}{N} \sum_{j=1}^N \langle c_j^\dag(t) c_j(t')  \rangle \, , \\
		D^>(t,t') &= -\frac{i}{M} \sum_{m=1}^M \langle x_m(t) x_m(t')  \rangle = D^<(t',t) \label{eq:boson_<_def} \, ,
	\end{align}
	where the imaginary bosonic quadrature has been integrated out, leaving only the real field $x_m = (a_m+a^\dag_m) / \sqrt{2}$~\footnote{Note that the imaginary part of the bosons does not take part in the interaction as given by Eq.~\eqref{eq:YSYK_H}. Hence, it is not only convenient but also natural to integrate out the imaginary part at the level of the path integral.}. From these, we can define the retarded and advanced propagators:
	\begin{align}
		G^R(t,t') =& \theta(t-t') \left[G^>(t,t')-G^<(t,t')\right] \, , \hspace{2cm} G^A(t,t') = [G^R(t',t)]^\dag \, ,  \\
		D^R(t,t') =& \theta(t-t') \left[D^>(t,t')-D^<(t,t')\right] \, , \hspace{2cm} D^A(t,t') = [D^R(t',t)]^\dag \, . \label{eq:boson_R_def}
	\end{align}
	As can be seen from their definitions, these propagators are not independent and it is sufficient to work with a minimal set. In our treatment, we choose the retarded and lesser Green’s functions. The retarded propagator encodes the spectral properties of the system, while the lesser one directly contains information about occupation and statistical properties. All other 2-point correlators can be reconstructed from this pair for both the bosonic and the fermionic fields. 
	
	As a further simplification, we restrict our analysis to the late-time
	stationary regime, where the system is time-translation invariant and all
	propagators depend only on time differences.
	We further consider the situation where the steady state is thermal at inverse temperature $\beta$. In this case, the fluctuation–dissipation theorem~\cite{Kubo1957,CallenWelton1951,Kubo1966} relates the lesser and retarded propagators as
	\begin{align}
		G^<(\omega) =& - 2 i \Im\left[G^R(\omega)\right] n_\mathrm{F}(\omega)\, , \qquad n_\mathrm{F} = \frac{1}{e^{\beta(\omega-\mu_C)} + 1}  \, , \label{eq:fluctuation_dissipation_G}\\ 
		D^<(\omega) =& 2 i \Im\left[D^R(\omega)\right] n_\mathrm{B}(\omega)\, , \qquad \quad \, \,  n_\mathrm{B} = \frac{1}{e^{\beta\omega} -1} \, . \label{eq:fluctuation_dissipation_D}
	\end{align}
	As a consequence, only the retarded propagators have to be calculated. In the following, we set the chemical potential to $\mu_C=0$. This choice makes the disorder-averaged action invariant under the particle--hole symmetry $c_j \to c_j^\dagger$, $x_m \to -x_m$. Since the system is pinned to half filling, the fermionic spectral function is even in frequency, and the conformal  solution has no spectral asymmetry parameter \cite{Sachdev:2015efa,Gu_2020}. 
	
	The fermionic ($\Sigma$) and bosonic ($\Pi$) self-energies associated with the two-point functions, introduced as Lagrange multipliers in the path-integral approach of Appendix~\ref{app:path_integral}, satisfy the Schwinger--Dyson equations,
	\begin{equation}
		G^R(\omega) = \frac{1}{G_0^R(\omega)^{-1} - \Sigma^R(\omega)} \, , \qquad
		D^R(\omega) = \frac{1}{D_0^R(\omega)^{-1} - \Pi^R(\omega)} \, , \label{eq:Schwinger_Dyson}
	\end{equation}
	where the free retarded propagators read
	\begin{equation}
		G^R_0(\omega) = \frac{1}{\omega+ i 0^+} \, , \qquad D^R_0(\omega) = \frac{\omega_0}{(\omega +i 0^+ )^2-\omega_0^2} \, .
	\end{equation}
	As anticipated, the structure of the Yukawa--SYK interaction closes this set of equations by providing explicit expressions for the self-energies in terms of the two-point functions: 
	\begin{align}
		\Sigma^R(t) &= i g^2 \left[G^<(t) D^R(t) + G^R(t) D^<(t) + G^R(t) D^R(t) \right] \label{eq:SigmaR_t}\,, \\
		\Pi^R(t) &= -i g^2 \frac{N}{M} \left[G^<(t) G^A(-t) + G^R(t) G^<(-t) \right] \, . \label{eq:PiR_t} 
	\end{align}
	The equation for the lesser components of the self-energies are also pure products in time, which ensures that interaction does not spoil thermality (see Appendix~\ref{app:SK_path_integral_YSYK}).
	
	In the remainder of this section, we analyze and solve this closed set of equations in different parameter regimes, both analytically and numerically. A summary of the resulting phase diagram is presented in Fig.~\ref{fig:phase-diagram}.
	
	\subsection{The conformal/non-Fermi liquid phase}
	\label{subsec:conf_phase}
	
	We first focus on the zero temperature case, $T=1/\beta =0$, where a conformal non-Fermi-liquid (NFL) phase emerges due to self-tuned criticality. The discussion in this section extends known results in the context of the SYK and YSYK models in imaginary time~\cite{Maldacena:2016hyu,Esterlis_2019} to the real-time formalism.
	
	In the long-time, or equivalently low-energy, limit the system develops a reparameterization symmetry that leaves the saddle point equations in Eqs.~\eqref{eq:fluctuation_dissipation_G}--\eqref{eq:PiR_t} invariant. Under this symmetry, the fermionic and bosonic Green’s functions transform as
	\begin{equation}
		G(t,t') \to \left[f(t)f(t')\right]^{\Delta_\mathrm{F}} \, G(f(t),f(t')) \,, \qquad D(t,t') \to \left[f(t)f(t')\right]^{\Delta_\mathrm{B}} D(f(t),f(t')) \,,\label{eq:conf_transform}
	\end{equation}
	where $f(t)$ is a monotonic smooth function and, as we will see below, self-consistency of the saddle-point equations imposes $\Delta_\mathrm{B} = 1- 2\Delta_\mathrm{F}$~\cite{Fu_2017,Bi:2017yvx, Kim_2020}. In Eq.~\eqref{eq:conf_transform}, we intentionally remain agnostic about the causal nature of the propagator, since this symmetry acts on fields defined on the full closed-time contour (see Appendix~\ref{app:conf_symmetry} for a more detailed discussion). 
	
	For this symmetry to emerge, the Schwinger--Dyson Eqs.~\eqref{eq:Schwinger_Dyson} at small $\omega$ must become dominated by the interaction self-energies, so that the free inverse propagators can be neglected and the low-energy solution does not carry any intrinsic scale. 
	For the fermions at zero chemical potential, we have $G_0^R(\omega)^{-1} = \omega$, which naturally vanishes at low energies.
	For the bosons, the bare inverse propagator is finite at low frequencies, $D_0^R(0)^{-1} = -\omega_0$ and the conformal regime can therefore only be reached if the bosonic self-energy compensates this term exactly,
	\begin{equation}
		\Pi^R(0) = -\omega_0 \label{eq:def_self_criticality} \, .
	\end{equation}
	Unlike its fermionic counterpart, the vanishing of the denominator in the bosonic Schwinger--Dyson equation is not guaranteed, but it is instead satisfied dynamically, an effect known as self-tuned quantum criticality~\cite{Esterlis_2019}.
	The boson--fermion interaction generates a bosonic self-energy whose zero-frequency value precisely cancels the bare mass term, so that the boson becomes effectively gapless at low energy without fine-tuning any microscopic parameter. Once the boson is critical, it feeds back into the fermionic self-energy and enforces anomalous fermionic scaling as well; the conformal solution is thus a mutually sustained critical state of the coupled system.
	It is then useful to introduce $\delta\Pi^R(\omega) = \Pi^R(\omega)-\Pi^R(0)$, so that the Dyson equation for the retarded bosonic propagator takes the form
	\begin{equation}
		D^R(\omega)=\frac{1}{D^R_0(\omega)^{-1}-\Pi^R(0)-\delta\Pi^R(\omega)} \, .
		\label{eq:bosonic_conformal_propagator}
	\end{equation}
	With this rewriting, $\Sigma^R$ and $\delta\Pi^R$ are the objects that transform covariantly under the reparameterization symmetry of Eq.~\eqref{eq:conf_transform}, with respect to the same function $f(t)$ but different conformal weights.

	The resulting low-energy solution is a pure power law at long times:
	\begin{equation}
		G^R(t) = -2ib \cos (\pi\Delta_\mathrm{F}) \frac{\theta(t)}{t^{2\Delta_\mathrm{F}}}  \, , \qquad D^R(t) = - 2 \tilde{b} \sin (\pi\Delta_\mathrm{B})\frac{\theta(t)}{t^{2\Delta_\mathrm{B}}} \, , \label{eq:conformal_ansatz_main}
	\end{equation}
	where $b$ and $\tilde{b}$ are real constants.\footnote{The trigonometric prefactors follow from analytically continuing the
		conformal ansatz from Euclidean to Lorentzian time, following the conventions of Ref.~\cite{Maldacena:2016hyu}. The different
		trigonometric function in the two cases reflects the fermionic versus bosonic statistics.} By plugging these ans\"atze into Eqs.~\eqref{eq:fluctuation_dissipation_G}--\eqref{eq:PiR_t}, we can also derive the expressions for the self-energies in real time:
	\begin{equation}
		\Sigma^R(t) = i \frac{(1-2 \Delta_\mathrm{F}) \sin (\pi  \Delta_\mathrm{F})}{\pi  b} \frac{\theta(t)}{t^{2-2 \Delta_\mathrm{F}}}\,, \qquad \delta \Pi^R(t) =  -2 b^2 g^2 \sin (2\pi  \Delta_\mathrm{F} ) \frac{N}{M} \frac{\theta(t)}{t^{4 \Delta_\mathrm{F}}} \label{eq:conformal_self_energies_T=0}\,,
	\end{equation}
	and
	\begin{equation}
		\tilde{b} = - \frac{(4 \Delta_\mathrm{F} -1) \cot (2 \pi  \Delta_\mathrm{F} )}{2\pi  b^2 g^2} \frac{M}{N} \, , \qquad \qquad \frac{(4 \Delta_\mathrm{F} -1) \cos (2 \pi  \Delta_\mathrm{F} )}{2(2 \Delta_\mathrm{F} -1)\sin^2(\pi  \Delta_\mathrm{F} )}= \frac{N}{M} \, , \label{eq:conformal_constant}
	\end{equation}
	where the second equation is a transcendental equation that determines $\Delta_\mathrm{F}$ as a function of the ratio $N/M$, as already discussed in Ref.~\cite{Bi:2017yvx,Kim_2020}. The scaling exponent is limited to the interval $\Delta_\mathrm{F} \in (1/4,1/2)$ and the model interpolates between an effective SYK$_2$ ($\Delta_\mathrm{F} =1/2$) for $M \ll N$ and an effective SYK$_4$ ($\Delta_\mathrm{F} =1/4$) for $M \gg N$ \cite{Bi:2017yvx,Kim_2020}. Following the conventions of Ref.~\cite{Esterlis_2019} and setting $N=2M$ to account for the absence of spin degrees of freedom in our model, one finds $\Delta_\mathrm{F}= 0.42037$, in agreement Ref.~\cite{Esterlis_2019}. 
	
	The constant $b$, in contrast, is fixed by the condition in~\cref{eq:def_self_criticality}. To calculate it, we need to regularize the conformal propagator, which is otherwise UV divergent. A crude approximation is to use the conformal expression in~\cref{eq:conformal_self_energies_T=0} for $\Sigma^R(\omega)$ in the Schwinger--Dyson equation:
	\begin{equation}
		G^R(\omega) = \frac{1}{\omega - \Sigma^R(\omega)}.
	\end{equation}
	With this approximation, we can equate the self-criticality condition with~\cref{eq:PiR_t}, written as a convolution:
	\begin{equation}
		\Pi^R(0) = -i g^2 \frac{N}{M} \int \frac{\dd \omega}{2\pi} \left[G^<(\omega) G^A(\omega) + G^R(\omega) G^<(\omega) \right]=-\omega_0 \, .
		\label{eq:self_fix_b}
	\end{equation}
	An explicit calculation gives the same functional dependence on $g$ as in Ref.~\cite{Esterlis_2019}:
	\begin{equation}
		b = c_1 \left( \frac{g^2}{\omega_0}\right)^{-2\Delta_\mathrm{F}}, \qquad c_1 = \frac{1}{2 \cos (\pi  \Delta_\mathrm{F} ) \Gamma (1-2 \Delta_\mathrm{F} )} \left(\frac{  (4 \Delta_\mathrm{F} -1)  \cos (2 \pi  \Delta_\mathrm{F} )}{8 \Delta_\mathrm{F} ^2 \sin \left(\frac{\pi }{2 \Delta_\mathrm{F} }\right) \sin ^2(\pi  \Delta_\mathrm{F} )}\right)^{-2 \Delta_\mathrm{F} }\, .
		\label{eq:constant_c1}
	\end{equation}
	For $N=2M$, the above expression gives $c_1 = 0.4156$. Although this constant has been obtained by a quite crude estimate, it is relatively close to the value obtained by an interpolation using numerical solutions, as discussed in Appendix~\ref{app:conf_symmetry}, which gives $c_1 \approx 0.315$.
	
	\begin{figure}[htbp]
		\centering
		\includegraphics[width=0.49\linewidth]{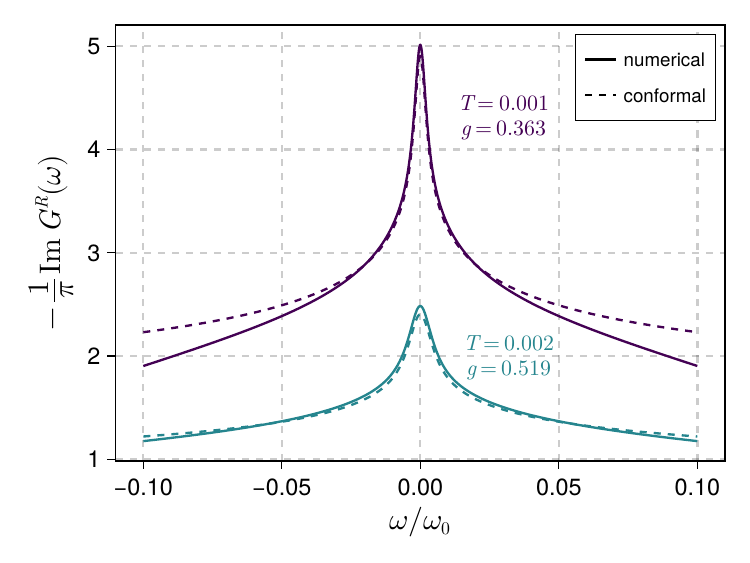}
		\includegraphics[width=0.49\linewidth]{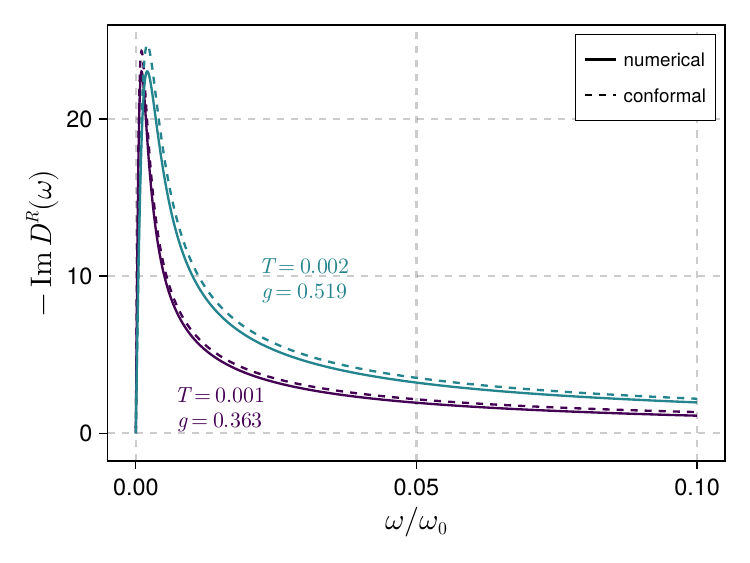}
		\caption{Left Panel: Comparison of the numerical solution of the fermionic spectral function $-\frac{1}{\pi}\Im G^R(\omega)$ with the finite-temperature expression in \cref{eq:GR_finiteT}. Right Panel: Same comparison for the bosonic propagator, using the analytical expression in \cref{eq:DR_finiteT}.
			In both cases, the numerical and analytical solution compares well in the small frequency regime, where the conformal solution applies. The temperature $T$ and coupling $g$ are expressed in units of $\omega_0$.
		}
		\label{fig:conformal_plot}
	\end{figure}
	
	A key feature of the conformal symmetry is that the finite-temperature propagator can be obtained from the zero-temperature solution via a reparametrization. The reparameterization function $f$ must compactify time along the imaginary direction, $t \sim t + i\beta$, ensuring that the Kubo--Martin--Schwinger (KMS)~\cite{Kubo1957,Martin:1959jp,Haag:1967sg} condition is satisfied. 
	Imposing that it also maps the unbroken SL(2,$\mathbb{R}$) symmetry given by the M\"{o}bius transformation to the isometries of the thermal circle uniquely fixes $f(t) = \tanh \frac{\pi t}{\beta}$.\footnote{The conformal saddle point solution spontaneously breaks the reparameterization symmetry of~\cref{eq:conf_transform} except for the M\"{o}bius transformations $f(t) = \frac{at+b}{ct+d} \, , ad-bc =1$, which therefore preserves the vacuum solution. In the thermal case $f(t) = \tanh \frac{\pi t}{\beta}$, these transformations become isometries of the thermal circle: $\frac{a f(t) + b}{c f(t) + d} = \alpha_1 (\tanh \frac{\pi (t+t_0)}{\beta} + \alpha_2)$ for appropriate constants $\alpha_{1,2},t_0$.} 
	As a result, we have the following forms for the propagators:
	\begin{equation}
		G^R(t) = -2i b \cos(\pi \Delta_\mathrm{F}) \frac{\theta(t)}{\left[\frac{\beta}{\pi }  \sinh \left(\frac{\pi  t}{\beta }\right)\right]^{2 \Delta_\mathrm{F} }} \, , \qquad D^R(t) = -2 \tilde{b} \sin\left[\pi \Delta_\mathrm{B}\right] \frac{\theta(t)}{\left[\frac{\beta}{\pi }  \sinh \left(\frac{\pi  t}{\beta }\right)\right]^{2 \Delta_\mathrm{B} }} \, , \label{eq:finite_T_conformal}
	\end{equation}
	and similarly for the self-energies.
	Fourier transforming these propagators to frequency space yields rather lengthy expressions, which we defer to Appendix~\ref{app:conf_symmetry}, Eqs.~\eqref{eq:GR_finiteT}--\eqref{eq:deltaPiR_finiteT}.
	Figure~\ref{fig:conformal_plot} compares these conformal expressions with the numerical solution obtained by directly solving the saddle-point equations. The agreement is good in the low-frequency regime, where the conformal approximation accurately captures the numerical results. Deviations become progressively more pronounced as the frequency increases, reflecting the expected breakdown of the conformal description away from the infrared limit.
	
	From the expression for the boson self-energy in frequency space, 
	\begin{equation}
		\delta \Pi^R(\omega) = \frac{(2 \pi )^{4 \Delta_\mathrm{F} -2} \beta ^{1-4 \Delta_\mathrm{F} }}{\tilde{b} \Gamma (4 \Delta_\mathrm{F} -1)\sin (2 \pi  \Delta_\mathrm{F} )} \sin \left(2 \pi  \Delta_\mathrm{F} +\frac{i \beta  \omega }{2}\right) \Gamma \left(2 \Delta_\mathrm{F} -\frac{i \beta  \omega }{2 \pi }\right) \Gamma \left(2 \Delta_\mathrm{F} +\frac{i \beta  \omega }{2 \pi }\right)\,, \label{eq:PiR_omega_finite_T}
	\end{equation}
	we can derive the renormalized mass at finite temperature: 
	\begin{equation}\label{eq:boson_mass_finite_T}
		\omega_r^2 = \omega_0 \, \delta \Pi^R(0) = - \omega_0^2 \, c_1^2 \, \frac{N}{M}\frac{\tan (2 \pi  \Delta_\mathrm{F} ) \Gamma (2 \Delta_\mathrm{F} )^2}{\Gamma (4 \Delta_\mathrm{F})} \left( \frac{g^2}{2 \pi T \omega_0} \right)^{1-4 \Delta_\mathrm{F} } \, .
	\end{equation}
	Again, while the $g$- and $T$-dependence is the same as in Ref.~\cite{Esterlis_2019}, the numerical coefficient is different.
	Since $\omega_r\neq 0$, this expression provides a first quantitative estimate of how far the conformal regime can extend in the phase diagram. At finite temperature, the boson is no longer gapless: thermal fluctuations generate an effective gap $\omega_r$, which cuts off the infrared conformal behavior. The conformal regime therefore requires this thermal mass to remain small compared to the interaction scale. From~\cref{eq:boson_mass_finite_T}, this translates parametrically into the condition
	$T \ll g^2$. When this condition is violated, the bosonic gap becomes large and the system is driven away from the conformal regime, as indicated in Fig.~\ref{fig:phase-diagram}.
	There is, however, a distinct regime that emerges when the bosonic gap is small, which we discuss in the next section.
	
	\subsection{SYK$_2$ (impurity) regime}
	\label{sec:impurity_phase}
	
\begin{figure}[htbp]
		\centering
		\includegraphics[width=0.49\linewidth]{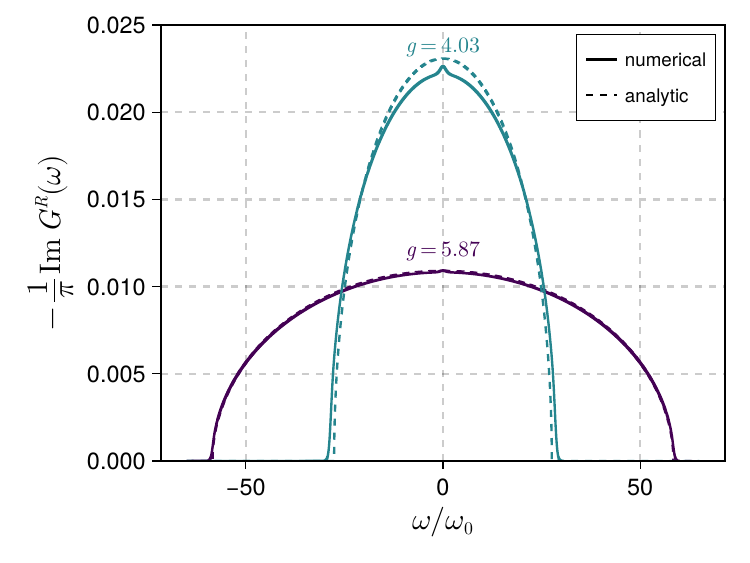}
		\includegraphics[width=0.49\linewidth]{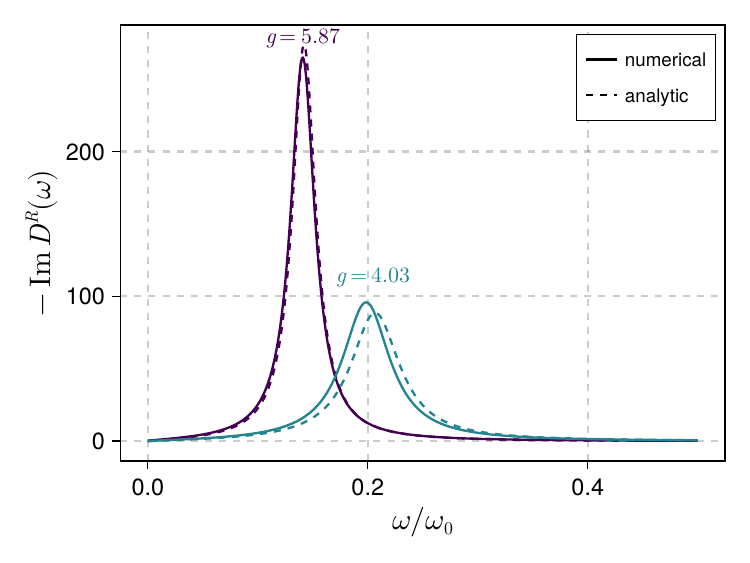}
		\caption{Left Panel: Comparison of the numerical solution of the fermionic spectral function with the analytic expressions given in Eqs.~\eqref{eq:impurity_green}--\eqref{eq:impurity_constants} for $T = 0.5$ (in units of $\omega_0$) and $N=2M$.
			Right Panel: Same comparison for the bosonic propagator. The deeper we are in the window $g^{-2} \ll T \ll g^{2}$, the sharper the boson peak becomes, the more the fermionic spectral density approaches a broad semicircle, and the more the agreement between the analytic and numerical solutions improves.}
		\label{fig:impurity_plot}
	\end{figure}

	As discussed in Ref.~\cite{Esterlis_2019} and summarized in Fig.~\ref{fig:phase-diagram}, at large coupling and finite temperature a window opens up in the YSYK phase diagram, parametrically defined by $ g^{-2} \ll T \ll g^2$, where the model transitions into a qualitatively different solution. Within these approximations, we can self-consistently find a solution where the boson behaves as a free quasi-particle with a small renormalized mass $\omega_r$, while the fermion spectrum becomes very broad:
	\begin{equation}
		D^R(\omega) = \frac{\omega_0}{(\omega+i \gamma)^2-\omega_r^2} \, , \qquad  G^R(\omega) = 2 \, \frac{\omega - \text{sign}(\omega) \sqrt{\omega^2-\Omega^2_0}}{\Omega_0^2} \, .
		\label{eq:impurity_green}
	\end{equation}
	Here, $\gamma$ denotes an interaction-induced line-width of the boson, while $\Omega_0$ sets the support of the fermionic spectral density. These parameters can be determined self-consistently from Eqs.~\eqref{eq:fluctuation_dissipation_G}--\eqref{eq:PiR_t}, yielding: 
	\begin{equation}
		\Omega_0 = \frac{8  g^2 N}{3 \pi M \omega_0} \, , \qquad \gamma = \frac{3 \omega_0^2}{4  \Omega_0} \, , \qquad \omega_r^2 = T \omega_0 \left(\frac{3 \pi M \omega_0}{4  g N}\right)^2 \, , \label{eq:impurity_constants}
	\end{equation}
	as derived in Appendix~\ref{app:impurity}. The expressions for $\Omega_0$ and $\omega_r$ agree with those originally obtained in Ref.~\cite{Esterlis_2019} in the imaginary-time formalism, while the real-time method also gives easy access to the boson broadening $\gamma$. Notably, $\gamma$ is inversely proportional to $\Omega_0$, indicating that the broader the fermionic spectrum becomes, the sharper the bosonic mode is. A direct comparison of these analytical predictions and the numerical solution, obtained by solving the saddle point equations \eqref{eq:Schwinger_Dyson}--\eqref{eq:PiR_t}, is shown in Fig.~\ref{fig:impurity_plot}, where we find excellent agreement up to a few small details, in particular deep inside the window $g^{-2} \ll T \ll g^{2}$.
	
	The renormalized bosonic gap $\omega_r$ in~\cref{eq:impurity_constants} is pushed toward criticality as the coupling increases, producing again a parametrically small gap. This makes it difficult to distinguish this regime from the conformal one based solely on the closing of the gap. For this reason, in Sec.~\ref{sec:observables} we introduce specific observables that are suited to diagnose the crossover between the conformal and the SYK$_2$ regime.
	
	The closing of the gap again signals the presence of an effective conformal symmetry, but in a much more trivial sense than in the previous section: in the limit $\omega_r \to 0$, the bosonic lesser propagator
	\begin{equation}
		D^<(t) = -\frac{i \omega_0 }{\beta  \omega_r^2} \cos (\omega_r t) \, ,
	\end{equation}
	dominates the fermionic self-energy in~\cref{eq:SigmaR_t}, while becoming approximately time-independent. As a consequence, one finds  $\Sigma^R(t) \propto G^R(t)$,  which is the characteristic structure of the quadratic SYK$_2$ model~\cite{Maldacena:2016hyu, Winer2020, Legramandi2024}. This suggests that in this regime the boson effectively drops out of the interaction, and the Hamiltonian in~\cref{eq:YSYK_H} reduces to an effective Fermi-liquid theory with interaction given by a quadratic fermionic vertex with random hopping, as also argued in Ref.~\cite{Solis_2026} for finite system sizes of the YSYK model.
	This provides a different interpretation from Ref.~\cite{Esterlis_2019} of the semicircular fermionic spectral density,
	\begin{equation}
		\rho(\omega) = - \frac{1}{\pi} \text{Im} \, G^R(\omega) = \frac{2}{\pi} \frac{\sqrt{\Omega_0^2-\omega^2}}{\Omega_0^2} , \label{eq:rho_impurity}
	\end{equation}
	as arising from the average over a random-matrix distribution of fermionic masses, rather than from dynamic interaction with the bosonic field.
	
	\subsection{The criticality budget}
	\label{sec:criticality_budget}
	
	Both regimes discussed so far rely on the interaction driving the boson close to criticality, $\omega_r^2 \ll 1$. In both cases, we found this mechanism to operate only as long as $T \ll g^2/\omega_0$. 
	In this section, we demonstrate that the underlying mechanism cannot extend to $T \gg g^2/\omega_0$: a fundamental bound on the bosonic self-energy $|\Pi^R(0)|$ (due to Kramers--Kronig relations) prevents the self-criticality condition~\eqref{eq:def_self_criticality} from being satisfied once $T\omega_0 /g^2$ becomes too large.

	Starting from \cref{eq:self_fix_b}, we have:
	\begin{align}
		\Pi^R(0)
		&= -i g^2 \frac{N}{M} \int \frac{\dd \omega}{2\pi}\,
		G^<(\omega)\left[G^A(\omega) + G^R(\omega)\right]
		= 2 g^2 \frac{N}{M}\int \dd\omega\, \rho(\omega)\, n_\mathrm{F}(\omega)\,
		\Re G^R(\omega) \nonumber \\[2pt]
		&= g^2\frac{N}{M}\iint \dd\omega\,\dd\omega'\,
		\rho(\omega)\rho(\omega')\,
		\frac{n_\mathrm{F}(\omega)-n_\mathrm{F}(\omega')}{\omega-\omega'}
		\;\le\; 0 \, ,
		\label{eq:PiR0_spectral}
	\end{align}
	where in the first step we introduced the fermionic spectral function $\rho = - \Im G^R / \pi$ . In the last step, we used the Kramers--Kronig relations 
	\begin{equation}
		\Re G^R(\omega)=\mathcal{P} \int  \dd\omega ' \, 
		\frac{\rho(\omega')}{\omega-\omega'} \, ,
	\end{equation}
	and we symmetrized the integral with respect to $\omega \leftrightarrow \omega'$. The self-energy $\Pi^R(0)$ expressed as in~\cref{eq:PiR0_spectral} is manifestly non-positive because $n_\mathrm{F}$ is
	monotonically decreasing, and therefore can only soften the boson, which is what makes self-criticality possible in the first place.  
	
	The magnitude of $\Pi^R(0)$, on the other hand, is bounded from above.  Apart from the fermionic spectral function, the integrand in~\cref{eq:PiR0_spectral} is given by the difference quotient of the fermion occupancy, which cannot exceed the maximal value of its derivative, $|n_\mathrm{F}'|=|\mathrm{d}n_\mathrm{F}/\mathrm{d}\omega|$. So, 
	\begin{equation}
		\left| \frac{n_\mathrm{F}(\omega)-n_\mathrm{F}(\omega')}{\omega-\omega'}
		\right| \le \sup_\omega |n_\mathrm{F}'(\omega)| = |n_\mathrm{F}'(0)|
		= \frac{\beta}{4} \, .
		\label{eq:nF_der_dis}
	\end{equation}
	Since the spectral density is normalized to one, we get
	\begin{equation}
		|\Pi^R(0)| \le \frac{g^2}{4 T} \frac{N}{M} \equiv \Pi_\mathrm{max} \, .
		\label{eq:criticaly_budget}
	\end{equation}
	The capacity of the system to become self-critical is thus bounded from above by a ``criticality budget'' $\Pi_\mathrm{max}$, saturated in the limit in which the fermion is completely free, $\rho(\omega) = \delta(\omega)$. 
	This bound reflects a competition between bosonic softening and fermionic broadening. On the one hand, the YSYK interaction drives the boson towards criticality, tending to spend the budget by softening it. On the other hand, spending the entire budget requires the fermion to remain nearly bare. As the boson softens, however, the fermionic self-energy $\Sigma^R$ broadens the fermion spectrum, thereby limiting how much of the budget can be spent.
	
	Three remarks are in order. First, the bound follows from the normalization of $\rho$ and from the Fermi function alone: it does not depend on the specific solution of the saddle-point equations and, as we will see in Sec.~\ref{sec:dissipation}, neither does it depend on the modifications of the spectrum due to the coupling to an external bath (dissipation does contribute, however, to saturating the criticality budget). 
	
	Second, the self-criticality condition~\cref{eq:def_self_criticality} requires $|\Pi^R(0)| = \omega_0$, which is possible only if
	\begin{equation}
		T \le \frac{N}{4M} \frac{g^2}{\omega_0} \, .
		\label{eq:conformal_window_budget}
	\end{equation}
	This inequality captures the parametric condition $T\ll g^2$ that gives rise to the conformal and SYK$_2$ regimes. 
	
	Lastly, although~\cref{eq:criticaly_budget} is a statement about $\omega=0$, $\Pi_\mathrm{max}$ controls the imaginary part of $\Pi^R$ at every frequency. 
	Indeed, starting from the Fourier transformed version of Eq.~\eqref{eq:PiR_t} and following similar steps as before, we have
	\begin{align}
		\Pi^R(\omega)
		&= -i g^2 \frac{N}{M} \int \frac{\dd \omega '}{2\pi}\,
		G^<(\omega')\left[G^A(\omega'-\omega) + G^R(\omega'+\omega)\right]  \\[2pt]
		&= g^2\frac{N}{M}\iint \dd\omega' \,\dd\omega'' \,
		\rho(\omega')\rho(\omega'')\, n_\mathrm{F}(\omega') \left[ \frac{1}{\omega ' - \omega '' - \omega - i 0^+}+ \frac{1}{\omega ' - \omega '' + \omega + i 0^+}\right] \nonumber \\[2pt]
		&= g^2\frac{N}{M}\iint \dd\omega' \,\dd\omega'' \,
		\rho(\omega')\rho(\omega'') \frac{n_\mathrm{F}(\omega')-n_\mathrm{F}(\omega'')}{\omega' - \omega''} \frac{(\omega ' - \omega '')^2}{(\omega ' - \omega '')^2 - (\omega + i 0^+)^2}\, , \nonumber
	\end{align}
	from which,  taking the imaginary part and using \eqref{eq:nF_der_dis}, we have
	\begin{equation}
		\big|\mathrm{Im}\,\Pi^R(\omega)\big| \le \pi\,\Pi_\mathrm{max}\,|\omega|\, \Phi(\omega)\,, \qquad
		\Phi(\omega) = \int \dd\omega' \,\rho(\omega')\,\rho(\omega'+\omega)\,,
	\end{equation}
	where $\Phi(\omega)$ is the autocorrelation function of $\rho$ and $\int \Phi(\omega) \dd \omega =1$.
	
	In the next section, we focus on the regimes that emerge at $T \gg g^2$, which are not captured by the analyses in sections \ref{subsec:conf_phase} and \ref{sec:impurity_phase}.
	
	\subsection{Bare boson regimes}
	\label{sec:high_t_regimes}
	
	In the limit $T \gg g^2$, the criticality budget $\Pi_{\mathrm{max}}$ becomes very small, and significantly suppresses the bosonic self-energy. 
	The condition $T \gg g^2$ can be realized in two distinct ways: either at finite temperature and sufficiently weak coupling, $g^2 \ll 1$, or at finite coupling and sufficiently high temperature. In the former case, the small coupling also suppresses the fermionic self-energy, resulting in an effectively decoupled theory of bare bosons and fermions, as shown in Fig.~\ref{fig:phase-diagram}. The latter case is more interesting: as we show below, it gives rise to a regime where the fermion is driven by the bare boson, which has not been considered in previous analyses~\cite{Esterlis_2019}.
	
	In the limit where the temperature is the largest scale in the problem, the bosonic self-energy is suppressed can be derived from a simple symmetry argument: the Fermi distribution in~\cref{eq:fluctuation_dissipation_G} can
	be approximated as $n_\mathrm{F}(\omega) = 1/2$, the fluctuation--dissipation
	relation simplifies to $G^<(\omega) = -i \Im G^R(\omega)$, and
	\begin{equation}
		\label{eq:PiR_highT}
		\begin{split}
			\Pi^R(\omega) &= -i g^2 \frac{N}{M} \int \frac{\dd \omega'}{2\pi} \left[G^<(\omega') G^A(\omega'-\omega) + G^R(\omega') G^<(\omega'-\omega) )\right] \\
			&= -i g^2 \frac{N}{M} \int \frac{\dd \omega'}{2\pi} \Im G^R(\omega') \left[\Re G^R(\omega'-\omega)+\Re G^R(\omega'+\omega)\right] = 0 \, ,
		\end{split}
	\end{equation}
	where the vanishing of the integral follows from parity: $\Im G^R$ is an even function, while  $\Re G^R$ is odd, making the integrand odd overall. 
	The boson is therefore essentially undressed, $D^R \sim D^R_0$, in contrast to the SYK$_2$ regime of Sec.~\ref{sec:impurity_phase}, where the boson is also a sharp, but strongly renormalized, mode.

	The high-temperature limit enhances the bosonic lesser propagator. Using the free form of the bosonic spectral function, one finds
	\begin{equation}
		D^<(\omega) = - i \pi \left[ \delta(\omega-\omega_0)-\delta(\omega+\omega_0) \right] n_\mathrm{b}(\omega) \sim - i \frac{\pi}{\beta \omega_0} \left[ \delta(\omega-\omega_0)+\delta(\omega+\omega_0) \right] \, ,
	\end{equation}
	which dominates the right-hand side of the fermionic self-energy in~\cref{eq:SigmaR_t}:
	\begin{equation}
		\Sigma^R(\omega) \sim i g^2 \int \frac{d \omega'}{2 \pi} G^R(\omega') D^<(\omega-\omega') = \frac{g^2}{2 \beta \omega_0} \left[G^R(\omega+\omega_0)+G^R(\omega-\omega_0)\right] \, .
		\label{eq:driven_SYK2_omega}
	\end{equation}
	Using the Schwinger-Dyson equation for the fermion we can express the equation above as a self-consistency equation for the fermion only
	\begin{equation}
		\frac{g^2 T}{2 \omega_0} \big(G^R(\omega+\omega_0)+G^R(\omega-\omega_0)\big) = \omega - G^R(\omega)^{-1} \, .
		\label{eq:new_regime}
	\end{equation}
	Equation~\eqref{eq:driven_SYK2_omega} becomes more insightful in the time domain, where it takes the structure of a fermion coupled to a monochromatic drive at frequency $\omega_0$
	\begin{equation}
		\Sigma^R(t) = \frac{g^2}{\beta \omega_0} G^R(t) \cos (\omega_0 t ) \, . \label{eq:driven_SYK2}
	\end{equation}
	Indeed, we can map this saddle-point equation back to an effective, time-dependent, quadratic Hamiltonian
	\begin{equation}
		H_\mathrm{eff} = \frac{1}{\sqrt{N}}\sum_{ij} J_{ij}(t) c_i^\dag c_j \, ,
		\label{eq:effective_H}
	\end{equation}
	where, crucially, the variance of $J_{ij}(t)$ must also preserve time-translation invariance. 
	This condition is satisfied by choosing
	\begin{equation}
		J_{ij}(t) = J_{ij}^1 \sin (\omega_0 t) + J_{ij}^2 \cos (\omega_0 t) \, ,
		\label{eq:effective_J}
	\end{equation}
	with $J^{1,2}$ independent GUE matrices with vanishing mean and variance $\langle |J^{1,2}_{ij}|^2 \rangle = \frac{g^2}{\beta \omega_0}$. Indeed, with these definitions
	\begin{equation}
		\langle \overline{J_{ij}(t) J_{kl}(t')} \rangle = \delta_{il} \delta_{jk} \frac{g^2}{\beta \omega_0} \cos(\omega_0 (t-t')) \, ,
	\end{equation}
	preserves time-translation symmetry and gives the correct functional dependence to reproduce Eq.~\eqref{eq:driven_SYK2}, as explicitly shown in Appendix~\ref{app:high_T_model}.  The same effective Hamiltonian \cref{eq:effective_H} can also be obtained by writing~\cref{eq:YSYK_H} in the rotating frame and replacing the bosons by their typical fluctuations, which at high temperature are purely thermal, as also detailed in Appendix~\ref{app:high_T_model}. The interpretation of~\cref{eq:effective_H} is then transparent: in this regime the boson has ceased to act as a dynamical degree of freedom and enters only as a classical monochromatic drive at the bare frequency $\omega_0$ with an effective coupling fixed by $g^2 T / 2 \omega_0$. As a consequence, the fermions are governed by a Floquet-like problem~\cite{Sambe:1973cnm,Bukov_2015}. 
	
	Since the interaction strength is determined by the effective coupling $g^2 T / 2 \omega_0$, every crossover in the bare boson regime is a line $T \propto g^{-2}$ in the phase diagram. As shown in Fig.~\ref{fig:phase-diagram}, we can distinguish three main regions: 
	
	\paragraph{Free Theory:} If, in addition to the high-temperature limit, the coupling is also small,  $g^{-2} \gg T$, then the self-energy is parametrically suppressed, $\Sigma^R(\omega) \sim 0$, and the fermion is approximately undressed. In this limit the boson and the fermion are approximately decoupled and the theory becomes free. 
	
	\begin{figure}[htbp]
		\centering
		\hspace*{-0.2cm}\includegraphics[width=0.5105\linewidth]{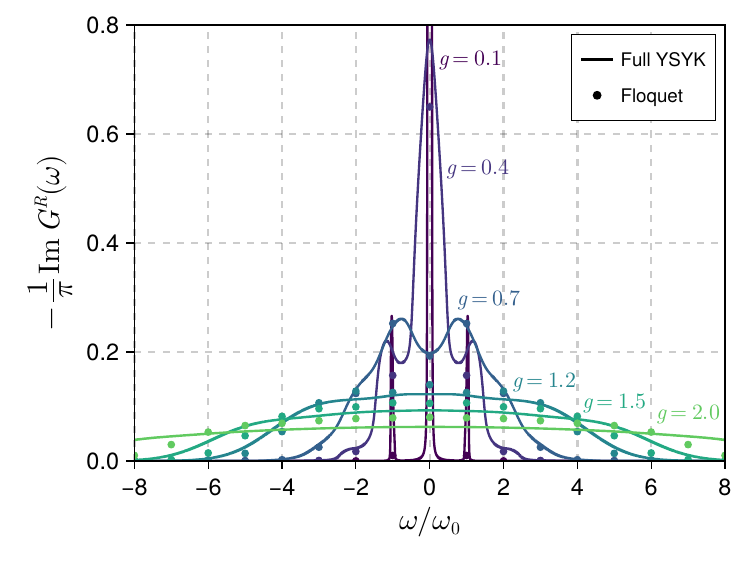}
		\hspace*{-0.3cm}\includegraphics[width=0.5105\linewidth]{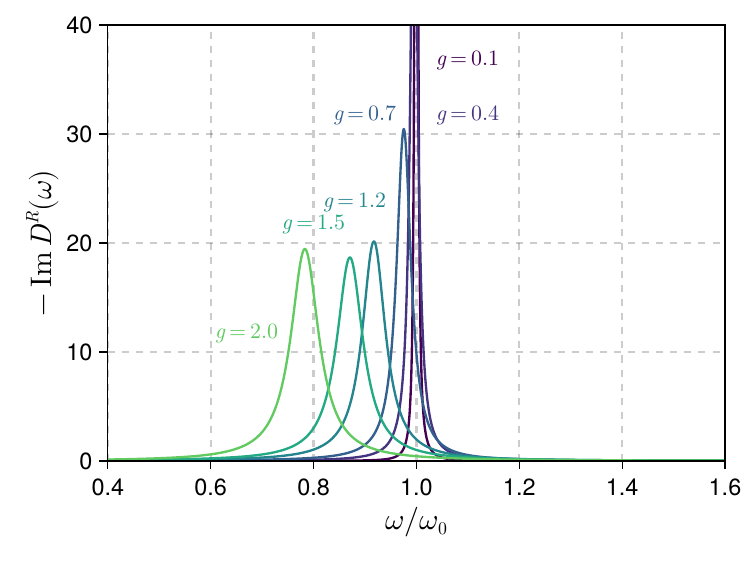}
		\caption{Left Panel: fermionic spectral function $-\frac{1}{\pi}\Im G^R(\omega)$ at temperature $T = 4$ for several values of the coupling $g$. The solid lines are the numerical solution of the full YSYK saddle-point equations. The dots indicate the solution of the simplified Floquet-like behavior in~\cref{eq:new_regime}, valid in the high-temperature limit. The agreement between the two confirms that the Floquet-like description correctly captures the salient physics of this regime. Right Panel: imaginary part of the bosonic propagator. As $g$ increases, the bosonic spectral peak broadens, and the bare frequency gets increasingly renormalized due to the proximity to the SYK$_2$ regime.}
		\label{fig:floquet_high_T}
	\end{figure}

	\paragraph{Resolved Sidebands} As $g^2 T / 2\omega_0$ increases, the fermion hybridizes more strongly with the bare boson, and its spectral density develops increasingly pronounced sidebands at $\omega=n\,\omega_0$, with $n\in\mathbb{Z}$, whose weight decreases as $(g^2 T / \omega_0)^{|n|}$ with increasing $|n|$. For small frequencies $\omega \ll \omega_0$, we can use Eq.~\eqref{eq:driven_SYK2_omega} to derive an approximate expression for the spectral density of central band: 
	\begin{equation}
		\Sigma^R(\pm \omega_0+\omega) \approx \frac{g^2 T}{2 \omega_0} G^R(\omega) \qquad \Longrightarrow \qquad G^R(\pm \omega_0+\omega) = \frac{1}{\pm \omega_0 - \frac{g^2 T}{2 \omega_0} G^R(\omega)}
	\end{equation}
	where in the first equation we neglected $G^R(\pm n \, \omega_0)$ for $n>1$. We can now use Eq.~\eqref{eq:new_regime} to write an equation for $G^R(\omega)$ alone:
	\begin{equation}
		\omega \, G^R(\omega) = \frac{1+ \left( \frac{g^2 T}{2 \omega_0^2} \right)^2 G^R(\omega)^2}{1-\left( \frac{g^2 T}{2 \omega_0^2} \right)^2 G^R(\omega)^2} \, . \label{eq:central_band}
	\end{equation}
	While it is possible to invert this equation, the resulting expression is not particularly illuminating. It is more instructive instead to write $\omega=\omega(G^R)$ and determine the band edge directly. For a purely real $G^R$, $\omega$ admits a solution up to a minimal value; below this value,~\cref{eq:central_band} can be solved just for complex $G^R$, indicating that $\Im G^R(\omega) \neq 0$ and the central band is opening. The band edge is therefore determined by:
	\begin{equation}
		\frac{\partial \omega}{\partial G^R} = 0 \qquad \Longrightarrow \qquad \omega_\mathrm{edge} = \sqrt{\frac{11+5\sqrt{5}}{8}} \, \,\frac{g^2 T}{\omega_0^2} \, .
	\end{equation}
	Taking the imaginary part of~\cref{eq:new_regime}, one can see that a finite support around $\omega =0$ induces an identical finite support at $\omega = \pm \omega_0$, reproducing the periodic structure of the bands anticipated before. Since the spectral density is also even, we have that the gap between the sidebands closes when $\omega_\mathrm{edge} = \omega_0 / 2$, i.e., for
	\begin{equation}
		T \approx 0.30 \, \frac{\omega_0^3}{g^2} \, , \label{eq:resolve_unresolved_transition}
	\end{equation}
	giving a rough estimate of the transition from the regime with resolved bands to the unresolved one, as shown in Fig.~\ref{fig:phase-diagram}.
	
	\paragraph{Unresolved Sidebands} At values of $T g^2$ larger than~\cref{eq:resolve_unresolved_transition}, the spectral bands merge and the density becomes progressively more flat at larger couplings. In this regime, we can approximate $G^R(\omega \pm \omega_0) \sim G^R(\omega)$, and~\cref{eq:new_regime} becomes a second-order equation for the retarded propagator, which leads to the same solution we saw in the SYK$_2$ regime~\cref{eq:impurity_green} with $\Omega_0^2 = 4g^2 T / \omega_0$. Notice that the two values of $\Omega_0$ become commensurate at $T \sim g^2$, i.e., exactly at the crossover predicted in section~\ref{sec:criticality_budget}, where the boson mass starts being renormalized. Lastly, we stress that, even if the semicircle spectral density~\cref{eq:rho_impurity} is common in both regimes, its origin here is quite different, as it is due to the envelope of broad sidebands at integer multiple of $\omega_0$, and does not require the boson mass to be drastically renormalized as in the SYK$_2$ regime.

	The features derived for each regime can now be seen explicitly by numerically solving the YSYK saddle-point equations~\eqref{eq:fluctuation_dissipation_G}--\eqref{eq:PiR_t} at $T = 4\,\omega_0$. At weak coupling, the left panel of Fig.~\ref{fig:floquet_high_T} shows a narrow central band accompanied by sidebands at $\omega = \pm\omega_0$, separated by gaps centered at $\omega = \pm\omega_0/2$. As $g$ increases, the bands broaden until the gaps close and the sidebands merge into a single broad, progressively flatter distribution, as expected in the unresolved regime. Throughout this evolution the bosonic spectral function (right panel) remains close to that of the bare mode, and deviates appreciably only on approach to the SYK$_2$ regime.
	The full numerical solution is in good agreement with the self-consistent solution of~\cref{eq:new_regime} for discrete multiples of $\omega_0$~\cite{tsuji_2008}.The residual deviations grow with $g$, reflecting the proximity of the SYK$_2$ regime, where the boson acquires a width and a renormalized frequency and~\cref{eq:new_regime} ceases to hold.

	In this section we have identified the regimes of the YSYK model in real time: the conformal and SYK$_2$ regimes, in which the boson is strongly renormalized, and the bare-boson regimes, in which the boson remains essentially free and the fermion ranges from essentially free to driven, with resolved or unresolved sidebands. Each regime leaves a distinct imprint on the two-point functions, but these imprints are not directly suited to locating the boundaries between regimes. In the next section we introduce observables, built from the bosonic and fermionic spectral functions, that mark the crossover lines between these regimes.

	\section{Measurable phase signatures}
	\label{sec:observables}
	
	Of particular interest in the YSYK model is the strongly coupled, holographic phase~\cite{Schmalian:2022web,inkof_quantum_2022} studied in Sec.~\ref{subsec:conf_phase}, characterized by the conformal behavior $|\omega|^{-2\Delta}$ of the two-point functions. The conformal dimension $\Delta$ provides a direct characterization of this regime, but extracting it experimentally requires resolving a sufficiently broad frequency window over which a clean power law is observed. This window lies at low frequencies, such that its observation generally requires long observation times. The situation is further complicated at finite temperature, where the simple zero-temperature power laws are replaced by the finite-temperature forms in Eqs.~\eqref{eq:GR_finiteT}--\eqref{eq:DR_finiteT}, and can be further obscured by crossovers with other regimes. Moreover, as we will discuss in Sec.~\ref{sec:dissipation}, dissipation, an unavoidable ingredient of experimental realizations, places an even stricter upper bound on the conformal window and deforms the Green's functions further away from their ideal conformal form. These limitations motivate the search for more robust experimental diagnostics of the different regimes, and in particular for signatures of holographic behavior that do not rely on directly resolving the conformal scaling over an extended frequency range.
	
	A natural starting point for this search is provided by the two-point functions themselves, which are experimentally accessible across a broad range of platforms. In solid-state systems, the local fermionic spectral function $\rho(\omega)=-\operatorname{Im}G^R(\omega)/\pi$ can be accessed by scanning tunneling spectroscopy, where the differential conductance is proportional to the local density of states~\cite{Tersoff1985,Fischer2007}, while angle-resolved photoemission spectroscopy accesses its occupied part, $\rho(\omega)n_{\rm F}(\omega)$~\cite{Damascelli2003}, i.e., the lesser propagator through Eq.~\eqref{eq:fluctuation_dissipation_G}. In cold-atom platforms, analogous information is accessed through radio-frequency spectroscopy~\cite{Chin2004}, which drives transitions to an initially empty internal state and, when momentum resolved, provides access to the occupied single-particle spectral function in direct analogy to photoemission~\cite{Stewart2008}. The retarded bosonic propagator, on the other hand, is directly related to the corresponding linear susceptibility and can be accessed through response measurements or, via its associated fluctuations, noise spectroscopy~\cite{Clerk2010}. In cavity-QED platforms, which are particularly promising for scaling implementation of the SYK and YSYK models \cite{cavity_proposal,Baumgartner:2024ysk,Baumgartner:2025uue,Solis_2026}, the cavity itself provides a direct output channel: through input--output relations, the field leaking from the cavity carries information about the intracavity dynamics and its fluctuations~\cite{GardinerCollett1985,Brennecke2013}. In Appendix~\ref{app:measurement}, we explicitly show how the retarded bosonic Green's function can be obtained in general cavity-QED experimental setups through homodyne detection.
	
	As stated earlier, the aim of this work is here not to mimic any of the above experimental settings exactly, but rather to provide general insights. 
	The goal is therefore to identify, starting from these experimentally accessible Green's functions, a set of general observables that can both discriminate between the different dynamical regimes and provide signatures of the emergence of holographic behavior without requiring a direct measurement of the conformal dimension. In Secs.~\ref{sec:markers} and~\ref{sec:QFI}, we introduce three such markers extracted from the retarded bosonic propagator: the bosonic gap $\alpha$, the curvature $\xi$ of the inverse propagator, and the quantum Fisher information of the bosonic quadrature. For this last one, we further derive upper bounds from sum rules obeyed by $D^R$, which we extend to the open system in Sec.~\ref{sec:QFI_diss}.
	Since the boson is essentially undressed in the bare-boson regimes of Sec.~\ref{sec:high_t_regimes}, we further introduce a marker in Sec.~\ref{sec:fermion_weight} based on the fermionic spectral weight, which distinguishes the free-fermion regime from the driven fermionic regimes, both with resolved and unresolved sidebands.
	
	\subsection{Bosonic gap and curvature}
	\label{sec:markers}
	
	The main mechanism that leads to the emergence of the conformal phase in the YSYK model is the boson being pushed to criticality by the interaction with the fermion. The two-point function of the real quadrature of a bosonic field and the associated self-energy always have imaginary parts that are odd functions in frequency space. That means, having an inverse propagator with vanishing real part is enough to trigger the static self-criticality. For this reason, we define the bosonic gap, or the inverse static susceptibility, as
	\begin{equation}
		\alpha = - D^R(0)^{-1} = \omega_0+ \Pi^R(0) \, .
		\label{eq:gapdef}
	\end{equation}
	The gap $\alpha$ is the inverse of the static susceptibility of the $x$ quadrature, and its decrease tracks the softening of the bosonic mode. The same softening, together with the growth of the corresponding fluctuations, is what cavity-QED experiments monitor when approaching the Dicke transition~\cite{Mottl2012,Brennecke2013}. At zero temperature, $\alpha = 0$ by virtue of~\cref{eq:def_self_criticality}. 
	At finite but sufficiently low temperatures, the gap is small but finite and can be expressed in terms of the renormalized boson mass
	\begin{equation}
		\alpha = \omega_r^2 /\omega_0 \, ,
		\label{eq:gapmass}
	\end{equation}
	derived in~\cref{eq:boson_mass_finite_T} for the conformal regime and~\cref{eq:impurity_constants} for the SYK$_2$ (impurity) one.
	As discussed in Sec.~\ref{sec:criticality_budget}, $\Pi^R(0)$ is always non-positive and therefore $\alpha$ is always smaller than $\omega_0$.
	As $T$ increases, the criticality budget $\Pi_\mathrm{max}$ introduced in~\cref{eq:criticaly_budget} decreases, leaving room for a larger renormalized boson mass, which approaches the free value $\alpha=\omega_0$ in the bare-boson regimes. As we show below, this value is reached only asymptotically, so that the boundary separating the conformal and SYK$_2$ regimes from the bare-boson ones is a crossover, which we locate at the end of this section.
	
\begin{figure}[t]
		\centering
		\includegraphics[width=0.6\textwidth]{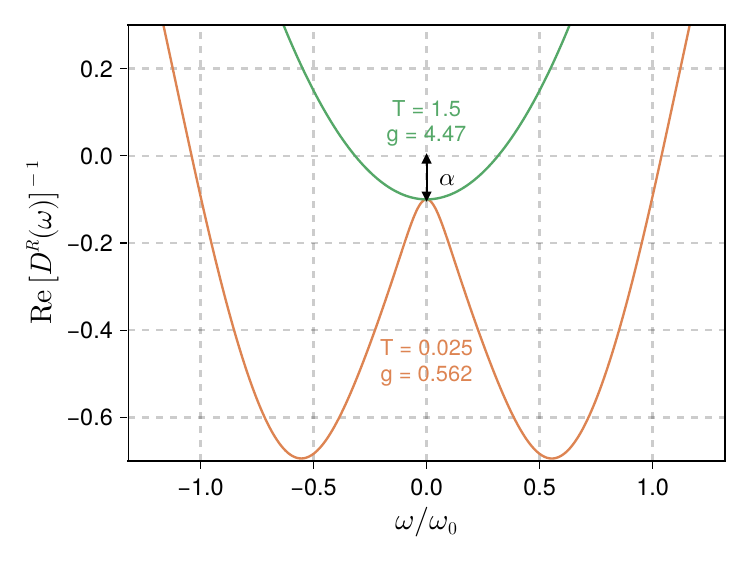}
		\caption{Real part of the inverse bosonic retarded propagator, $\operatorname{Re}[D^R(\omega)]^{-1}$, at two representative points of the phase diagram, illustrating the two markers defined in Eqs.~\eqref{eq:gapdef} and~\eqref{eq:curvature}, for $\alpha=0.1$. In the SYK$_2$ regime ($T=1.5$, $g=4.47$) the boson is a renormalized but otherwise free mode and the inverse propagator is a parabola with positive second derivative in $\omega = 0$. In the conformal regime ($T=0.025$, $g=0.562$) self-criticality generates a sharp low-frequency feature in $\Pi^R(\omega)$, pushing the bottom of the parabola upward, and inverting the sign of the second derivative.}
		\label{fig:Gap_plot}
	\end{figure}

	Having introduced a marker that separates the bare boson regimes from the rest of the phase diagram, we next seek a quantity capable of distinguishing the conformal and SYK$_2$ regimes. A convenient diagnostic is the curvature of the inverse bosonic retarded Green’s function,
	\begin{equation}
		\xi = \partial_\omega^2 \Re[D^R(\omega)^{-1}]|_{\omega=0} \, .
		\label{eq:curvature}
	\end{equation}
	In the SYK$_2$ regime, the boson behaves as a strongly renormalized but otherwise free mode, such that $D^R(\omega)^{-1}$ retains an approximately parabolic shape with positive curvature, $\xi>0$. By contrast, in the conformal regime, self-criticality strongly modifies the low-frequency behavior of the propagator. The self-energy $\delta \Pi^R(\omega)$ generates a sharp low-frequency feature that pushes the bottom of the otherwise parabolic inverse propagator upward, producing a negative curvature, $\xi<0$, as one can also explicitly check by taking the second derivative of the expression in~\cref{eq:PiR_omega_finite_T}. This behavior is illustrated in Fig.~\ref{fig:Gap_plot}.
	The sign of $\xi$ therefore provides a simple and robust criterion to distinguish the small-$\alpha$ phases: a negative second derivative is typical of the conformal regime, whereas a positive second derivative indicates the SYK$_2$ regime.

	The gap $\alpha$ also plays an important role in the numerical manifestation of the self-critical behavior when solving the saddle-point equations. The real-time approach requires integrating sharply peaked functions over a continuous frequency interval. Discretization therefore introduces numerical noise that grows as these functions approach a delta-function form. This is particularly problematic near criticality, where the self-consistent iterative scheme becomes unstable. The dominant instability arises from small shifts in the position of the delta-like feature near $\omega \sim 0$ in the retarded boson propagator. Its height grows inversely with its distance from $\omega=0$; consequently, even small fluctuations in its position produce large changes in its amplitude and hence in convolutions involving the propagator. To suppress this instability, we seek converged solutions at fixed $\alpha$ and adjust the coupling $g$ during the iterations to maintain this value. Besides substantially reducing the number of iterations required for convergence, this stabilization scheme enables access to parameter regimes that are otherwise numerically inaccessible. Our approach is inspired by related methods for critical and near-critical systems~\cite{Pini2019}; further numerical details are provided in Appendix~\ref{app:numerics}.
	
	\subsubsection*{Constant-$\alpha$ curves}
	
\begin{figure}[ht!]
		\centering
		\includegraphics[width=0.75\textwidth]{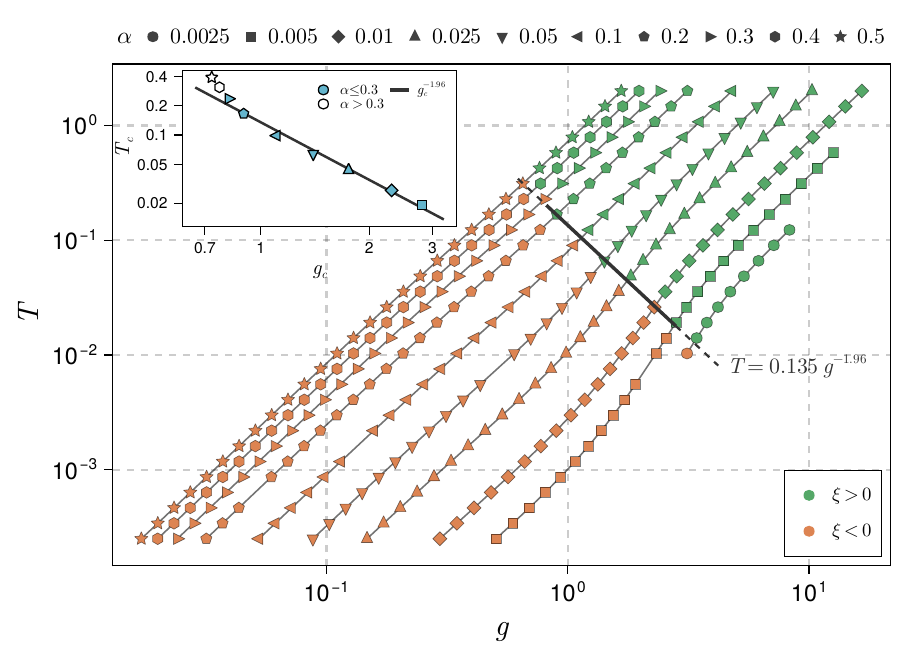}
		\caption{Constant-gap curves in the $(g,T)$-plane, on logarithmic axes, for ten values of the gap $\alpha$ between $0.0025$ and $0.5$ (in units of $\omega_0$), distinguished by marker shape. Each curve is a temperature sweep performed at fixed $\alpha$, the corresponding coupling $g$ is determined self-consistently at every temperature $T$ as described in Appendix~\ref{app:numerics}. The green markers indicate $\xi >0$, corresponding to the SYK$_2$ regime, while the orange markers with $\xi < 0$, indicate the conformal regime, respectively. The curves become straight and mutually parallel for increasing values of the gap $\alpha$, motivating the power law dependence in Eq.~\eqref{eq:constantalphafit}. We restrict our numerical analysis to values of the gap up to $\alpha=0.5$, as we are interested in the crossover from the SYK$_2$ to the conformal regime, and values too close to $\omega_0=1$ will inevitably be contaminated by the bare boson regimes. The effect of this can be seen from the inset panel that shows the critical points $(g_c, T_c)$ where the curvature $\xi = 0$. The fitting window is restricted to values of $\alpha \leq 0.3$, far enough away from the bare-boson boundary, as for larger gaps the characteristic behavior gets modified and they clearly deviate from the expected behavior. The power law fit gives a boundary of $T_c = (0.135 \pm 0.007) \, g^{-1.96 \pm 0.09}$ with $R^2=0.990$, plotted also as the boundary between the conformal and SYK$_2$ regime in the main panel. The exponent is an effective one and depends on the fitting window (Appendix~\ref{app:const_alpha_sweeps}). This numerically determines the boundary displayed in Fig.~\ref{fig:phase-diagram}.}
		\label{fig:Phase_diagram_plot}
	\end{figure}

	Because $g$, rather than $\alpha$, is the output of the iteration, a temperature sweep at fixed $\alpha$ traces out a curve in the $(g,T)$-plane along which the boson is held at a prescribed distance from criticality. Each such sweep therefore samples the phase diagram along a line of constant gap, and every point on it can be labeled by the sign of the curvature $\xi$. The resulting family of curves is shown in the main panel of Fig.~\ref{fig:Phase_diagram_plot}. Two features stand out. 
	First, the sign of the curvature changes across the same line, indicating that the constant-$\alpha$ curves cross over from the SYK$_2$ to the conformal regime. This allows us to numerically determine the boundary between both regimes by interpolating the critical point $(g_c,T_c)$ where $\xi=0$ (see Appendix~\ref{app:numerics} for more details on the procedure). The boundary between the conformal and SYK$_2$ regimes can then be inferred by fitting the resulting set of critical points, as shown in the inset of Fig.~\ref{fig:Phase_diagram_plot}. The largest values of $\alpha$ are excluded from the fit because they lie too close to the boundary with the bare-boson regimes, causing deviations from the expected scaling behavior. The extracted numerical boundary, $T = (0.135 \pm 0.007) \, g^{-1.96 \pm 0.09}$, is consistent with the analytically expected boundary of $T \sim g^{-2}$ between the conformal and SYK$_2$ regime.\footnote{The exponent is, however, an effective one: the local exponent between neighboring gaps is $-1.70$ to $-1.84$ for $\alpha \leq 0.1$ and steepens for larger gaps (Appendix~\ref{app:const_alpha_sweeps}).}
	
	Second, the constant-$\alpha$ curves are approximately parallel straight lines in the log-log plane. This is expected from the analytical solutions of Sec.~\ref{sec:phase_diagram}. Rewriting~\cref{eq:boson_mass_finite_T} and~\cref{eq:impurity_constants} in terms of $\alpha$ using~\cref{eq:gapmass}, we find
	\begin{equation}
		T_\mathrm{conf} = \frac{g^2}{2\pi\omega_0}\left[-\frac{M}{N}\,\frac{\Gamma(4\Delta_\mathrm{F})}{c_1^2\,\tan(2\pi\Delta_\mathrm{F})\,\Gamma(2\Delta_\mathrm{F})^2}\,\frac{\alpha}{\omega_0}\right]^{1/(4\Delta_\mathrm{F}-1)} , \qquad T_\text{SYK$_2$} = \frac{64}{9\pi^2} \frac{\alpha}{\omega_0^2}\, g^2 \, .
		\label{eq:T_g2_conf_imp}
	\end{equation}
	Both regimes give $T\propto g^2$ at fixed gap, but with different amplitudes. Each sweep therefore consists of two parallel branches, the low-temperature conformal one and the high-temperature SYK$_2$ one, smoothly joined at the crossover around $\xi=0$, as shown in Fig.~\ref{fig:Phase_diagram_plot}. The ratio of the two amplitudes, $T_\text{SYK$_2$}/T_\mathrm{conf}\propto\alpha^{1-1/(4\Delta_\mathrm{F}-1)}$ is larger for small $\alpha$, so that the two branches are clearly separated for small gaps, while they coincide at $\alpha\approx0.07\,\omega_0$ (for $N=2M$), after which each sweep is well described by a single straight line.\footnote{For larger gaps, Eq.~\eqref{eq:T_g2_conf_imp} predicts that the two branches separate again, with the conformal amplitude becoming the larger of the two. Although this effect is present, it is barely visible to the naked eye and is less pronounced than predicted by the analytical formulas, both of which assume $\alpha\ll\omega_0$.}
	
\begin{figure}[ht!]
		\centering
		\begin{minipage}[t]{0.49\textwidth}\centering
			\dissipationpanel{\bf a)}{}{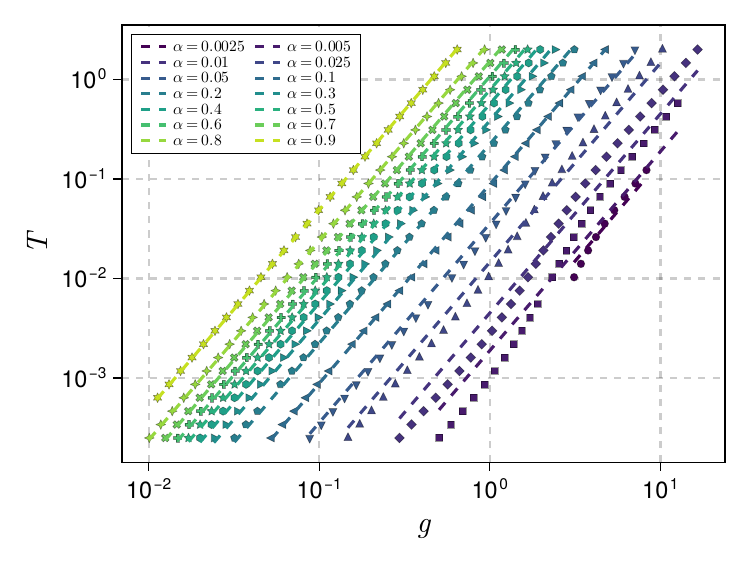}
		\end{minipage}\hfill
		\begin{minipage}[t]{0.49\textwidth}\centering
			\dissipationpanel{\bf b)}{}{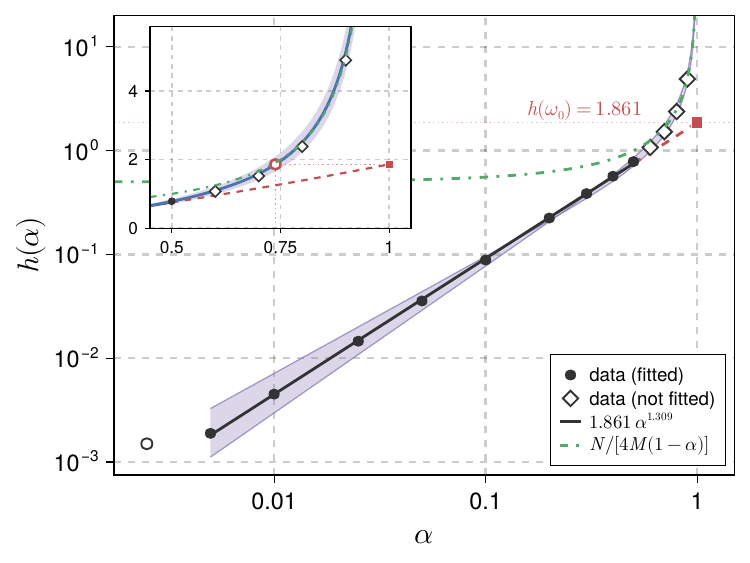}
		\end{minipage}
		\caption{Determination of the bare-boson boundary of Fig.~\ref{fig:phase-diagram}. \textbf{a)} Constant-$\alpha$ sweeps of Fig.~\ref{fig:Phase_diagram_plot}, extended up to $\alpha=0.9$ (markers), together with the fixed-slope fits $T=h(\alpha)g^2$ (dashed lines). For each sweep, $h(\alpha)$ is the geometric mean of fixed-exponent fits to its six lowest- and six highest-temperature points, corresponding to the conformal and SYK$_2$ branches respectively (a single fit for $\alpha=0.0025$). Non-converged points are omitted, including the three lowest temperatures of the $\alpha=0.9$ sweep. \textbf{b)} Amplitude $h(\alpha)$, the geometric mean of the low- and high-temperature branch amplitudes in {\bf a)}. The power law of Eq.~\eqref{eq:halphafit} (solid line) is fitted to all gaps $0.005 \leq \alpha \leq 0.5$ (filled circles). The open circle ($\alpha=0.0025$) has no low-temperature branch and shows the single fixed-exponent amplitude. The open diamonds ($\alpha \geq 0.6$, not fitted) show the departure of $h$ from the power law, while the red dashed line extrapolates the power-law fit to $\alpha=\omega_0$, where it reaches $h(\omega_0) = 1.861$ (red square). The shaded band lies between fits of $a \tan(\pi\alpha^b/2)$ to the two branch amplitudes separately. The green line is the criticality-budget bound of Eq.~\eqref{eq:budget_finite_gap}. Inset: large gaps on linear axes. The blue line $0.884 \tan(\pi \alpha^{1.131}/2)$ fits the divergent behavior for large gaps. It reaches $h(\omega_0)$ at $\alpha \approx 0.75\,\omega_0$ (open red circle), which defines the boundary of the bare-boson regimes.
		}
		\label{fig:fits}
	\end{figure}

	All sweeps can therefore be parametrized by
	\begin{equation}
		T \;=\; h(\alpha)\, \frac{g^2}{\omega_0}\, ,
		\label{eq:constantalphafit}
	\end{equation}
	with one amplitude $h(\alpha)$ per gap, obtained from a fit with fixed exponent. The low-temperature and high-temperature branch are fitted separately and $h(\alpha)$ is obtained as the geometric mean of these two amplitudes (see Appendix~\ref{app:const_alpha_sweeps} for more details).\footnote{Fitting the branches separately and weighting them equally removes the dependence of a single fit on the temperature grid and places $h(\alpha)$ on the line midway between the two branches in the log-log plane.} The resulting fit is given by the dashed lines for each gap in Fig.~\ref{fig:fits}{\bf a)}, while the resulting values of the amplitudes are shown in Fig.~\ref{fig:fits}{\bf b)}. For small and intermediate gaps, $h(\alpha)$ follows a power law, and a regression over $0.005\le\alpha\le0.5$ gives
	\begin{equation}
		h(\alpha) \;=\; A \, \alpha^B ,
		\qquad A = 1.861 \pm 0.036 \, , \qquad B = 1.309 \pm 0.007 \, .
		\label{eq:halphafit}
	\end{equation}
	The exponent lies between the linear dependence of the SYK$_2$ amplitude and the exponent $1/(4\Delta_\mathrm{F}-1)\approx1.467$ of the conformal one, as expected for an amplitude that interpolates between the two.
	
	Taken at face value, Eq.~\eqref{eq:halphafit} would reach the free value $\alpha=\omega_0$ at the finite amplitude $h(\omega_0)=A$, suggesting a line $T=A\,g^2/\omega_0$ beyond which the boson is no longer renormalized. However, for $\alpha\gtrsim0.6\,\omega_0$, the data in Fig.~\ref{fig:fits}{\bf b)} shows that $h(\alpha)$ bends away from the power-law behavior, and diverges as $\alpha\to\omega_0$ (inset of Fig.~\ref{fig:fits}{\bf b)}. This divergence can be seen as a consequence of the criticality budget. First, at any finite temperature, $\Pi^R(0)$ is strictly negative, and the free value $\alpha=\omega_0$ is attained only for $g=0$ or $T\to\infty$. Second, combining \cref{eq:gapdef} with Eq.~\eqref{eq:criticaly_budget} gives 
	\begin{equation}
		T \le \frac{N}{4M}\,\frac{g^2}{\omega_0-\alpha} \, .
		\label{eq:budget_finite_gap}
	\end{equation}
	The bound is increasingly saturated as more of the fermionic spectral weight lies within the thermal window, as occurs in the bare-boson regime of Sec.~\ref{sec:high_t_regimes}, where $T$ is the largest energy scale.
	Consequently, for $\alpha\to\omega_0$, the power-law behavior of \cref{eq:halphafit} is replaced by a singularity $h(\alpha)\simeq (N/4M)\,\omega_0/(\omega_0-\alpha)$, which indicates that in the bare-boson regimes the boson is therefore always renormalized, albeit weakly.
	
	The boundary separating the bare-boson regimes from the conformal and SYK$_2$ ones is thus a crossover from a strongly to a weakly renormalized boson, and locating it requires a convention. 
	We place it where $h(\alpha)$ leaves the small-gap power law by comparing the steep large-gap branch with the extrapolation of the smooth small-gap one. We extrapolate Eq.~\eqref{eq:halphafit} to $\alpha=\omega_0$, and identify the curve
	\begin{equation}
		T = 1.861\; \frac{g^2}{\omega_0} \, ,
		\label{eq:hightempboundary}
	\end{equation}
	as the boundary of the bare-boson regimes drawn in Fig.~\ref{fig:phase-diagram}. Along this line the bare frequency is renormalized by only about $14\%$, $\omega_r=\sqrt{\omega_0\alpha}\approx0.86\,\omega_0$, even though the fermions already use about $95\%$ of their criticality budget. The boson is weakly renormalized not because the fermions stop softening it, but because the budget itself has become small, $\Pi_{\rm max} \approx 0.27\, \omega_0$. Any other criterion formulated at fixed gap would change only the prefactor in Eq.~\eqref{eq:hightempboundary}, not the scaling $T\propto g^2$, which follows from Eq.~\eqref{eq:constantalphafit}.
	
	\subsection{Quantum Fisher information}
	\label{sec:QFI}
	
	The two markers introduced above are local properties of the bosonic propagator:
	both $\alpha$ and $\xi$ are determined by the behavior of $D^R(\omega)^{-1}$ in an
	infinitesimal neighborhood of $\omega=0$. This makes them sharp diagnostics, but requires resolving
	the inverse propagator and its curvature at the smallest accessible frequencies.
	It is therefore useful to complement them with a marker that is built out of the
	same response function but integrated over frequency---the quantum Fisher information (QFI)~\cite{Braunstein_Caves1994}. Moreover, the QFI carries a clear operational meaning in the context of quantum metrology. 
	
	\subsubsection{Background}
	
	The QFI is defined
	relative to a generator: it quantifies how distinguishable the states $\rho$ and
	$\rho_\lambda=e^{-i\lambda\hat O}\rho\,e^{+i\lambda\hat O}$ are for infinitesimal $\lambda$, and therefore how
	precisely the parameter $\lambda$ can be estimated~\cite{Paris_2009, Pezze2009, Toth_2012, Hyllus_2012}. In the present case, a suitable choice for $\hat O$ is the uniform bosonic quadrature
	\begin{equation}
		X \;=\; \frac{1}{\sqrt{M}}\sum_{m=1}^{M} x_m \, ,
		\label{eq:collectiveX}
	\end{equation}
	for which $\lambda$ is a coherent displacement of the whole bosonic field. Its Kubo
	response function reads,
	\begin{equation}
		\chi_{XX}(t-t')\;=\;-i\,\theta(t-t')\big\langle [X(t),X(t')]\big\rangle
		\;=\;\frac{1}{M}\sum_{m,n=1}^{M}\chi_{mn}(t-t') \, .
		\label{eq:chiXX}
	\end{equation}
	After disorder averaging, the off-diagonal susceptibilities vanish, while the diagonal ones reproduce, by the
	definitions in Eqs.~\eqref{eq:boson_<_def} and~\eqref{eq:boson_R_def} of the bosonic propagator,
	\begin{equation}
		\frac{1}{M}\sum_{m}\chi_{mm}(t-t')
		=-\frac{i}{M}\,\theta(t-t')\sum_{m}\big\langle [x_m(t),x_m(t')]\big\rangle
		=\theta(t-t')\big[D^{>}(t,t')-D^{<}(t,t')\big]
		=D^{R}(t,t')\, .
		\label{eq:chidiagonal}
	\end{equation}
	The uniform quadrature is thus the collective mode whose susceptibility is exactly the bosonic propagator $\chi_{XX}(\omega)=D^R(\omega)$.
	
	For a system in thermal equilibrium, the QFI associated with the generator $X$ can
	then be extracted from the dissipative part of that response \cite{Hauke_2016},
	\begin{equation}
		F_Q(T) \;=\; - \frac{4}{\pi}\int_{0}^{\infty}\!\mathrm{d}\omega\;
		\tanh\!\left(\frac{\omega}{2T}\right)\,\operatorname{Im}D^{R}(\omega) \, .
		\label{eq:QFIdef}
	\end{equation}
	
	The normalization in Eq.~\eqref{eq:collectiveX} makes $F_Q$ intensive: it is a QFI density, the QFI of the extensive quadrature $\sum_m x_m$ per bosonic mode.\footnote{The QFI density of a spin system, whose generator is a sum of bounded local operators, obeys a separability bound: for large systems, $f_Q>k$ witnesses $(k+1)$-partite entanglement~\cite{Hauke_2016,Pezze2009}. No such bound exists for the unbounded quadrature $X$ and, the susceptibilities being diagonal in the mode index, our $F_Q$ in fact coincides with the QFI of a single quadrature $x_m$. We therefore use it not as an entanglement witness, but as a measure of the genuinely quantum part of the quadrature fluctuations of one bosonic mode dressed by the $N$ fermions, equivalently of the metrological sensitivity of the steady state to displacements generated by $X$.} The relation~\eqref{eq:QFIdef} has permitted extraction of the QFI from scattering data in a variety of correlated materials~\cite{Mathew2020,Scheie2021,Laurell2021,Laurell2025,Scheie2025Tutorial,Balut2025IXS,TenHuisen2026}, and in particular from quantum critical and strange metals~\cite{Fang2025,Mazza2026,Balut2025}, which makes it a natural diagnostic for the non-Fermi-liquid physics we are after here. In the present setting, Eq.~\eqref{eq:QFIdef} is especially convenient, since $\operatorname{Im} D^R$ is exactly the quantity given by the homodyne measurement of Appendix~\ref{app:measurement}.
	
	Figure~\ref{fig:Phase_diagram_QFI} shows $F_Q$ across the phase diagram. Along every sweep of constant gap it first grows as the temperature is lowered, then reaches a maximum and drecreases again. The maxima form a ridge that runs alongside the conformal/SYK$_2$ boundary of Fig.~\ref{fig:Phase_diagram_plot}. The rest of this section explains this ridge. We first derive an upper bound on $F_Q$ from two sum rules, and then show that the SYK$_2$ boson saturates it while the conformal one does not.
	
	\begin{figure}[t]
		\centering
		\includegraphics[width=0.75\textwidth]{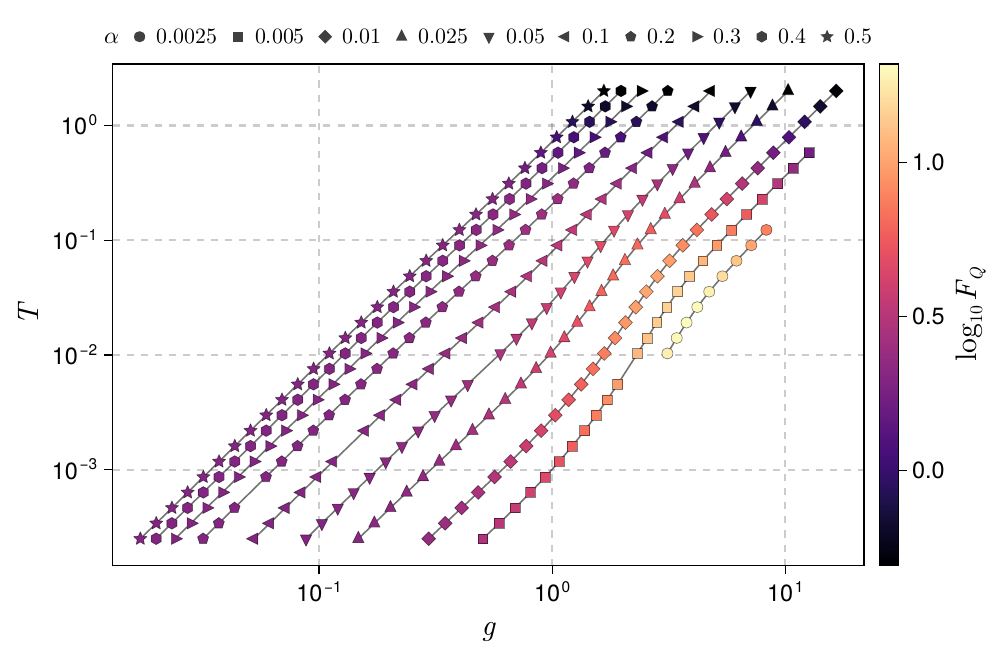}
		\caption{Quantum Fisher information $F_Q$ of the uniform bosonic quadrature, Eq.~\eqref{eq:QFIdef}, across the phase diagram. The data are the same constant-gap sweeps as in Fig.~\ref{fig:Phase_diagram_plot}, ten values of the gap $\alpha$ between $0.0025$ and $0.5$ (in units of $\omega_0$), distinguished by marker shape, with the markers colored by $\log_{10} F_Q$. Along each sweep, $F_Q$ grows as the temperature is lowered, saturating the bound of Eq.~\ref{eq:QFI_bound_sharp} while the boson is a sharp mode lying below the thermal scale, and decreases again at lower temperature as the sweep approaches the conformal regime, where the weight is spread over a power law extending above the thermal scale. The maxima form the light-colored ridge crossing the diagram, which runs closely alongside the conformal/SYK$_2$ boundary of Fig.~\ref{fig:Phase_diagram_plot}. The turnover is resolved sweep by sweep in Fig.~\ref{fig:QFI_bounds_collapse}.}
		\label{fig:Phase_diagram_QFI}
	\end{figure}

	\subsubsection{Bounds}
	
	A key structural property of~\cref{eq:QFIdef} is that the spectral weight integral is fixed by a sum rule~\cite{Thomas1925, Kuhn1925, Reiche1925, Bethe1930}, which is the bosonic analog of the fact that the fermionic spectral density is normalized. Indeed, $-\omega \Im D^R(\omega)$ is an even non-negative function which obeys
	\begin{equation}
		-\frac{2}{\pi} \int_0^\infty \dd{\omega} \, \omega \, \operatorname{Im} D^R(\omega) = i\partial_t \langle \comm{X(t)}{X(0)} \rangle |_{t=0} = \omega_0 \, .
		\label{eq:f-sum_rule}
	\end{equation}
	The sum rule follows from the bosonic commutation relation, but it can also be derived by expanding the Kramers--Kronig relation
	\begin{equation}
		\mathrm{Re}\,D^R(\omega)=\frac{1}{\pi}\mathcal{P}\!\int_{-\infty}^{\infty}\!\mathrm{d}\omega'\,
		\frac{\mathrm{Im}D^R(\omega')}{\omega'-\omega}\,
		\label{eq:KK_D^R}
	\end{equation}
	in the large-$\omega$ limit, where $D^R(\omega)\to\omega_0/\omega^2$ since the free propagator $D^R_0(\omega)$ dominates over the self-energy\footnote{The self-energy itself decays for $\omega \to \infty$, by the Kramers--Kronig relation for $\Pi^R(\omega)$.} in the Schwinger--Dyson equation~\eqref{eq:Schwinger_Dyson}.
	
	Using $\tanh x\le x$ in~\cref{eq:QFIdef}, the sum rule gives the following first bound on the quantum Fisher information~\cite{Hauke_2016}
	\begin{equation}
		F_Q \le -\frac{2}{\pi T}\int_0^\infty \dd{\omega}\,\omega\,\operatorname{Im}D^R(\omega) = \frac{\omega_0}{T} \, ,
		\label{eq:QFI_bound1}
	\end{equation}
	which is saturated whenever the entire spectral weight lies below the thermal scale, $\omega\ll T$. $D^R(\omega)$ obeys a second sum rule, which follows from the Kramers--Kronig relation~\eqref{eq:KK_D^R} at $\omega=0$ and involves the gap $\alpha$: 
	\begin{equation}
		-\frac{2}{\pi}\int_0^\infty \dd{\omega}\,\frac{\operatorname{Im}D^R(\omega)}{\omega} = -\Re D^R(0) = \frac{1}{\alpha} .
		\label{eq:gap_sum_rule}
	\end{equation}
	Therefore, we can introduce a normalized probability measure on $\omega>0$, 
	\begin{equation}
		\mathrm{d}\mu(\omega)=-\frac{2\alpha}{\pi}\, \frac{\mathrm{Im}D^R(\omega)}{\omega}\,\mathrm{d}\omega\,,
		\qquad \int\mathrm{d}\mu=1 \, .
		\label{eq:spectralmeasure}
	\end{equation}
	We can combine both sum rules to obtain:
	\begin{equation}
		\left\langle \omega^2 \right\rangle_\mu = \alpha \, \omega_0 = \omega_r^2 \, ,
	\end{equation}
	and express the quantum Fisher information as
	\begin{equation}
		F_Q = \frac{2}{\alpha}\left\langle \varphi(\omega^2) \right\rangle_\mu \, , \qquad \varphi(x) = \sqrt{x} \tanh \left( \frac{\sqrt{x}}{2 T} \right) \, . 
	\end{equation}
	Since $\varphi(x)$ is a concave function, Jensen's inequality $\langle \varphi(\omega^2) \rangle_\mu \le \varphi(\langle \omega^2 \rangle_\mu) = \varphi( \omega^2_r)$ gives
	\begin{equation}
		F_Q \le \frac{2 \omega_0}{\omega_r} \tanh \left( \frac{\omega_r}{2T} \right) \, , \label{eq:QFI_bound_sharp}
	\end{equation}
	where the equality is saturated for an effectively free boson. The bound is tighter than the sum-rule bound~\eqref{eq:QFI_bound1} at all temperatures and reduces to it for $\omega_r\ll T$. For $\omega_r\gg T$ it approaches the ceiling
	\begin{equation}
		F_Q \;\le\; 2\sqrt{\frac{\omega_0}{\alpha}} \, ,
		\label{eq:QFI_bound2}
	\end{equation}
	which, since $\tanh x\le1$, holds at all temperatures and is saturated only by a single infinitely sharp line above the thermal scale. The crossover between the two limits takes place at $T=\omega_r/2$.
	
	Up to its small spectral width, the boson in the SYK$_2$ regime exactly saturates~\cref{eq:QFI_bound_sharp}, as we show in Sec.~\ref{sec:QFI-impurity}; this solution ceases to apply once $\omega_r$ becomes comparable to $T$, below which the conformal boson spreads its weight above the thermal scale, where the $\tanh$ kernel discards it. Along each constant-$\alpha$ sweep $F_Q$ therefore develops a maximum near $T=\omega_r/2$, corresponding to the point of maximal metrological sensitivity, and broadly aligning with the conformal/SYK$_2$ crossover. 
	
	\subsubsection{Quantum Fisher information in different regimes}
	\label{sec:QFI-impurity}
	
	\paragraph{The SYK$_2$ regime}
	In the SYK$_2$ regime, the QFI can be computed in closed form, and the
	moment structure of the previous section becomes completely explicit.
	Neglecting the interaction-induced broadening $\gamma$, the bosonic propagator in~\cref{eq:impurity_green} reduces to a single undamped mode,
	\begin{equation}
		\operatorname{Im} D^R(\omega)=-\frac{\pi\omega_0}{2\omega_r}\big[\delta(\omega-\omega_r)-\delta(\omega+\omega_r)\big] \, .
		\label{eq:imp-Im_DR}
	\end{equation}
	The neglect of $\gamma$ is controlled: eliminating $\Omega_0$ between the two
	expressions in Eq.~\eqref{eq:impurity_constants} gives
	$\gamma/\omega_r = 3\,\omega_0^{3/2}/(8\,g\sqrt{T})$,
	which is parametrically small inside the window
	$g^{-2}\ll T\ll g^{2}$ where the regime is defined.
	Inserting Eq.~\eqref{eq:imp-Im_DR} into Eq.~\eqref{eq:QFIdef}, we find
	\begin{equation}
		F_Q(g,T) = \frac{2\omega_0}{\omega_r}\,\tanh\!\left(\frac{\omega_r}{2T}\right)
		= \frac{16\,g}{3\pi\sqrt{T}}\,
		\tanh\!\left(\frac{3\pi}{16\,g\sqrt{T}}\right) \, ,
		\label{eq:imp-FQ}
	\end{equation}
	where in the second equality we rewrote $\omega_r$ using
	Eq.~\eqref{eq:impurity_constants} for $N=2M$. Equation~\eqref{eq:imp-FQ} exactly saturates the bound in \cref{eq:QFI_bound_sharp}, as expected from the fact that the bosons are approximately free. For the numerical analysis, it is convenient to express the QFI in terms of the variable 
	\begin{equation}
		u =\frac{\omega_r}{2T}= \frac{\sqrt{\omega_0\alpha}}{2T} \,, \qquad F_Q=(\omega_0/T) \frac{\tanh u}{u}   \, .
		\label{eq:imp-FQ-alpha}
	\end{equation}
	Deep inside the impurity window, one has $u\ll1$ and the QFI saturates the bound in~\cref{eq:QFI_bound1} $F_Q \;\to\; \frac{\omega_0}{T}$.
	
	Along a sweep of constant $\alpha$, $\omega_r$ is fixed and the whole temperature dependence of Eq.~\eqref{eq:imp-FQ-alpha} is in $\tanh u/u$. Lowering $T$, $F_Q$ would then grow from $\omega_0/T$ towards the ceiling~\eqref{eq:QFI_bound2}. However, this limit is never reached, because it requires $\omega_r \gg T$, while the SYK$_2$ solution assumes $\omega_r \ll T$. The sweep leaves the SYK$_2$ regime around $u \sim 1$, before the $\tanh$ can saturate. Below that temperature the boson spreads its weight above the thermal scale, so $F_Q$ falls below Eq.~\eqref{eq:QFI_bound_sharp} and decreases again.
	
\begin{figure}[t!]
		\centering
		\twopanelfig{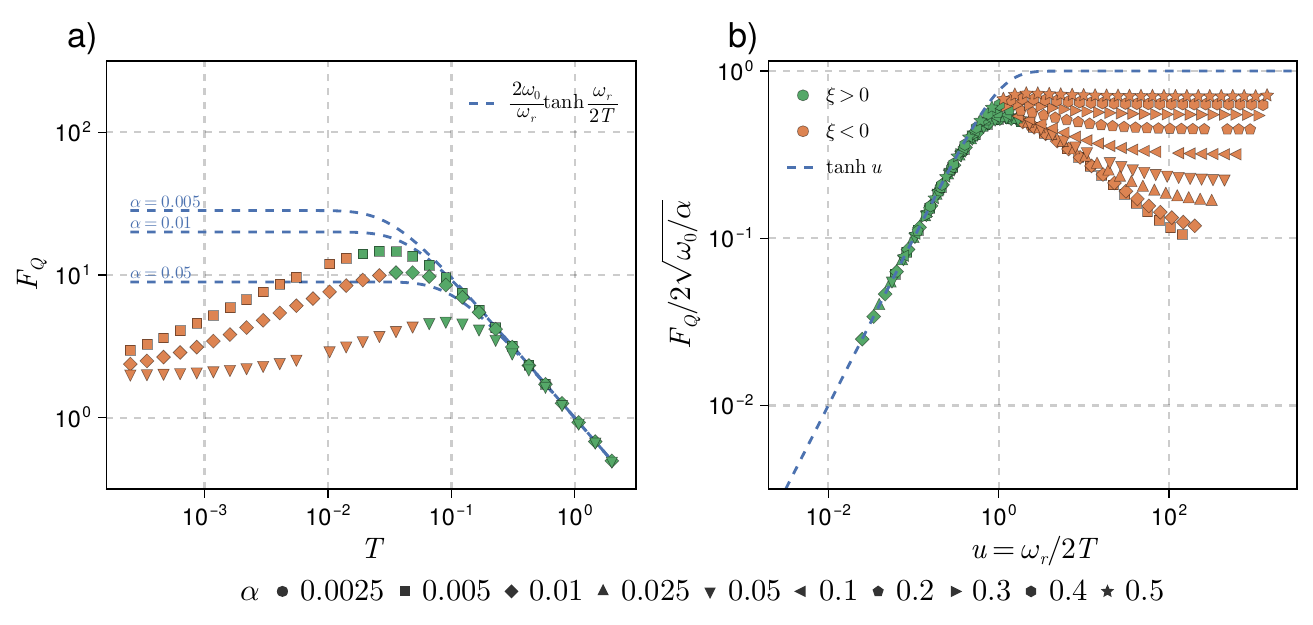}{\bf a)}{\bf b)}
		\caption{\textbf{a)} $F_Q$ as a function of temperature along three constant-gap sweeps, $\alpha = 0.005$, $0.01$ and $0.05$ (in units of $\omega_0$), together with the bound of Eq.~\eqref{eq:QFI_bound_sharp} for each gap (blue dashed curve). Marker shape labels $\alpha$ and marker color the sign of the curvature $\xi$, as in Fig.~\ref{fig:Phase_diagram_plot}. At high temperature the three bounds merge into the sum-rule value $\omega_0/T$, which the data saturate. Around $T=\omega_r/2$ each bound bends over towards its ceiling $2\sqrt{\omega_0/\alpha}$, while the data fall below it, reach a maximum and decrease as the sweep approaches the conformal regime. 
			\textbf{b)} The same quantity for all ten gaps, but rescaled for the variables of~\cref{eq:imp-FQ-alpha}, in which Eq.~\eqref{eq:QFI_bound_sharp} becomes $\tanh u$ for every $\alpha$ (blue dashed curve). For $u \lesssim 0.3$, deep in the SYK$_2$ regime ($\xi>0$) the data saturate the bound. At larger $u$ they fall below it, before $\xi$ changes sign around $u \simeq 1.85$ to the conformal regime: the self-critical boson spreads its weight over a power law extending above the thermal scale, which lowers $F_Q$ below the single-line value. For $\alpha \lesssim 0.05$ the data nicely collapse onto one curve, with its maximum at $u\simeq 1.2$ and $F_Q \simeq 0.52 \times 2\sqrt{\omega_0/\alpha}$. For larger gaps the maximum moves to larger $u$, owing to the bare boson regimes already visible in the inset of Fig.~\ref{fig:Phase_diagram_plot}. 
		}
		\label{fig:QFI_bounds_collapse}
	\end{figure}

	Figure~\ref{fig:QFI_bounds_collapse} shows this sweep by sweep. In panel \textbf{a)}, $F_Q$ saturates Eq.~\eqref{eq:QFI_bound_sharp} at high temperature, falls below it from $u \approx 0.3$ and turns over. In the variables of Eq.~\eqref{eq:imp-FQ-alpha}, the bound becomes $\tanh u$ for every $\alpha$ (panel \textbf{b)}), and for $\alpha \lesssim 0.05$ the data collapse onto a single curve. The maximum therefore sits at the same point for all these gaps, $u \simeq 1.2$ and $F_Q \simeq 0.52 \times 2\sqrt{\omega_0/\alpha}$, close along the crossover as given by the sign change of $\xi$. The height of the QFI peaks grow as $\alpha^{-1/2}$ as criticality is approached. These maxima form the ridge of Fig.~\ref{fig:Phase_diagram_QFI}. Note that they are a property of the constant-$\alpha$ sweeps: at fixed $g$ or fixed $T$, $F_Q$ is monotonic across the crossover.

	\paragraph{The conformal regime}
	In the SYK$_2$ regime, the boson concentrates the entirety of its spectral weight into a single line, and the QFI saturates its bound as a consequence. In contrast, in the conformal regime the weight is spread over a power law and thus has a longer tail, and it is the failure to saturate the bound that becomes the diagnostic. Although the pure conformal propagator is sufficient to determine infrared quantities such as $\alpha$, it cannot determine the QFI, since the latter is UV-sensitive and the integral in Eq.~\eqref{eq:QFIdef} diverges. However, the propagator can be regularized by substituting the self-criticality condition of Eq.~\eqref{eq:def_self_criticality} into the expression for retarded bosonic propagator of Eq.~\eqref{eq:bosonic_conformal_propagator}, giving
	\begin{equation}
		D^R(\omega) = \frac{1}{\omega^2/\omega_0 - \delta\Pi^R(\omega)} \, ,
		\label{eq: D^r_conf_self_crit}
	\end{equation}
	This form of the propagator allows us to find a closed expression for the QFI at $T=0$. 
	Fourier transforming the time-domain expression for $\delta\Pi^R$ given in \cref{eq:conformal_self_energies_T=0} (i.e., Eq.~\eqref{eq: delta_Pi^R_freq} of Appendix \ref{app:regimes}) we find 
	\begin{equation}
		F_Q(T\to 0) = -\frac{4}{\pi}\int_0^{\infty}\!\! d\omega\,\bigl(\operatorname{Im}D^R(\omega)\bigr) =  \mathcal{A}\left(\frac{g^2}{\omega_0^2}\right)^{\frac{4\Delta_F-1}{3-4\Delta_F}} \, ,
		\label{eq:FQ-T0}
	\end{equation}
	with
	\begin{equation}
		\mathcal{A} = \frac{4}{(3-4\Delta_F)\, \sin\!\left[\dfrac{\pi}{3-4\Delta_F}\right]} \left( \frac{\pi c_1^{2}\,(N/M)}{\Gamma(4\Delta_F)\, \sin\!\left[\frac{\pi}{2}(4\Delta_F-1)\right]} \right)^{-1/(3-4\Delta_F)} \, .
	\end{equation}
	
	Therefore, the QFI in the conformal window is a pure power of the coupling $g$, fixed entirely by the conformal dimension $\Delta_F$. The closed-form expression in Eq.~\eqref{eq:FQ-T0} can be tested directly against the saddle-point data of Fig.~\ref{fig:Phase_diagram_QFI}. Moreover, as a consistency-check, we also computed the integral in Eq.~\eqref{eq:QFIdef} numerically using the specific choice of bosonic propagator given in Eq.~\eqref{eq: D^r_conf_self_crit}. Both results are in very good agreement with the saddle-point solutions as long as we are deep in the conformal regime, but are expected to fail close to the crossover lines, where the analytic conformal solution is contaminated by the other regimes. This effect is particularly noticeable for very small $g$, since~\cref{eq:FQ-T0} predicts a very small QFI, but the propagator develops a quasi-particle excitation at $\omega_0$ which makes $F_Q \to 2$ for the full saddle point solutions, the value of Eq.~\eqref{eq:QFI_bound_sharp} for an undressed oscillator ($\omega_r=\omega_0$) at $T=0$.
	
	\subsection{Fermionic spectral weight}
	\label{sec:fermion_weight}
	
	The observables introduced in the previous sections provide criteria to delimit the conformal and SYK$_2$ regimes, but they are all constructed from the bosonic two-point function. This strategy necessarily fails for bare boson regimes of Sec.~\ref{sec:high_t_regimes}, where the bosonic propagator carries no information about the state of the system. To characterize these regimes, we must therefore turn to the fermionic two-point function.
	
	A natural candidate is suggested by the structure of Eq.~\eqref{eq:new_regime}, which describes a Floquet-like problem in which the fermion hybridizes with sidebands displaced by integer multiples of $\omega_0$. Weight develops first at the fundamental resonance, higher harmonics being generated only at higher order in the effective coupling. A direct measure of the onset of hybridization is therefore the fermionic spectral weight in a narrow window around the first sideband,
	\begin{equation}
		W_{\omega_0}(g) = \frac{1}{2\delta}\int_{\omega_0-\delta}^{\omega_0+\delta}
		\rho(\omega)\,d\omega ,
		\label{eq:spectral_weight}
	\end{equation}
	with $\delta \ll \omega_0$. The prefactor makes $W_{\omega_0}$ the average spectral density in the bin. In the free limit, where $\rho(\omega)=\delta(\omega)$, the weight vanishes identically for any $\delta<\omega_0$, so $W_{\omega_0}$ is a sensitive diagnostic of the Floquet resonance.

	\begin{figure}[t]
		\centering
		\begin{minipage}[t]{0.49\textwidth}\centering
			\dissipationpanel{\bf a)}{}{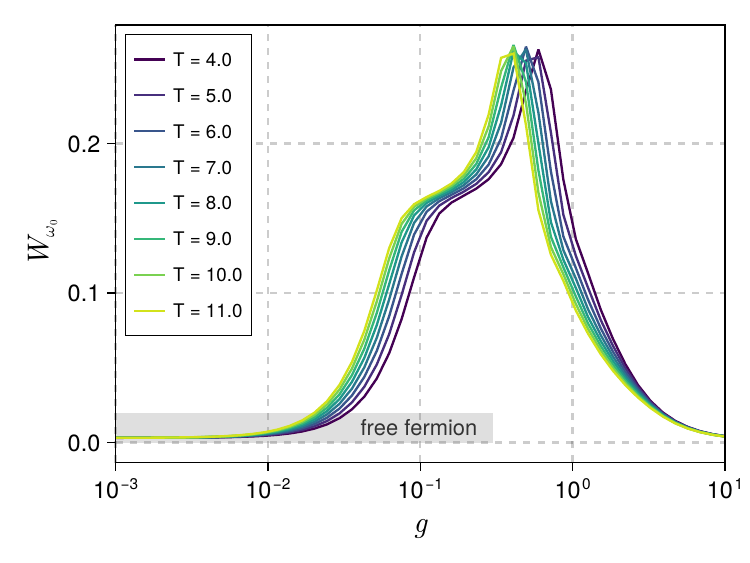}
		\end{minipage}\hfill
		\begin{minipage}[t]{0.49\textwidth}\centering
			\dissipationpanel{\bf b)}{}{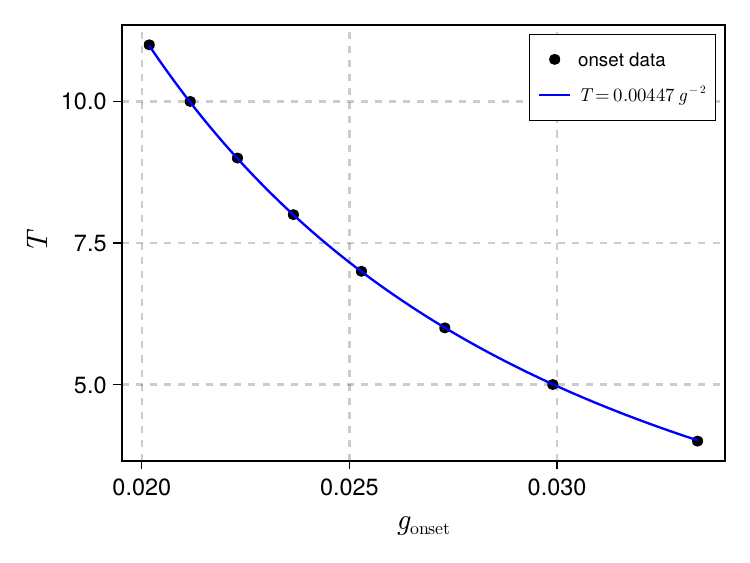}
		\end{minipage}
		\caption{
			{\bf a)} Spectral weight
			$W_{\omega_0}(g)$
			for $\omega_0 = 1$, and $\delta = 0.05$ plotted against the coupling $g$ for several temperatures $T$
			(colored lines). The shaded gray region marks $W_{\omega_0} < \delta_{\mathrm{dev}}
			= 0.02$, a guide to the eye delimiting the approximate free-fermion
			regime, where the coupling has not yet produced appreciable spectral weight around $\omega_0$. At larger coupling, the fermion hybridizes with the bosonic mode, transferring spectral weight into the resonance, as shown in Fig.~\ref{fig:floquet_high_T}. As $g$ increases, we move closer to the SYK$_2$ regime, where the spectral density becomes a broad semicircle which leads to a drop of the spectral weight around $\omega_0$. Since in the SYK$_2$ regime the semicircle just depends on $g$ and it is $T$ independent, the spectral weight curves tend to collapse on the same line.
			{\bf b)} Onset coupling $g_{\mathrm{onset}}(T)$ at which the weight shown on the left
			first exceeds the deviation threshold $\delta_{\mathrm{dev}} = 0.02$ (black circles),
			as a function of temperature $T$. Each $g_{\mathrm{onset}}$ is extracted from the
			left-panel data by a shape-preserving monotone cubic Hermite interpolation
			of the discretely sampled $W_{\omega_0}(g)$ curve, followed by a
			bisection to locate the crossing $W_{\omega_0}(g_{\mathrm{onset}}) =
			\delta_{\mathrm{dev}}$. The
			blue line is a fit to the fixed power law $T = C\, g_{\mathrm{onset}}^{-2}$, with the
			exponent held fixed, which follows from the analysis of Sec.~\ref{sec:high_t_regimes}. These results have been checked to be robust after changing for $\delta$ and $\delta_{\mathrm{dev}}$, and for the fit for $C$ is in good agreement with the analytical prediction in \cref{eq:free_boundary_prediction}, which gives $C=0.004$;
		}
		\label{fig:onset_and_weight}
	\end{figure}
	
	Along a constant-temperature scan of the coupling, $W_{\omega_0}$ is expected to be non-monotonic: at small $g$ the fermion is essentially free and $W_{\omega_0}\simeq 0$. As $g$ increases, the fermion hybridizes with the bosonic mode and spectral weight is transferred from the central quasiparticle peak into the resonance at $\omega_0$, cf.\ Fig.~\ref{fig:floquet_high_T}, so that $W_{\omega_0}$ grows. As we move into the unresolved sideband regime, however, the spectral density flattens into the broad semicircle until it enters the SYK$_2$ regime, where the semicircle radius is given by $\Omega_0 \propto g^2$. The weight in the bin then decreases as
	$W_{\omega_0} \simeq \rho(\omega_0) \simeq 2/(\pi\Omega_0) \propto g^{-2}$,
	and, since $\Omega_0$ is temperature independent to leading order, the curves at different temperatures collapse onto a common envelope at large coupling. The numerical results of Fig.~\ref{fig:onset_and_weight}{\bf a)} confirm this picture.

	Since the free-fermion and resolved sideband regimes are separated by a smooth crossover rather than a sharp transition, delimiting them requires a convention. We therefore introduce a threshold $\delta_{\mathrm{dev}}$ and define the onset coupling $g_{\mathrm{onset}}(T)$ as the smallest coupling at which $W_{\omega_0}(g)=\delta_{\mathrm{dev}}$. 
	Its temperature dependence can be estimated from the results of Sec.~\ref{sec:high_t_regimes}. Starting from
	\begin{equation}
		G^R(\omega_0+\omega) = \frac{1}{\omega_0-\frac{g^2T}{2 \omega_0} G^R(\omega)},
	\end{equation}
	the spectral weight of the sideband is fixed by the analytic structure of $G^R(\omega)$, namely, by the coefficient of the $1/\omega$ at frequencies large compared with the width of the central band $\omega \to \infty$. There $G^R(\omega) \to 1/\omega$, so that
	\begin{equation}
		G^R(\omega_0+\omega) \sim \frac{1}{\omega_0} + \frac{g^2T}{2 \omega_0^3} \frac{1}{\omega}.
	\end{equation}
	Provided $\delta$ is large enough to contain the whole sideband, this gives $W_\mathrm{\omega_0} = \frac{g^2T}{4 \delta \omega_0^3}$, and setting $W_{\omega_0} = \delta_{\mathrm{dev}}$ yields
	\begin{equation}
		T = 4 \delta \delta_{\mathrm{dev}} \frac{\omega_0^3}{g_\mathrm{onset}^2} \, . \label{eq:free_boundary_prediction}
	\end{equation}
	The numerical value of $g_{\mathrm{onset}}$ thus depends on the choice of $\delta$ and $\delta_{\mathrm{dev}}$, while the scaling of the resulting boundary does not. The numerics are consistent with this power law: as shown in Fig.~\ref{fig:onset_and_weight}{\bf b)}, the data are well described by the fixed-exponent fit $T = C\, g_{\mathrm{onset}}^{-2}$, and the result is stable under variations of both $\delta$ and $\delta_{\mathrm{dev}}$. For $\delta = 0.05$, $\delta_\mathrm{dev}=0.02$ and $\omega_0=1$, \cref{eq:free_boundary_prediction} predicts $C=0.004$, which is relatively close to the fitted value $C=0.00447$\footnote{The deviation is due to the fact that the fermionic and bosonic $i0^+$ regulator must be numerically finite and leaks sideband weight out of the bin, leading to a larger $C$. Using \eqref{eq:free_boundary_dissipation} with the numerical regulator $\kappa = 0.01$ to account for this effect, we find $C=0.00457$.}.

	\begin{figure}[t]
		\centering
		\begin{minipage}[t]{0.49\textwidth}\centering
			\dissipationpanel{\bf a)}{}{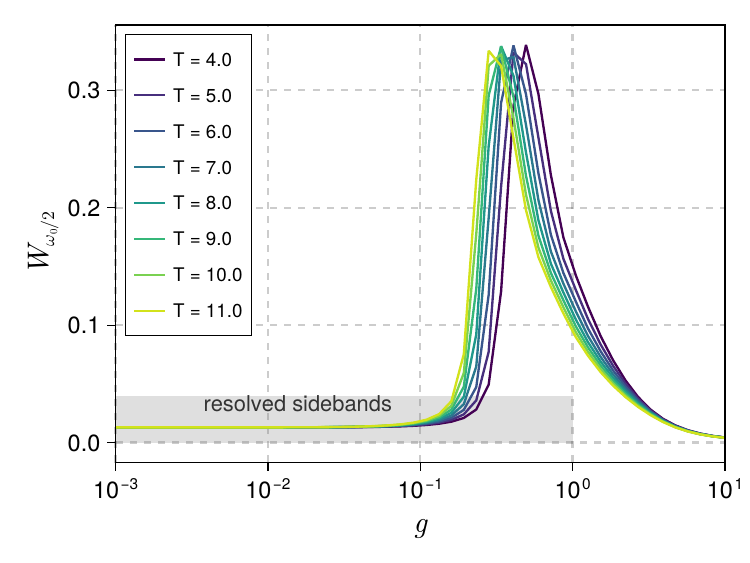}
		\end{minipage}\hfill
		\begin{minipage}[t]{0.49\textwidth}\centering
			\dissipationpanel{\bf b)}{}{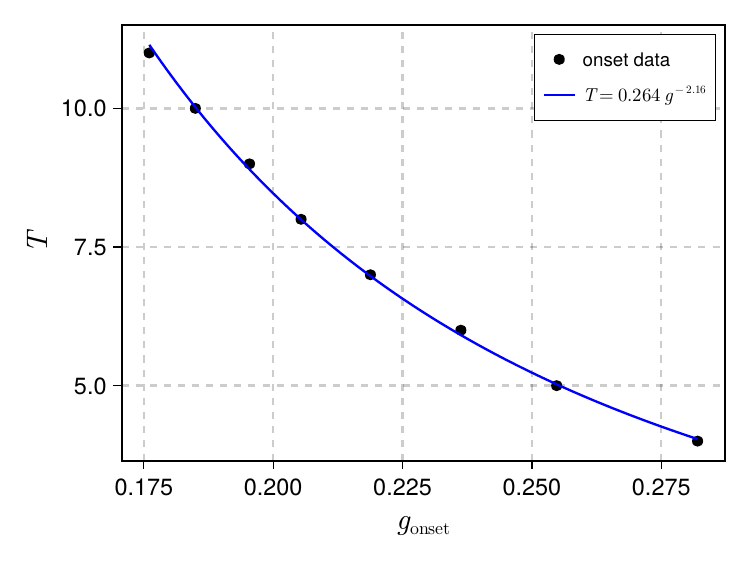}
		\end{minipage}
		\caption{
			{\bf a)} Spectral weight
			$W_{\omega_0/2}(g)$
			for $\omega_0 = 1$ and $\delta = 0.05$ plotted against the coupling $g$ for several temperatures $T$
			(colored lines). The shaded gray region marks $W_{\omega_0/2} < \delta_{\mathrm{dev}}
			= 0.04$, which delimits the resolved sideband
			regime from the unresolved one.  As  for Fig.~\ref{fig:onset_and_weight}, the large $g$ decay is conditioned by the proximity to the SYK$_2$ regime.
			{\bf b)} Onset coupling $g_{\mathrm{onset}}(T)$ at which the weight shown on the left
			first exceeds the deviation threshold $\delta_{\mathrm{dev}} = 0.04$ (black circles),
			as a function of temperature $T$. The point has been extracted following the same method as in Fig.~\ref{fig:onset_and_weight}. The
			blue line is a power-law fit $T = C\, g_{\mathrm{onset}}^{a}$. The fit the analytical prediction $T=0.3 \, g^{-2}$ of \cref{{eq:resolve_unresolved_transition}};
		}
		\label{fig:onset_and_weight_0p5}
	\end{figure}

	The same observable detects the crossover between the resolved and unresolved sideband regimes, once the window is centered at $\omega_0/2$, where the gap between the central band and the first sideband closes (see Sec.~\ref{sec:high_t_regimes}).
	As shown in Fig.~\ref{fig:onset_and_weight_0p5}{\bf a)}, $W_{\omega_0/2}$ stays small while the gap is open, starts to grow as the gap closes, reaches a maximum, and finally decreases as $g^{-2}$ on approaching the SYK$_2$ regime. To delimit the two regimes we again introduce a threshold $\delta_{\mathrm{dev}}$ and define $g_{\mathrm{onset}}(T)$ as the smallest coupling at which $W_{\omega_0/2}(g)=\delta_{\mathrm{dev}}$. The fitted crossover line $T=0.264\, g^{-2.16}$ is in good agreement with the estimate of \cref{eq:resolve_unresolved_transition}\footnote{It is worth stressing that, differently from the free-theory/resolved-sideband crossover, the point where the sidebands merge is independent on the definition of $\delta_\mathrm{dev}$, and to extract it one should extrapolate for $\delta_\mathrm{dev} \to 0$ while also sending the $i 0^+$ regulator of the free fermionic and bosonic two-point function to zero. The regulator is also responsible for the fact that, in Fig.~\ref{fig:onset_and_weight_0p5}, $W_{\omega_0/2}$ does not vanish for small $g$.}.
	
	Together with the inverse static susceptibility $\alpha$ and the curvature $\xi$ introduced above, the spectral weight $W$ completes a minimal set of observables capable of discriminating all the regimes of the phase diagram. Being a frequency-integrated quantity over a finite window, it is moreover robust against finite spectral resolution, and is thus accessible through spectroscopic probes such as those discussed at the beginning of this section.
	
	In the next section, we investigate how the regimes studied in Sec.~\ref{sec:phase_diagram}, as well as the markers introduced in this section, are modified when the system is coupled to a bath at the same temperature.

	\section{Open system dynamics}
	\label{sec:dissipation}
	
	One of the main advantages of the Keldysh real-time formalism is that the coupling to an environment is straightforward to incorporate, so that the previous analysis can be extended to include open-system effects. The most interesting question is to understand if the conformal phase, the most relevant one for the holographic duality, survives once the system is opened, as dissipation is inevitable in every realistic implementation of the YSYK model, and therefore a necessary ingredient to account for in order to realize holographic quantum matter in tabletop experiments.
	
	As discussed in Refs.~\cite{Almheiri:2019jqq,Zhang2019,Hosseinabadi:2023mid}, a contact coupling between the fermionic field and an environment produces a relevant deformation of the theory, which destroys the conformal phase in the IR.  We therefore couple the environment to the bosonic sector instead. We further assume that the environment is much larger than the system, so that its back-reaction can be neglected, and that each boson is coupled to its own independent set of oscillators $b_{k,m}$, which prevents the bath from generating off-diagonal correlations $\langle a_m a_n\rangle$ that would spoil the solvability of the model. The full model Hamiltonian then becomes 
	\begin{equation}
		H_\mathrm{tot} = H + \sum_{m=1}^M \left( \sum_{k} \epsilon_k b^\dag_{k,m} b_{k,m} + g_k a_m b^\dag_{k,m} + \bar{g}_k b_{k,m} a^\dag_m  \right) \, ,
		\label{eq:H_total}
	\end{equation}
	where $\epsilon_k$ are the internal energy levels of the environment. The above linear coupling amounts to a boson loss, mimicking in generic form a typical dissipation channel that appears, e.g., in cavity-QED experiments where photons unavoidably leak out of the cavity or in the solid state where phonons leak into the substrate. 
	
	We first identify the infrared properties of a generic bath that constrain self-criticality.
	We then show that weak dissipation preserves all regimes but shifts most crossovers to larger couplings and erodes the conformal and resolved-sideband regimes, while strong dissipation collapses the fermion onto a sharp quasiparticle while the boson is bath-dominated.
	Finally, we follow the markers of Sec.~\ref{sec:observables} in the presence of dissipation; the resulting deformation of the phase diagram is summarized in Fig.~\ref{fig:diss-phase-diagrams}.

	\newcommand{\pdWa}{16.0cm}   
	\newcommand{\pdHa}{10.5cm}
	\newcommand{\pdFadea}{8mm}   
	\newcommand{\pdNsa}{160}
	\newcommand{\pdPanel}{0.49}  
	
	\def\Mx{0.9}
	\def\xRz{0.9}
	\def\xUz{0.23}
	\def\xLz{-0.02}
	\def\xSz{-0.22}
	\def\MxB{0.9}     
	\def\MLB{1.2}     
	\def\xRzB{1.08}   
	\def\xUzB{0.22}   
	\def\xLzB{0.0}    
	
	\tikzset{
		crossoverline/.style={black!55, line width=1.2pt, dash pattern=on 5.5pt off 4pt},
		eqlabel/.style={inner sep=1pt, font=\large},
		regime/.style={align=center, inner sep=1pt, font=\bfseries\LARGE, text=#1!70!black},
		dissarrow/.style={-{Latex[length=3.4mm,width=4.2mm]}, line width=2.8pt, black!75},
		panellab/.style={anchor=south west, font=\bfseries\LARGE, inner sep=0pt},
	}
	
	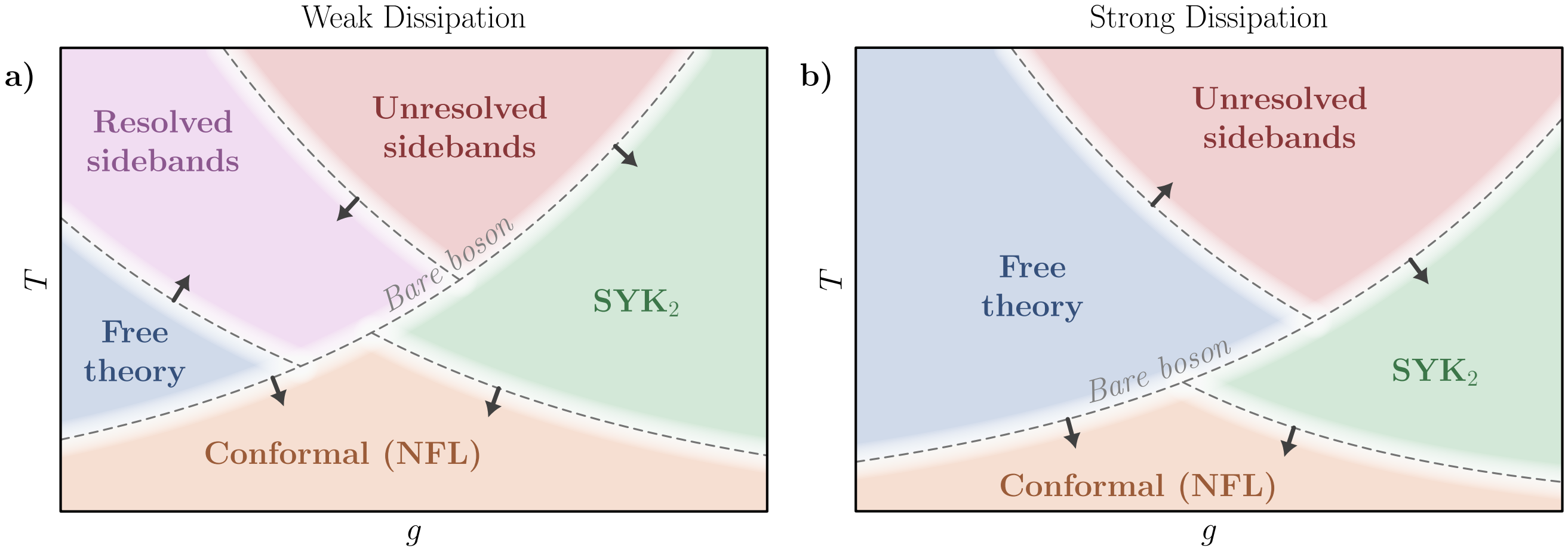
\begin{figure*}[t]
		\centering
		\begin{minipage}{\pdPanel\textwidth}\centering
			\resizebox{\textwidth}{!}{%
				\begin{tikzpicture}[
					x=\pdWa, y=\pdHa, line join=round, line cap=round,
					declare function={
						yR(\x) = pow(10,  \Mx*(\x-\xRz));
						yU(\x) = pow(10, -\Mx*(\x-\xUz));
						yL(\x) = pow(10, -\Mx*(\x-\xLz));
						yS(\x) = pow(10, -\Mx*(\x-\xSz));
					}]
					\pgfmathsetmacro{\xA}{0.5*(\xRz+\xUz)}
					\pgfmathsetmacro{\xB}{0.5*(\xRz+\xLz)}
					\pgfmathsetmacro{\xC}{0.5*(\xRz+\xSz)}
					\def\pR#1#2{plot[domain=#1:#2,samples=\pdNsa] (\x,{yR(\x)})}
					\def\pU#1#2{plot[domain=#1:#2,samples=\pdNsa] (\x,{yU(\x)})}
					\def\pL#1#2{plot[domain=#1:#2,samples=\pdNsa] (\x,{yL(\x)})}
					\def\pS#1#2{plot[domain=#1:#2,samples=\pdNsa] (\x,{yS(\x)})}
					\newcommand{\halo}[1]{\foreach \k in {1,...,12}{\draw[white,opacity=0.15,line width={(13-\k)/12*\pdFadea}] #1;}}
					\begin{scope}
						\clip (0,0) rectangle (1,1);
						\fill[fillFree]   (0,{yR(0)})
						-- plot[domain=0:\xA    ,samples=\pdNsa] (\x,{yR(\x)})
						-- plot[domain=\xA:\xUz ,samples=\pdNsa] (\x,{yU(\x)})
						-- (0,1) -- cycle;
						\fill[fillDriven] (\xUz,1)
						-- plot[domain=\xUz:\xA ,samples=\pdNsa] (\x,{yU(\x)})
						-- plot[domain=\xA:\xRz ,samples=\pdNsa] (\x,{yR(\x)}) -- cycle;
						\fill[fillImp]    (\xRz,1)
						-- plot[domain=\xRz:\xB ,samples=\pdNsa] (\x,{yR(\x)})
						-- plot[domain=\xB:1    ,samples=\pdNsa] (\x,{yL(\x)}) -- (1,1) -- cycle;
						\fill[fillConf]   (0,{yR(0)})
						-- plot[domain=0:\xB    ,samples=\pdNsa] (\x,{yR(\x)})
						-- plot[domain=\xB:1    ,samples=\pdNsa] (\x,{yL(\x)}) -- (1,0) -- (0,0) -- cycle;
						\fill[fillSide]   (0,{yR(0)})
						-- plot[domain=0:\xC   ,samples=\pdNsa] (\x,{yR(\x)})
						-- plot[domain=\xC:0   ,samples=\pdNsa] (\x,{yS(\x)}) -- cycle;
						\halo{\pR{0}{\xRz}} \halo{\pU{\xUz}{\xA}} \halo{\pL{\xB}{1}} \halo{\pS{0}{\xC}}
						\draw[crossoverline] \pR{0}{\xRz};
						\draw[crossoverline] \pU{\xUz}{\xA};
						\draw[crossoverline] \pL{\xB}{1};
						\draw[crossoverline] \pS{0}{\xC};
					\end{scope}
					\draw[line width=1.6pt] (0,0) rectangle (1,1);
					\node[below=5pt] at (0.5,0) {\LARGE $g$};
					\node[left=5pt]  at (0,0.5) {\rotatebox{90}{\LARGE $T$}};
					\node[above=5pt]  at (0.5,1.0) {\LARGE Weak Dissipation};
					\node[panellab]  at (-0.08,0.9) {\bf a)};
					\node[regime=csidebands] at (0.105,0.34) {Free\\theory};
					\node[regime=cFree]      at (0.145,0.80) {Resolved\\sidebands};
					\node[regime=cDriven]    at (0.565,0.83) {Unresolved\\sidebands};
					\node[regime=cImp]       at (0.815,0.45) {SYK$_2$};
					\node[regime=cConf]      at (0.40,0.12)  {Conformal (NFL)};
					\path[decorate, decoration={text along path, text align=center, raise=6pt,
						text={|\LARGE\itshape\color{gray}| Bare boson}}] \pR{0.4}{0.70};
					\def\arrowlen{7mm}
					\def\arrowrot{90}
					\newcommand{\shiftarrow}[3]{%
						\coordinate (sa) at ({#2-0.02},{#1(#2-0.02)});
						\coordinate (sb) at ({#2+0.02},{#1(#2+0.02)});
						\pgfmathanglebetweenpoints{\pgfpointanchor{sa}{center}}{\pgfpointanchor{sb}{center}}%
						\pgfmathsetmacro{\dirang}{\pgfmathresult+#3*\arrowrot}%
						\draw[dissarrow] ({#2},{#1(#2)}) -- ++(\dirang:\arrowlen);}
					\shiftarrow{yS}{0.16}{1}
					\shiftarrow{yU}{0.42}{-1}
					\shiftarrow{yR}{0.785}{-1}
					\shiftarrow{yR}{0.3}{-1}
					\shiftarrow{yL}{0.62}{-1}
			\end{tikzpicture}}
		\end{minipage}\hfill
		\begin{minipage}{\pdPanel\textwidth}\centering
			\resizebox{\textwidth}{!}{%
				\begin{tikzpicture}[
					x=\pdWa, y=\pdHa, line join=round, line cap=round,
					declare function={
						yR(\x) = pow(10,  \MxB*(\x-\xRzB));
						yU(\x) = pow(10, -\MxB*(\x-\xUzB));
						yL(\x) = pow(10, -\MLB*(\x-\xLzB));
					}]
					\pgfmathsetmacro{\xA}{0.5*(\xRzB+\xUzB)}                       
					\pgfmathsetmacro{\xB}{(\MxB*\xRzB+\MLB*\xLzB)/(\MxB+\MLB)}     
					\def\pR#1#2{plot[domain=#1:#2,samples=\pdNsa] (\x,{yR(\x)})}
					\def\pU#1#2{plot[domain=#1:#2,samples=\pdNsa] (\x,{yU(\x)})}
					\def\pL#1#2{plot[domain=#1:#2,samples=\pdNsa] (\x,{yL(\x)})}
					\newcommand{\halo}[1]{\foreach \k in {1,...,12}{\draw[white,opacity=0.15,line width={(13-\k)/12*\pdFadea}] #1;}}
					\begin{scope}
						\clip (0,0) rectangle (1,1);
						\fill[fillSide]   (0,{yR(0)})                                
						-- plot[domain=0:\xA     ,samples=\pdNsa] (\x,{yR(\x)})
						-- plot[domain=\xA:\xUzB ,samples=\pdNsa] (\x,{yU(\x)})
						-- (0,1) -- cycle;
						\fill[fillDriven] (\xUzB,1)                                  
						-- plot[domain=\xUzB:\xA ,samples=\pdNsa] (\x,{yU(\x)})
						-- plot[domain=\xA:\xRzB ,samples=\pdNsa] (\x,{yR(\x)}) -- cycle;
						\fill[fillImp]    (\xRzB,1)                                  
						-- plot[domain=\xRzB:\xB ,samples=\pdNsa] (\x,{yR(\x)})
						-- plot[domain=\xB:1     ,samples=\pdNsa] (\x,{yL(\x)}) -- (1,1) -- cycle;
						\fill[fillConf]   (0,{yR(0)})                                
						-- plot[domain=0:\xB     ,samples=\pdNsa] (\x,{yR(\x)})
						-- plot[domain=\xB:1     ,samples=\pdNsa] (\x,{yL(\x)}) -- (1,0) -- (0,0) -- cycle;
						\halo{\pR{0}{\xRzB}} \halo{\pU{\xUzB}{\xA}} \halo{\pL{\xB}{1}}
						\draw[crossoverline] \pR{0}{\xRzB};
						\draw[crossoverline] \pU{\xUzB}{\xA};
						\draw[crossoverline] \pL{\xB}{1};
					\end{scope}
					\draw[line width=1.6pt] (0,0) rectangle (1,1);
					\node[below=5pt] at (0.5,0) {\LARGE $g$};
					\node[left=5pt]  at (0,0.5) {\rotatebox{90}{\LARGE $T$}};
					\node[above=5pt]  at (0.5,1.0) {\LARGE Strong Dissipation};
					\node[panellab]  at (-0.08,0.9) {\bf b)};
					\node[regime=csidebands] at (0.25,0.48) {Free\\theory};
					\node[regime=cDriven]    at (0.60,0.85) {Unresolved\\sidebands};
					\node[regime=cImp]       at (0.82,0.30) {SYK$_2$};
					\node[regime=cConf]      at (0.4,0.05) {Conformal (NFL)};
					\path[decorate, decoration={text along path, text align=center, raise=6pt,
						text={|\LARGE\itshape\color{gray}| Bare boson}}] \pR{0.25}{0.6};
					\def\arrowlen{7mm}
					\def\arrowrot{90}
					\newcommand{\shiftarrow}[3]{%
						\coordinate (sa) at ({#2-0.02},{#1(#2-0.02)});
						\coordinate (sb) at ({#2+0.02},{#1(#2+0.02)});
						\pgfmathanglebetweenpoints{\pgfpointanchor{sa}{center}}{\pgfpointanchor{sb}{center}}%
						\pgfmathsetmacro{\dirang}{\pgfmathresult+#3*\arrowrot}%
						\draw[dissarrow] ({#2},{#1(#2)}) -- ++(\dirang:\arrowlen);}
					\shiftarrow{yU}{0.42}{1}
					\shiftarrow{yR}{0.785}{-1}
					\shiftarrow{yR}{0.3}{-1}
					\shiftarrow{yL}{0.62}{-1}
			\end{tikzpicture}}
		\end{minipage}
		\caption{\label{fig:diss-phase-diagrams}%
			Effect of linear coupling to a bosonic bath on the regimes of the YSYK model.
			{\bf a)} Weak dissipation: arrows indicate the direction in which each crossover is
			displaced as the dissipation is increased.
			{\bf b)} Strong dissipation: the resolved-sideband regime has been squeezed out
			between the two lines that bound it, leaving a single crossover between the free
			and the unresolved-sideband regimes, while the bare-boson line and the
			SYK$_2$/conformal line move down, the latter becoming steeper, so that both the
			SYK$_2$ and the conformal regimes shrink.}
	\end{figure*}

	\subsection{Self-tuned criticality in the presence of a bath}
	\label{sec:bath_IR}

	As discussed in Appendix~\ref{app:boson_environment}, the net effect of integrating out the bath operators $b_{k,m}$ is a modification of the Schwinger--Dyson equations for the system boson \eqref{eq:Schwinger_Dyson}. 
	In this section, we will always assume the environment to be in thermal equilibrium with the fermions and bosons. The effect of the environment coupling therefore simplifies to an induced spectral broadening. This manifests as an additional self-energy contribution only in the retarded boson propagator: 
	\begin{equation}
		D^R(\omega) = \frac{1}{D_0^R(\omega)^{-1} - \Pi^R(\omega)-\Pi^R_\mathrm{E}(\omega)} \label{eq:Schwinger_Dyson_bath} \, .
	\end{equation}
	It is worth emphasizing that this is the only place where the environment enters: everything that the bath in the present setting does to the fermionic sector is therefore mediated by $D^R$, and the rest of the equations~\eqref{eq:fluctuation_dissipation_D}--\eqref{eq:PiR_t} are unmodified.
	
	We split the environment-induced self-energy into a static and a dynamical piece,
	$\Pi^R_\mathrm{E}(\omega) = \Pi^R_\mathrm{E}(0)
	+ \delta\Pi^R_\mathrm{E}(\omega)$; the exact form of $\delta \Pi^R_\mathrm{E}(\omega)$ depends on the details of the bath. We take the following generic ansatz \cite{Caldeira_Legget_1981,Leggett_1987,Hosseinabadi:2023mid}, 
	\begin{equation}
		\Im \Pi^R_\mathrm{E}(\omega) \propto \omega_0 \sin \left(\frac{\pi s}{2} \right) \,
		\mathrm{sgn}(\omega)\left|\frac{\omega}{\omega_0}\right|^{s} e^{-\omega / \Lambda} \, ,
		\label{eq:diss_param}
	\end{equation}
	where $\Lambda$ is a UV cutoff, assumed to be larger than all other energy scales in the system, while the exponent $s$ classifies the bath in Ohmic ($s=1$), sub-Ohmic ($s<1$), and super-Ohmic ($s>1$).  Ohmic behavior is the generic case in the IR: whenever the environment has a finite and smooth density of states at low frequency, $\Im \Pi^R_\mathrm{E}$ is odd and analytic at $\omega=0$ and therefore starts linearly.
	Super-Ohmic baths exhibit a stronger low-frequency suppression, whereas sub-Ohmic behavior requires an environment that is itself critical. 
	The real part of $\delta\Pi^R_\mathrm{E}$ is fixed by the imaginary one through the Kramers--Kronig relations. For a power-law bath as in Eq.~\eqref{eq:diss_param}, we have $\delta\Pi^R_\mathrm{E}(\omega)
	\propto \omega_0\,(-i\omega/\omega_0)^{s}$ for $\omega \ll \Lambda$, so that real and imaginary parts carry the same power of $\omega$.

	Reproducing the conformal solution of Sec.~\ref{subsec:conf_phase} (see also Appendix~\ref{app:conf_symmetry}) in the presence of an environment requires two separate ingredients. The first is the self-criticality of the system boson, which in the presence of dissipation takes the form 
	\begin{equation}
		\Pi^R(0) = - \omega_0 - \Pi^R_\mathrm{E}(0) \, ,
		\label{eq:self_criticality_bath}
	\end{equation}
	and the second is that $\delta \Pi^R(\omega)$ remains the leading contribution to Eq.~\eqref{eq:Schwinger_Dyson_bath} in the infrared.  We treat them in turn.

	\paragraph{ (i) Criticality and criticality budget:} The condition in Eq.~\eqref{eq:self_criticality_bath} requires the fermions to produce a stronger back-reaction on the boson than in the isolated case whenever $\Pi^R_\mathrm{E}(0)>0$. Their ability to do so, however, is limited by the criticality budget introduced in Sec.~\ref{sec:criticality_budget}. By
	\cref{eq:PiR0_spectral}, the interaction can only soften the boson, $\Pi^R(0) \le 0$, which together with \cref{eq:self_criticality_bath} already implies that a self-critical solution can exist only if $\omega_0 + \Pi^R_\mathrm{E}(0) > 0$. By \cref{eq:criticaly_budget}, its
	magnitude is in addition bounded by the criticality budget, so that the conformal phase requires
	\begin{equation}
		0 < \omega_0 + \Pi^R_\mathrm{E}(0) \le \Pi_\mathrm{max}
		= \frac{g^2}{4T}\frac{N}{M} \, .
		\label{eq:budget_with_bath}
	\end{equation}
	Importantly, in thermal equilibrium, the upper bound $\Pi_\mathrm{max}$ is a property of the system that carries no dependence on the environment, and it cannot be enlarged by engineering the bath. At $\Pi^R_\mathrm{E}(0) = 0$, \cref{eq:budget_with_bath} reduces to the closed-system condition \cref{eq:conformal_window_budget}, and a finite static shift of the bath simply tightens it, shrinking the window at which criticality is allowed to $T \le \frac{g^2}{\omega_0+\Pi^R_\mathrm{E}(0)}\frac{N}{4M}$.
	
	If $\Pi^R_\mathrm{E}(0) \gg \Pi_\mathrm{max}$, no amount of tuning can make the system self-critical.
	In terms of the competition between bosonic softening and fermionic broadening discussed in Sec.~\ref{sec:criticality_budget}, the bath keeps the boson stiff, so that the fermion spends nearly the entire budget in an attempt to soften the boson at the price of becoming nearly bare. Even the full budget, however, cannot compensate the stiffening induced by the bath, and the boson is driven away from criticality.

	\paragraph{(ii) Infrared relevance of the bath:} Even in the regimes where the dissipation lies below the criticality budget, the second condition for the conformal symmetry to emerge is that
	\begin{equation}
		\delta \Pi^R (\omega) \gg \delta \Pi^R_\mathrm{E} (\omega)
		\qquad \text{for} \qquad \omega \to 0 \, .
		\label{eq:IR_dominance}
	\end{equation}
	Since $\delta \Pi^R (\omega) \propto |\omega|^{4 \Delta_\mathrm{F}-1}$, the comparison is controlled by  weather $s \lessgtr 4\Delta_\mathrm{F}-1$, with
	the threshold $s_c = 4\Delta_\mathrm{F}-1$ being the one obtained in Ref.~\cite{Hosseinabadi:2023mid} by demanding that the system--bath coupling be irrelevant at the conformal fixed point.  For $s <  4\Delta_\mathrm{F}-1$, the dissipation is
	relevant and the conformal symmetry is lost in the infrared, while for $s >  4\Delta_\mathrm{F}-1$ it is irrelevant and there exists a scale under which the conformal description holds.
	Since $\Delta_\mathrm{F} \in (1/4,1/2)$ depending on the ratio $M/N$, one has $s_c \in (0,1)$, so that Ohmic and super-Ohmic environments preserve the
	conformal phase for any value of $M/N$, and only sufficiently sub-Ohmic ones destroy it. In the limit $N \gg M$, $s_c \to 1$, and Ohmic dissipation becomes exactly marginal.\\

	The two conditions above show that only two features of an arbitrary environment survive in the infrared. Taking the low-frequency exponent to $s=1$, i.e., working at the marginal point, the only key quantity that remains is the static shift $\Pi^R_\mathrm{E}(0)$. This justifies modeling the environment through the single-parameter Markovian kernel
	\begin{equation}
		\Pi^R_\mathrm{E}(\omega) \overset{\rm Markov}{=} \frac{\kappa^2}{\omega_0}
		-\frac{2i\kappa\,\omega}{\omega_0} \, ,
		\label{eq:diss_markov_kernel}
	\end{equation}
	which distills the influence of the bath into a single knob $\kappa$, which we use to tune the strength of dissipation.  Equation~\eqref{eq:diss_markov_kernel} is equivalent to giving a finite width to the bare bosonic field, $D_0^R(\omega) \to \omega_0/[(\omega+i\kappa)^2-\omega_0^2]$, which is also convenient for the numerical analysis. The Markovian kernel is moreover a conservative choice on both counts.  It is Ohmic, that is, the least irrelevant among the generic environments that preserve the conformal phase; and it carries $\Pi^R_\mathrm{E}(0) = \kappa^2/\omega_0 > 0$, which spends criticality budget rather than freeing it. Consequently, the analysis based on  Eq.~\eqref{eq:diss_markov_kernel} provides an upper bound on the effect of the coupling to the bath.
	
	Equation~\eqref{eq:diss_markov_kernel} can appear from a physical environment, rather than merely yielding an upper bound on the effect of
	more general baths, if one replaces the coupling of Eq.~\eqref{eq:H_total} with a linear coupling between real quadratures, $\sum_{m,k} \,(b_{k,m}+b_{k,m}^\dagger)(a_m+a_m^\dagger)$. 
	For a bath in thermal equilibrium at the temperature $T$ of the system, the bosonic Schwinger--Dyson equations are again given by Eq.~\eqref{eq:Schwinger_Dyson_bath}. For an Ohmic spectral density, the dynamical part of $\Pi^R_E$ gives an imaginary term that is linear in the frequency, while the static part, which depends on the UV cutoff, only renormalizes $\omega_0$, reproducing an expression identical to Eq.~\eqref{eq:diss_markov_kernel} for the renormalized boson mass (see, e.g., Ref.~\cite{Pini_2025}).
	In this case, the thermal state at temperature $T$ becomes a self-consistent stationary solution of the full saddle-point equations.
	The situation is different for boson losses described by the coupling~\eqref{eq:H_total} in the Markov limit, which are compatible with a far-detuned optical cavity in the rotating-wave approximation (see Appendix \ref{app:boson_environment}). There, the bath is effectively at zero temperature, and the steady state is not thermal~\cite{Pelliconi2026}. Treating this case requires promoting the lesser propagators to independent variables in the saddle-point equations. We leave this to future work and restrict in the next sections to characterizing how weak and strong interactions to the thermal bath deform the regimes studied in Sec.~\ref{sec:phase_diagram} and the markers defined in Sec.~\ref{sec:observables}. As a consequence, the phase diagram is also modified, as shown in Fig.~\ref{fig:diss-phase-diagrams}.

	\subsection{Weak dissipation}
	
	As long as $\Pi^R_\mathrm{E}$ is small compared to all the other scales, the phases studied in Sec.~\ref{sec:phase_diagram} survive, albeit with some modifications. As we show below, in this regime one of the main effects of dissipation is to induce an effective renormalization of the interaction strength,
	\begin{equation}
		\frac{g^2}{\omega_0} \longrightarrow \frac{g^2_\mathrm{eff}}{\omega_0} = \frac{g^2}{\Pi^R_\mathrm{E}(0)+\omega_0} \, ,
		\label{eq:effective_coupling_dissipation}
	\end{equation}
	which applies uniformly across all regimes and consequently shifts some of the crossover lines towards larger values of the coupling, as shown in Fig.~\ref{fig:diss-phase-diagrams}.

	\paragraph{Conformal regime:} As long as the dissipation is an irrelevant deformation, the line of argument discussed in Sec.~\ref{subsec:conf_phase} and Appendix~\ref{app:conf_symmetry} applies at sufficiently small frequencies. Dissipation then enters only through the self-criticality condition in Eq.~\eqref{eq:self_criticality_bath}, leading to a rescaling of the overall coefficient $b$ of $G^R$ in \cref{eq:conformal_ansatz_main}:
	\begin{equation}
		b = c_1 \left( \frac{g^2}{\omega_0 + \Pi^R_\mathrm{E}(0)}\right)^{-2\Delta_\mathrm{F}} \overset{\rm Markov}{=} c_1 \left( \frac{\omega_0 g^2}{\omega_0^2 + \kappa^2}\right)^{-2\Delta_\mathrm{F}} \, ,
	\end{equation}
	where $c_1$ takes the same numerical value as in Sec.~\ref{subsec:conf_phase}. A larger value of $\Pi^R_\mathrm{E}(0)$ decreases $G^R$ and, since in the conformal limit $\Sigma^R(\omega) \sim - G^R(\omega)^{-1}$, thus suppresses $\Sigma^R(\omega)$. 
	Since the conformal behavior requires $\Sigma^R(\omega) \gg G_0^{-1}(\omega) =\omega$, an increased $\Pi^R_\mathrm{E}(0)$ restricts the conformal frequency window, which is satisfied for
	\begin{equation}
		\omega \ll \frac{g^2}{\Pi^R_\mathrm{E}(0)+\omega_0} \overset{\rm Markov}{=} \frac{\omega_0 \, g^2}{\omega_0^2 + \kappa^2}\, .
	\end{equation}
	
	At finite temperature, the renormalized mass $\omega_r^2 = \omega_0 \alpha$ of \cref{eq:boson_mass_finite_T} is also affected by the coupling with the environment:
	\begin{equation}\label{eq:gap_T_dependence_with_diss}
		\alpha = \delta \Pi^R(0) = - \big(\omega_0 + \Pi^R_E(0)\big)\, c_1^2\, \frac{N}{M}\,\frac{\tan (2 \pi \Delta_\mathrm{F})\, \Gamma (2 \Delta_\mathrm{F})^2}{\Gamma (4 \Delta_\mathrm{F})} \left( \frac{g^2}{2 \pi T \big(\omega_0 + \Pi^R_E(0)\big)} \right)^{1-4 \Delta_\mathrm{F}} 
	\end{equation}
	Hence, at fixed $T$ and $g$, a larger static self-energy $\Pi^R_\mathrm{E}(0)>0$ leads to a larger gap and therefore to a reduced self-criticality of the boson. 
	
	The boson is, however, also strongly affected by the dynamical part of the environment self-energy $\delta \Pi^R_\mathrm{E}(\omega)$. Imposing $\delta \Pi^R(\omega) \gg \delta \Pi^R_\mathrm{E}(\omega)$ yields
	\begin{equation}
		\omega \overset{\rm Markov}{\ll} \omega_0 \left(\frac{\omega_0}{\kappa}\right)^{\frac{1}{2-4\Delta_\mathrm{F}}} .
		\label{eq:dissipation_boson_conf}
	\end{equation}
	In the limit in which the number of fermionic modes is far greater than that of the bosons, $N \gg M$, one has $\Delta_\mathrm{F} \to 1/2$, so that the right-hand side of Eq.~\eqref{eq:dissipation_boson_conf} vanishes for every $\kappa > \omega_0$, completely destroying the conformal regime even at moderate dissipation strength. This is expected, since in this limit the deformation due to the bath becomes marginal. Even for $N=2M$, however, one has $1/(2-4\Delta_\mathrm{F}) \approx 3.14$, which still produces a rapid erosion of the conformal window. 
	
\begin{figure}[th!]
		\centering
		\begin{minipage}[t]{0.49\textwidth}\centering
			\dissipationpanel{\bf a)}{Criticality budget}{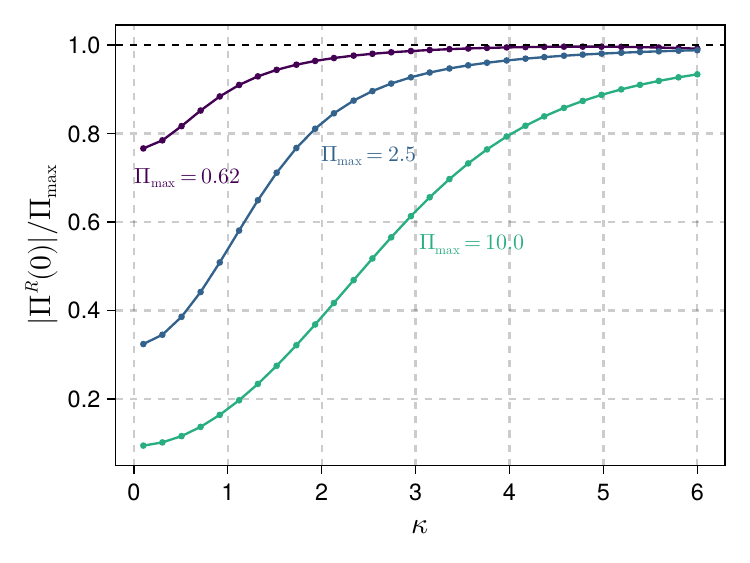}
		\end{minipage}\hfill
		\begin{minipage}[t]{0.49\textwidth}\centering
			\dissipationpanel{\bf b)}{Conformal regime}{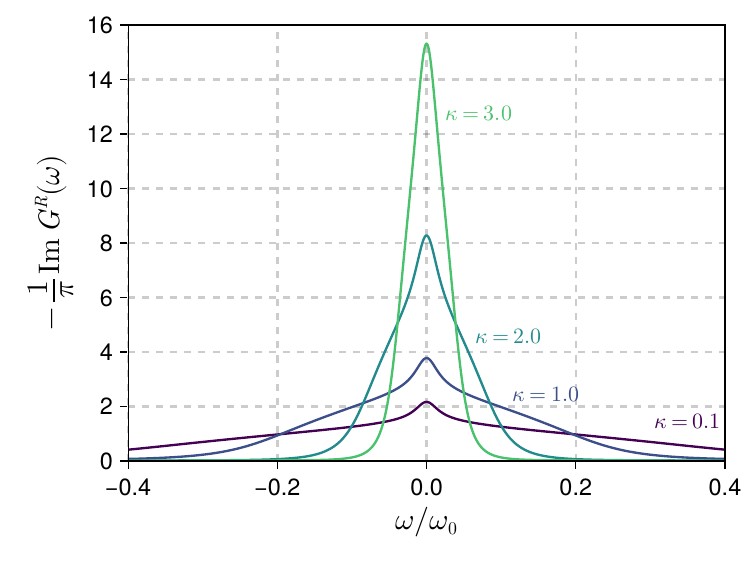}
		\end{minipage}
		
		\vspace{6pt}
		
		\begin{minipage}[t]{0.49\textwidth}\centering
			\dissipationpanel{\bf c)}{SYK$_2$ regime}{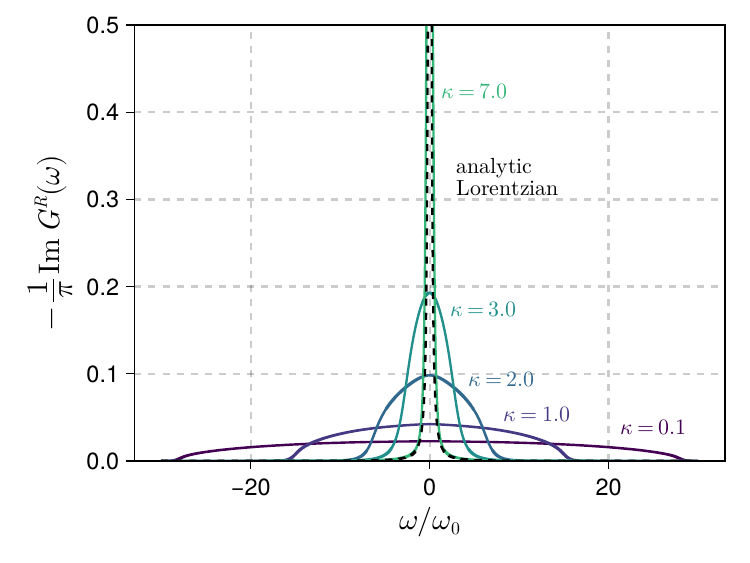}
		\end{minipage}\hfill
		\begin{minipage}[t]{0.49\textwidth}\centering
			\dissipationpanel{\bf d)}{Unresolved sidebands regime}{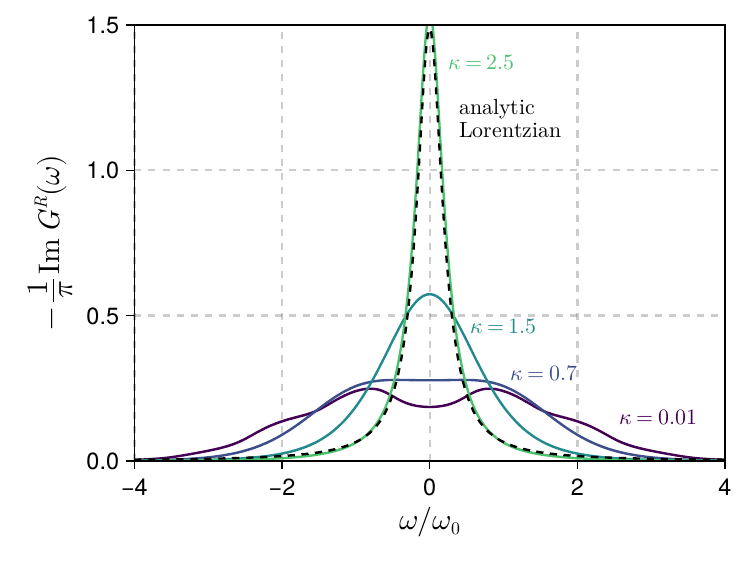}
		\end{minipage}
		
		\caption{Effect of dissipation on the different regimes of the phase diagram. {\bf a)} Static self-energy $\Pi^R(0)$ as a function  of the dissipation strength, in units of the criticality budget $\Pi_\mathrm{max}$ at fixed temperature $T=0.2$ and $N=2M$. As $\kappa$ grows the fermions saturate the budget, therefore becoming closer and closer to a free particle, while the boson is driven away from criticality. {\bf b)}--{\bf d)} Fermionic spectral function for increasing dissipation in the conformal ($T=0.007$, $g=0.5$), impurity ($T=1$, $g=4$) and driven-fermion regimes ($T=9$, $g=0.5$), showing how all three interpolate from the characteristic shape of each regime to a free-fermion solution at large $\kappa$. Black dashed lines are the analytic large dissipation result of \cref{eq:free_fermion_dissipation} with $\eta$ given by \cref{eq:eta_diss_def}. No such comparison is shown in the conformal regime, since that phase has the largest criticality budget, so overcoming it requires a much stronger dissipation, and the resulting fermionic width becomes so small that resolving it would demand a frequency grid both very fine and very wide, since the boson becomes more broad as the fermion becomes more sharp.}
		\label{fig:dissipation_regimes}
	\end{figure}

	Dissipation therefore pushes the conformal behavior towards progressively smaller frequencies. Combined with a finite temperature, this may cause the conformal window, which extends over
	\begin{equation}
		T \ll \omega \overset{\rm Markov}{\ll} \omega_0  \, \, \min \left(\left(\frac{\omega_0}{\kappa}\right)^{\frac{1}{2(1-2\Delta_\mathrm{F})}} ,\frac{g^2}{\omega_0^2 + \kappa^2} \right),
		\label{eq:conformal_window_diss}
	\end{equation}
	to close entirely. This makes the observables introduced in Sec.~\ref{sec:observables} even more relevant,  since they diagnose the onset of the conformal phase without requiring a direct fit of the conformal window, which may be very hard to resolve in realistic implementations.
	Figure~\ref{fig:dissipation_regimes}{\bf b)} shows numerically how dissipation destabilizes the conformal phase.

	\paragraph{SYK$_2$ regime:} Weak dissipation does not change the functional form of the two-point functions in Eq.~\eqref{eq:impurity_green}; it only modifies the constants defined in Eq.~\eqref{eq:impurity_constants}, as we now determine. As in the conformal case, dissipation affects the propagators through the self-criticality condition \eqref{eq:self_criticality_bath}, which, repeating the derivation of the dissipationless case \eqref{eq:impurity_self_appendix}, gives
	\begin{equation}
		\Omega_0=\frac{8g^2N}{3\pi M\,(\omega_0 + \Pi^R_\mathrm{E}(0))} \, .
		\label{eq:Omega_T_dependence_diss}
	\end{equation}
	The renormalized boson mass is again related to $\Omega_0$ by
	\begin{equation}
		\omega_r^2=\frac{4g^2T\omega_0}{\Omega_0^2}\propto (\omega_0 + \Pi^R_\mathrm{E}(0))^2 \, .
		\label{eq:omega_r_T_dependence_diss}
	\end{equation}
	Since $\Omega_0$ is the radius of the semicircular fermionic spectral density of Eq.~\eqref{eq:rho_impurity}, we conclude that dissipation makes the fermionic spectrum narrower while leaving the boson mass less renormalized, and therefore further from criticality than in the dissipationless case. This is also confirmed by the numerical simulation in Fig.~\ref{fig:dissipation_regimes}{\bf c)}. Finally, in the Markovian case, it is simple to account for the effect of dissipation on the boson broadening, as it adds to the one produced by the interaction:
	\begin{equation}
		\gamma \overset{\rm Markov}{=} \frac{3 \omega_0^2}{4  \Omega_0} + \kappa \, .
	\end{equation}

	\paragraph{Bare boson regimes:} At high temperature, the statement $\Pi^R\equiv0$ of Eq.~\eqref{eq:PiR_highT} survives the introduction of dissipation, so that the boson becomes entirely controlled by the environment. 
	If we work in the regime where the boson broadening can be ignored, we can still write $D^<(\omega) \propto\delta(\omega-\omega_0)+\delta(\omega+\omega_0)$, 
	\begin{equation}
		\Sigma^R(\omega) \sim i g^2 \int \frac{d \omega'}{2 \pi} G^R(\omega') D^<(\omega-\omega') = i \frac{g^2}{2} I [G^R(\omega+\omega_0)+G^R(\omega-\omega_0)] \, ,
	\end{equation}
	with the proportionality constant $I$ fixed by
	\begin{equation}
		I = \int \frac{\dd \omega}{2 \pi} D^<(\omega) = i T \int \frac{\dd \omega}{\pi} \frac{\Im D^R(\omega)}{\omega} = i T \Re D^R(0) = -i \frac{T}{\omega_0 + \Pi^R_\mathrm{E}(0)} \, ,
	\end{equation}
	where in the first step we expanded the boson occupation $n_B(\omega)$ for large temperature, while in the second we used Kramers--Kronig relations.
	We therefore have
	\begin{equation}
		\Sigma^R(\omega) = \frac{g^2 T}{2(\omega_0 + \Pi^R_\mathrm{E}(0))} [G^R(\omega+\omega_0)+G^R(\omega-\omega_0)] \, , 
		\label{eq:high_T_generic_diss}
	\end{equation}
	which means that the free-fermion boundary $T \propto g^{-2}$ is pushed to larger $g$ accordingly.
	
	In the Markovian case, we can be more explicit and calculate how the dissipation broadens the fermion. For $t>0$, \cref{eq:SigmaR_t} reduces to
	\begin{equation}
		\Sigma^R(t) = i g^2 G^R(t) D^<(t) \qquad D^<(t) \overset{\rm Markov}{=} -i\frac{T}{2} \left[ \frac{e^{i (\omega_0+i \kappa) t}}{\omega_0+i \kappa}  + \frac{e^{-i (\omega_0-i \kappa) t}}{\omega_0-i \kappa}\right]
	\end{equation}
	which translates in frequency to 
	\begin{equation}
		\Sigma^R(\omega)  \overset{\rm Markov}{=} \frac{g^2 T}{2} \left[\frac{G^R(\omega+\omega_0+ i \kappa)}{\omega_0+ i \kappa}+\frac{G^R(\omega-\omega_0+i \kappa)}{\omega_0-i \kappa}\right] \, .
		\label{eq:SigmaR_markov_highT}
	\end{equation}
	This equation become equal to \eqref{eq:high_T_generic_diss} if we ignore the shift $+i\kappa$ of the arguments and set $i \kappa [G^R(\omega+\omega_0)-G^R(\omega-\omega_0)]\approx 0$, which are the terms responsible for the broadening of the fermion. At small coupling, i.e. in the resolved sidebands regime, we can approximate $G^R(\omega)\sim 1/\omega$ in \cref{eq:SigmaR_markov_highT} and find\footnote{Since we are taking the free-fermion limit for the central band, this coincides with $\eta$ in the case of strong dissipation considered in~\cref{sec:strong_diss}.}
	\begin{equation}
		\Im\Sigma^R(0) \sim -g^2T\frac{2 \kappa\,\omega_0}{(\omega_0^2+\kappa^2)^2} \, .
	\end{equation}
	This suggests that the central peak in the band structures becomes a Lorentzian of width $\sim \kappa$ and the gaps between sidebands turn into pseudogaps. Therefore, the resolved-sideband regime is no longer sharply defined and disappears at strong dissipation (see Fig.~\ref{fig:diss-phase-diagrams}).  
	As $\kappa$ grows, the Floquet resonances discussed in Sec.~\ref{sec:high_t_regimes} are progressively less pronounced until they are completely washed out at larger coupling (see Fig.~\ref{fig:dissipation_regimes}{\bf d)}).

	\subsection{Strong dissipation}
	\label{sec:strong_diss}
	
	In this section, we consider the opposite limit to that discussed above, in which the coupling to the environment becomes the dominant scale in the problem. Since the bosonic dynamics is then completely dominated by the environmental self-energy, we fix the bath to the Markovian case, which allows for closed-form expressions.
	The large-$\kappa$ limit has two important consequences. First, since $\Pi^R(0)$ is bounded from above by the $\kappa$-independent criticality budget $\Pi_\mathrm{max}$, the boson approaches a free damped particle
	\begin{equation}
		D^R(\omega) \sim \frac{\omega_0}{(\omega +i \kappa )^2-\omega_0^2} \, , \qquad \kappa \gg 1 \, . \label{eq:free_boson_dissipation}
	\end{equation}
	Second, the tendency of the YSYK interaction towards criticality drives $\Pi^R(0)$ towards saturation of the criticality budget.
	This is shown in Fig.~\ref{fig:dissipation_regimes}{\bf a)}, where $|\Pi^R(0)|$ approaches $\Pi_\mathrm{max}$ as $\kappa$ grows.
	As discussed in Sec.~\ref{sec:criticality_budget}, this  can occur only when the fermion becomes a very narrow quasiparticle:
	\begin{equation}
		G^R \sim \frac{1}{\omega+ i \eta} \, , \qquad \eta \ll 1 \, .
		\label{eq:free_fermion_dissipation}
	\end{equation}
	Under these assumptions, the remaining saddle-point equations can be solved self-consistently. The imaginary part of the fermion self-energy follows from
	\begin{equation}
		\begin{split}
			\Im \Sigma^R(\omega) &= - g^2 \int \frac{\dd \omega'}{2 \pi} [\Im G^R(\omega') D^<(\omega' -\omega)- G^<(\omega') \Im D^R(\omega' -\omega)] \\
			&= g^2 \coth\left( \frac{\omega}{2T} \right) \Im D^R(-\omega) \sim -  g^2 T \frac{2\kappa \omega_0}{\left(\kappa ^2+\omega_0^2\right)^2}\,,
		\end{split}
	\end{equation}
	where, in the second equality, we have approximated the fermionic spectral density to be a Dirac delta compared to the scale set by the boson two-point functions, while in the last step we took the $\omega \ll \kappa$ limit. This yields 
	\begin{equation}
		\eta = g^2 T \frac{2\kappa \omega_0}{\left(\kappa ^2+\omega_0^2\right)^2}\,,
		\label{eq:eta_diss_def}
	\end{equation}
	which is indeed rapidly suppressed as $\kappa$ increases.
	The real part of $\Sigma^R$ can be obtained from the Kramers--Kronig relations and scales as $\Re \Sigma^R \sim \kappa^{-3}$. It is therefore subleading compared to the inverse free-fermion propagator and can be neglected, confirming the ansatz \eqref{eq:free_fermion_dissipation}. The tendency of the fermionic propagator to collapse onto this same free-quasiparticle form throughout the phase diagram is illustrated in Fig.~\ref{fig:dissipation_regimes}{\bf c)} and Fig.~\ref{fig:dissipation_regimes}{\bf d)}. This is reflected in the progressive expansion of the free-fermion regime across the phase diagram a strong coupling, as schematically illustrated in Fig.~\ref{fig:diss-phase-diagrams}.
	
	We next consider the bosonic self-energy. While we saw that dissipation suppresses $D^R$ and therefore $\Sigma^R$, $\Pi^R$ is built out of fermion propagators only and becomes larger as the fermionic quasiparticle sharpens. Starting from \cref{eq:PiR_t} in frequency space, we obtain
	\begin{equation}
		\Im \Pi^R(\omega) = - g^2 \frac{N}{M} \int \frac{\dd \omega'}{\pi} \Im G^R(\omega') \Im G^R(\omega'-\omega) (n_\mathrm{F}(\omega'-\omega)-n_\mathrm{F}(\omega')) \sim - \Pi_\mathrm{max} \frac{2 \omega \eta}{\omega^2 + (2\eta)^2}\,,
		\label{eq:Im_PI^R_dissipation}
	\end{equation}
	where we have expanded the Fermi distribution for small $\omega$ and $\omega'$. The corresponding retarded self-energy can then be obtained by a residue integral,
	\begin{equation}
		\Pi^R(\omega) = \int \frac{\dd \omega'}{\pi} \frac{\Im \Pi^R(\omega')}{\omega'-\omega-i0^+} = - \frac{\Pi_\mathrm{max}}{1-i \omega / 2\eta} \, ,
		\label{eq:Drude_self-energy_diss}
	\end{equation}
	and takes the form of a Drude-like response \cite{Leggett_1987,Hu_1992}.
	Since $\Pi_\mathrm{max} \ll \kappa$, the ansatz \eqref{eq:free_boson_dissipation}  is self-consistent as well, and the saddle-point equations are therefore solved consistently in the large-$\kappa$ limit.

	\subsection{Phase signatures under dissipation}
	\label{sec:Phase_signatures_diss}
	
	We now ask how the markers introduced in Sec.~\ref{sec:observables} respond to dissipation, and whether they still separate the regimes of Fig.~\ref{fig:diss-phase-diagrams}. This is where they are most needed: the bath narrows the conformal window from above (Eq.~\eqref{eq:conformal_window_diss}) while the temperature cuts it from below, such that resolving the conformal power laws becomes increasingly difficult. We first follow the crossover lines through $\alpha, \xi$ and $W$ (Sec.~\ref{sec:phase_diagram_diss}), and then turn to the QFI (Sec.~\ref{sec:QFI_diss}).
	
	\subsubsection{Phase Diagram}\label{sec:phase_diagram_diss}
	
	We repeat the constant-$\alpha$ analysis of Sec.~\ref{sec:markers} for the Markovian kernel~\eqref{eq:diss_markov_kernel} with $\kappa=0.1$ and $\kappa=1$; the complete set of results is shown in Figs.~\ref{fig:h_fits_dissipation} and~\ref{fig:h_collapse_dissipation}, and Fig.~\ref{fig:QFI_results_dissipation} for the QFI, of Appendix~\ref{app:additional_results_diss}. Its static shift $\Pi^R_E(0)=\kappa^2/\omega_0$ enters every quantity that probes the boson only at $\omega=0$ through the effective coupling $g_{\rm eff}$ of Eq.~\eqref{eq:effective_coupling_dissipation}, and displaces the corresponding crossover lines rigidly. Its dynamical part, $-2i\kappa/\omega_0$, instead broadens the spectral features and reshapes the lines located by markers that probe finite frequencies.
	
\begin{figure}[t]
		\centering
		\begin{minipage}[t]{0.49\textwidth}\centering
			\dissipationpanel{\bf a)}{}{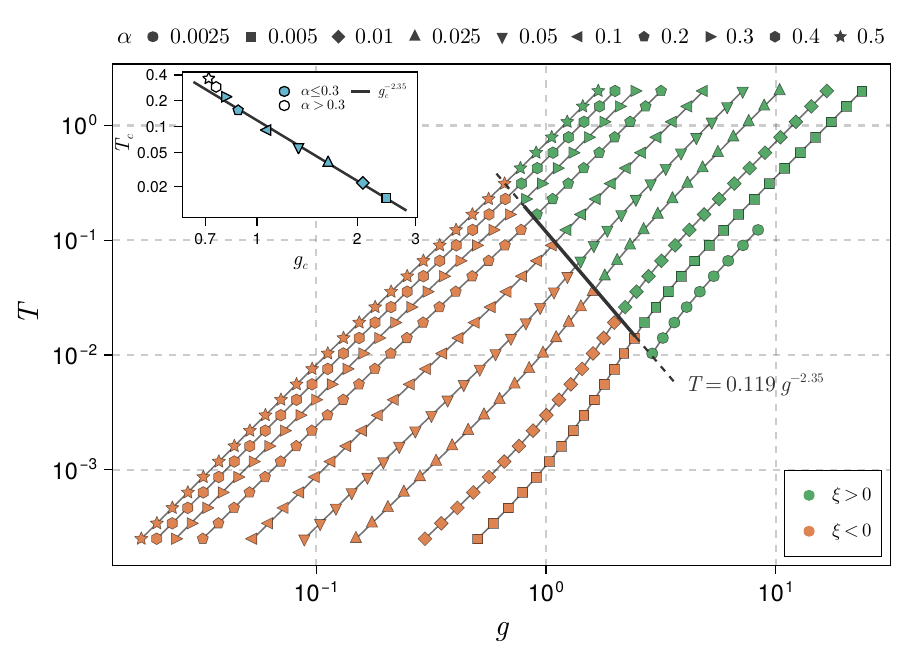}
		\end{minipage}\hfill
		\begin{minipage}[t]{0.49\textwidth}\centering
			\dissipationpanel{\bf b)}{}{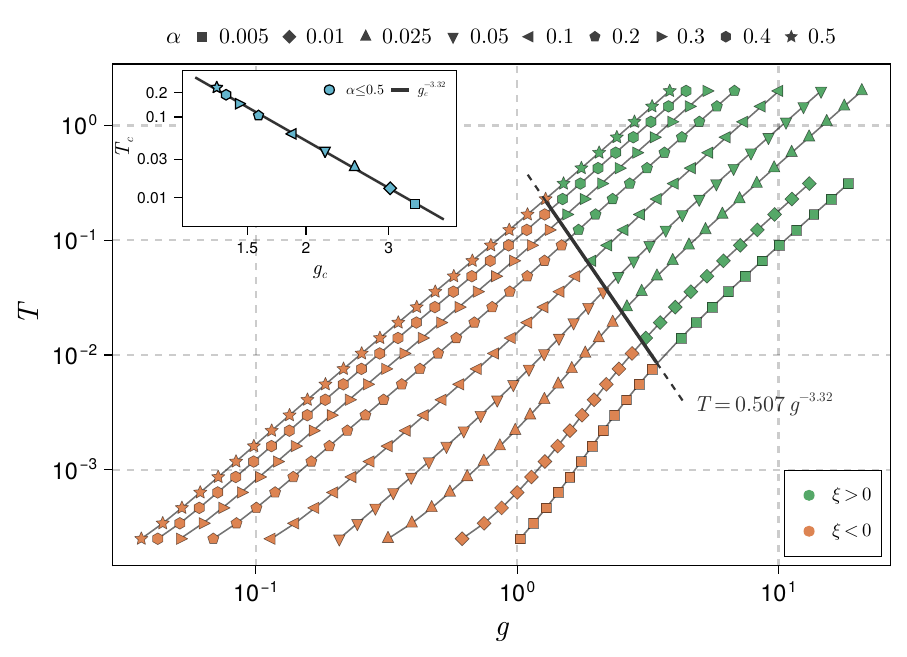}
		\end{minipage}
		
		\caption{Constant-gap curves in the $(g,T)$-plane with dissipation, $\kappa=0.1$ in \textbf{a)} and $\kappa=1$ in \textbf{b)} (in units of $\omega_0$), to be compared with the closed system in Fig.~\ref{fig:Phase_diagram_plot}. Each curve is a temperature sweep at fixed gap $\alpha$, distinguished by marker shape. Green markers indicate $\xi>0$ (SYK$_2$ regime) and orange markers $\xi<0$ (conformal regime). The insets show the critical points $(g_c,T_c)$ where $\xi=0$, together with the power-law fits, $T_c=(0.119\pm 0.004)g_c^{-2.35\pm 0.07}$ with $R^2=0.995$ in \textbf{a)} and $T_c=(0.507\pm 0.016)g_c^{-3.32\pm 0.04}$ with $R^2=0.999$ in \textbf{b)}, which are also drawn as the conformal/SYK$_2$ boundary in the main panels (solid over the fitted range, dashed outside). In both panels, the boundary steepens, so that at large coupling the conformal regime shrinks in favor of the SYK$_2$ one.}
		\label{fig:Phase_diagram_xi_dissipation}
	\end{figure}

	\paragraph{Bare-boson boundary.} With a bath, the definition~\eqref{eq:gapdef} generalizes to 
	\begin{equation}
		\alpha=-D^R(0)^{-1} = \omega_0 + \Pi^R_E(0) + \Pi^R(0) \,, 
		\label{eq:alpha_diss}
	\end{equation}
	which is the quantity held fixed in the numerics, as discussed in Sec.~\ref{sec:markers} (see also Appendix~\ref{app:numerics}). Since Eqs.~\eqref{eq:PiR0_spectral} and~\eqref{eq:criticaly_budget} do not involve the environment, and the static susceptibility $1/\alpha$ is positive by Eq.~\eqref{eq:gap_sum_rule}, the gap is confined to 
	\begin{equation}
		\max\!\Big(0,\,1-\frac{T_b}{T}\Big)\le\frac{\alpha}{\omega_0 + \Pi^R_E(0)}\le1,\qquad T_b\equiv\frac N{4M}\frac{g^2}{\omega_0 + \Pi^R_E(0)} = \frac{N}{4M}\frac{g^2_{\rm eff}}{\omega_0} \, .
		\label{eq:rescaled_gap_diss}
	\end{equation}
	The upper end corresponds to an undressed boson, while the lower one is the criticality budget, reappearing here as a statement about a measurable quantity: $T_b$ is the temperature above which the fermions can no longer close the gap, i.e., Eq.~\eqref{eq:budget_with_bath} solved for $T$. The natural marker is therefore the normalized gap $\alpha/[\omega_0+\Pi^R_E(0)]$.\footnote{Since $\omega_0 + \Pi^R_E(0) = -D^R(0)^{-1}|_{g=0}$, it is simply the ratio of the static responses measured without and with the Yukawa coupling, and thus requires neither the calibration discussed in Appendix~\ref{app:homodyne} nor any knowledge of the bath.} Because $\alpha$ is a static quantity, its temperature dependence in the conformal and SYK$_2$ regimes follows directly from Eqs.~\eqref{eq:gap_T_dependence_with_diss} and~\eqref{eq:Omega_T_dependence_diss}--\eqref{eq:omega_r_T_dependence_diss}. Expressed through $g_{\rm eff}$ and the normalized gap, the constant-gap curves~\eqref{eq:T_g2_conf_imp} keep their closed-system form,
	\begin{equation}
		T_{\rm conf} \propto \frac{g^2_{\rm eff}}{\omega_0} \left( \frac{\alpha}{\omega_0 + \Pi^R_E(0)} \right)^{1/(4\Delta_F-1)} \,, \qquad T_\text{SYK$_2$} = \frac{64}{9\pi^2} \frac{g^2_{\rm eff}}{\omega_0} \frac{\alpha}{\omega_0 + \Pi^R_E(0)} \,,
		\label{eq:T_conf_imp_diss}
	\end{equation}
	so that at fixed gap and temperature they are displaced towards larger g. Figure~\ref{fig:Phase_diagram_xi_dissipation} shows the constant-$\alpha$ sweeps, labeled by the sign of the curvature $\xi$, for $\kappa=0.1$ in panel \textbf{a)} and for $\kappa=1$ in panel \textbf{b)}. For $\kappa=1$ the displacement with respect to Fig.~\ref{fig:Phase_diagram_plot} is clearly visible. For the power law~\eqref{eq:halphafit}, Eq.~\eqref{eq:T_conf_imp_diss} changes only the amplitude $A$, and indeed the fit in Fig.~\ref{fig:h_fits_dissipation}{\bf b)} gives $B=1.304 \pm 0.008$ for $\kappa=0.1$, close to the closed-system value $B=1.309 \pm 0.007$. Correspondingly, the extrapolation towards the bare-boson limit $\alpha \to (\omega_0^2+\kappa^2)/\omega_0$ turns the boundary~\eqref{eq:hightempboundary} into
	\begin{equation}
		T = 1.861 \; \frac{g^2\omega_0}{\omega_0^2 + \kappa^2} = 1.861 \; \frac{g^2_{\rm eff}}{\omega_0} \, ,
		\label{eq:T_g^2_diss}
	\end{equation}
	which predicts the extrapolated amplitudes $1.842$ and $0.930$ for $\kappa=0.1$ and $1$, respectively. For $\kappa=0.1$ this agrees with Fig.~\ref{fig:h_fits_dissipation}b), while for $\kappa=1$ the extrapolation in Fig.~\ref{fig:h_fits_dissipation}{\bf d)} lies below it, owing to the smallest gaps, where the dynamical part of the bath enters (Appendix~\ref{app:additional_results_diss}).\footnote{This deviation is amplified by the longer extrapolation, from $\alpha=0.5$ to $2\omega_0$ instead of $\omega_0$ (cf. Sec.~\ref{sec:markers}).} Moreover the budget line $T_b$ and the extrapolated bare-boson boundary are both of the form $T \propto g^2_{\rm eff}/\omega_0$, so dissipation displaces them rigidly towards larger coupling, by a factor $\sqrt{1 + \kappa^2/\omega^2_0}$ in $g$ at fixed temperature.
	
	\paragraph{Conformal/SYK$_2$ boundary.}
	Repeating the analysis of Fig.~\ref{fig:Phase_diagram_plot}, we locate on each constant-$\alpha$ sweep the point $(g_c,T_c)$ where $\xi$ changes sign and fit the resulting boundary (insets of Fig.~\ref{fig:Phase_diagram_xi_dissipation}), obtaining
	\begin{equation}
		T_c = (0.119\pm0.004)\,g_c^{-2.35\pm0.07}\ (\kappa=0.1)\,,\qquad
		T_c = (0.507\pm0.016)\,g_c^{-3.32\pm0.04}\ (\kappa=1)\,,
		\label{eq:conf_SYK_boundary_diss}
	\end{equation}
	to be compared with $T_c = (0.135 \pm 0.007) g_c^{-1.96 \pm 0.09}$ without dissipation. The rescaling in Eq.~\eqref{eq:effective_coupling_dissipation} alone would leave the exponent unchanged and only raise the amplitude. Instead, the boundary steepens and lies below this prediction, i.e., beyond the static shift, the conformal regime shrinks in favor of the SYK$_2$ one. This is most likely due to the dynamical part of the bath, Eq.~\eqref{eq:diss_markov_kernel}: although an Ohmic bath is irrelevant at the conformal fixed point, its weight relative to $\delta\Pi^R(\omega)$ decays only as $|\omega|^{2-4\Delta_F} \approx |\omega|^{0.32}$, so that it still reduces to the conformal regime at the accessible temperatures, consistent with the deviation of the low-temperature amplitudes from the static rescaling in Fig.~\ref{fig:h_collapse_dissipation}.
	
	Notice that something peculiar happens in the strong-dissipation limit of Sec.~\ref{sec:strong_diss}. The Drude-like self-energy~\eqref{eq:Drude_self-energy_diss} gives 
	\begin{equation}
		\xi = \frac{2}{\omega_0}-\frac{\Pi_\mathrm{max}}{2 \eta^2} \, ,
	\end{equation}
	which is negative whenever $4\eta^2<\Pi_{\rm max}\omega_0$. Since $\eta \propto \kappa^{-3}$ (Eq.~\eqref{eq:eta_diss_def}) while $\Pi_{\rm max}$ does not depend on the bath, this condition is always met at sufficiently strong dissipation, even though the boson is then far from criticality: the budget is saturated, $\alpha/[\omega_0 + \Pi^R_E(0)] \simeq 1 - T_b/T$, which is close to one since $\Pi_{\rm max} \ll \omega_0 + \Pi^R_E(0)$. Here the negative curvature originates from the sharp fermionic quasiparticle, which produces a narrow peak of width $2\eta$ in $\Pi^R(\omega)$ around $\omega=0$, and not from self-criticality. A negative curvature therefore identifies the conformal regime only in combination with a small gap $\alpha$.
	
	\paragraph{Fermionic markers.}
	In the bare-boson regimes, dissipation acts in two ways: the static shift $\Pi^R_E(0)$ rescales the effective coupling as in \cref{eq:high_T_generic_diss}, while the boson linewidth is transferred to the fermion through \cref{eq:SigmaR_markov_highT}. To see how the latter reshapes the weight $W$ of Sec.~\ref{sec:fermion_weight}, we adopt the Markov approximation and replace $G^R(z)\simeq1/z$ on the right-hand side of \cref{eq:SigmaR_markov_highT} in the small coupling limit, which gives
	\begin{equation}
		\Im\Sigma^R(\omega) \simeq -\frac{2g^2T\kappa\,\omega_0}{\left[(\omega-\omega_0)^2+\kappa^2\right]\left[(\omega+\omega_0)^2+\kappa^2\right]} \, .
	\end{equation}
	As long as $g^2T / \omega_0^2 \ll\kappa \ll \omega_0$, the self-energy for $\omega \sim \omega_0$ is small, $\rho(\omega)\simeq-\Im\Sigma^R(\omega)/(\pi\omega^2)$, and the sideband becomes a Lorentzian of half-width $\kappa$,
	\begin{equation}
		\rho(\omega_0+\omega)\simeq \frac{g^2T}{2\omega_0^3}\,\frac{\kappa/\pi}{\omega^2+\kappa^2} \, .
	\end{equation}
	Integrating the Lorentzian over the bin of \cref{eq:spectral_weight} gives
	\begin{equation}
		W_{\omega_0}=\frac{g^2T}{2\pi\,\delta\,\omega_0^3}\arctan\frac{\delta}{\kappa} \, ,
	\end{equation}
	and setting $W_{\omega_0}=\delta_{\mathrm{dev}}$ yields
	\begin{equation}
		T = \frac{2\pi\,\delta\,\delta_{\mathrm{dev}}}{\arctan(\delta/\kappa)}\,\frac{\omega_0^3}{g_{\mathrm{onset}}^2} \, ,
		\label{eq:free_boundary_dissipation}
	\end{equation}
	which reduces to \cref{eq:free_boundary_prediction} for $\kappa=0$. At finite $\kappa$, the broadening of the fermion reduces the support of the spectral density around $\omega_0$, $W_{\omega_0} \le W_{\omega_0}|_{\kappa=0}$, and the free-theory regime therefore extends to larger couplings.
	It is worth noticing that, for small $\kappa$, the rescaling of the coupling constant \cref{eq:effective_coupling_dissipation} enters only at order $\kappa^2/\omega_0^2$, so it is the broadening that produces the boundary shift in Fig.~\ref{fig:diss-phase-diagrams}.
	
	At the same time, since $\rho(\omega_0/2)$ is finite already at weak dissipation, $W_{\omega_0/2}$ no longer vanishes in the resolved regime and its onset moves to smaller couplings. In particular, the threshold-independent definition of the sideband merging is lost: as $\delta_{\mathrm{dev}}\to0$ the onset is pushed to $g\to0$, and the resolved/unresolved boundary becomes a convention, like the free-fermion one. Squeezed from both sides, the resolved sideband regime survives only for weak dissipation, as shown in Fig.~\ref{fig:diss-phase-diagrams}. 
	
	The unresolved regime, where the bands are already broader than $\kappa$, is instead affected mainly by the static rescaling, which reduces the semicircle radius to $\Omega_0^2=4g^2_{\rm eff}T/\omega_0$; its boundary with the SYK$_2$ regime is the bare-boson line discussed above, displaced rigidly to larger coupling.

	\subsubsection{Quantum Fisher information}\label{sec:QFI_diss}
	
\begin{figure}[t]
		\centering
		\includegraphics[width=1\linewidth]{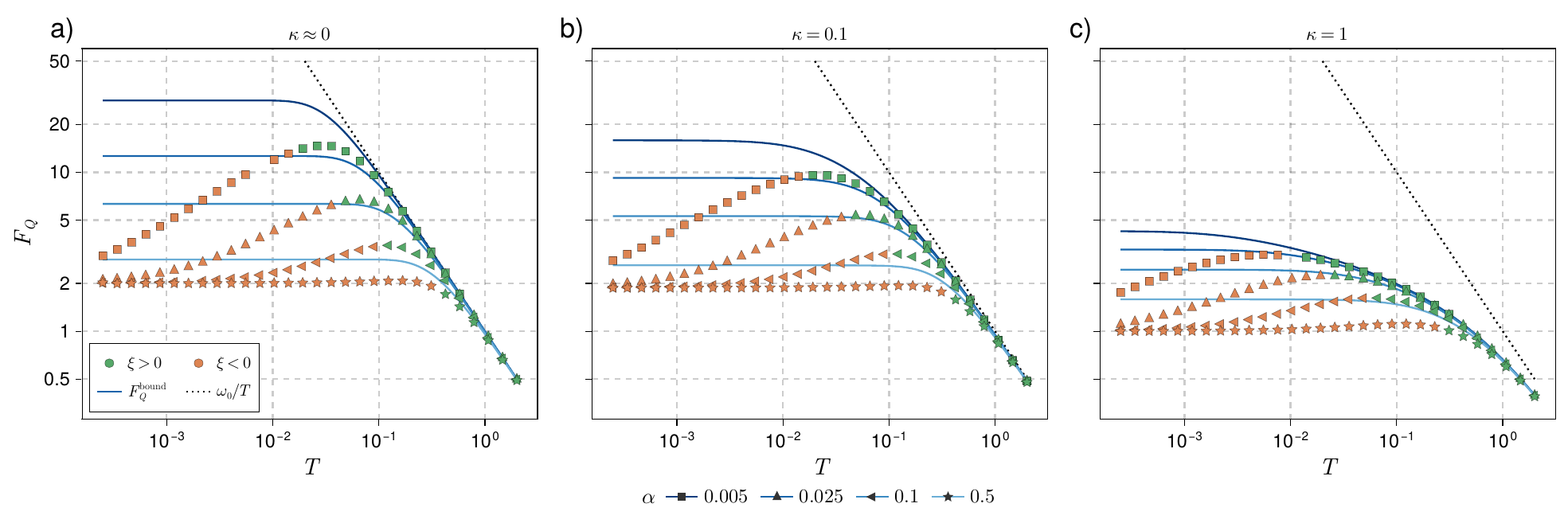}
		\caption{Quantum Fisher information $F_Q$ of the uniform bosonic quadrature as a function of temperature along four constant-gap sweeps, $\alpha=0.005, 0.025, 0.1$ and $0.5$ (marker shape, in units of $\omega_0$), for increasing dissipation: {\bf a)} $\kappa \approx 0$, i.e., only the small numerical broadening (see Appendix~\ref{app:numerics}, corresponding to the closed-system data of Sec.~\ref{sec:QFI}, {\bf b)} $\kappa=0.1$, and {\bf c)} $\kappa=1$. The marker color gives the sign of the curvature $\xi$ (green: $\xi>0$, SYK$_2$; orange: $\xi<0$, conformal). Solid lines show the bound $F_Q^{\rm bound}(\omega_r,\kappa;T)$, from dark to light blue with increasing $\alpha$. The sum-rule bound $\omega_0/T$ is plotted as a dotted line for reference. In all three panels, the curves deep in the SYK$_2$ regime saturate the bound and the maximum of each sweep remains close to the sign change of $\xi$, i.e., at the conformal/SYK$_2$ crossover. Dissipation lowers and flattens this maximum, but does not fully remove it. Moreover, at high temperature the data approach $\omega_0/T$ only as long as $T\gg\kappa$, clearly showing the deviation of the bound from the closed-system case for increasing loss.}
		\label{fig:QFI_vs_dissipative_bound}
	\end{figure}

	The sum rule in Eq.~\eqref{eq:f-sum_rule}, which holds for any number-conserving bath such as the one in Eq.~\eqref{eq:H_total}, and the Kramers-Kronig relation~\eqref{eq:gap_sum_rule}, with $\alpha$ given by Eq.~\eqref{eq:alpha_diss}, remain valid in the presence of the bath. Consequently, the representation of $F_Q$ as a moment of the measure $\mu$, and with it the bound~\eqref{eq:QFI_bound_sharp}, continues to hold for any environment.\footnote{With a bath, Eq.~\eqref{eq:QFIdef} evaluated with the full $D^R$ is the QFI of the global equilibrium state of system and environment. Since $X$ acts on the system only, monotonicity of the QFI under the partial trace makes it an upper bound on the QFI of the reduced bosonic state.} Dissipation, however, makes this bound far from tight, since the bath spreads the spectral weight of the boson over a width of order $\kappa$. A sharper bound follows from rewriting Eq.~\eqref{eq:QFIdef} on the imaginary axis. Using $\tanh(\omega/2T) = \sum_{n\geq 0} 4T\omega / (\omega^2+\nu_n^2)$ and the spectral representation $D^R(i\nu)=(2/\pi) \int_0^\infty \dd{\omega}\,\omega \operatorname{Im}D^R(\omega)/(\omega^2+\nu^2)$, we find
	\begin{equation}
		F_Q = -8T\sum_{n\ge0} D^R(i\nu_n) \, , \qquad \nu_n = (2n+1)\pi T \, .
		\label{eq:QFI_in_complex_freq}
	\end{equation}
	For the Markovian kernel~\eqref{eq:diss_markov_kernel}, we have 
	\begin{equation}
		-D^R(i\nu)^{-1} = \frac{(\nu^2+2\kappa\nu)}{\omega_0} + \alpha + \delta\Pi^R(i\nu) \, , \qquad \delta\Pi^R(i\nu) \equiv \Pi^R(i\nu) - \Pi^R(0) \geq 0 \,.
		\label{eq:D^R_complex_freq}
	\end{equation}
	Since $\operatorname{Im}\Pi^R(\omega) \leq 0$ for $\omega > 0$ due to Eq.~\eqref{eq:Im_PI^R_dissipation}, which is expected as this corresponds to the fermions damping the bosons, and using Kramers-Kronig we find that $\delta\Pi^R(i\nu) = \Pi^R(i\nu) - \Pi^R(0) \geq 0$. Consequently, we find the following bound on the QFI
	\begin{align}
		F_Q \leq F_Q^{\rm bound}(\omega_r, \kappa ; T) &\equiv 8T \sum_{n\geq 0} \frac{\omega_0}{\nu^2_n + 2\kappa\nu_n + \omega_r^2} \\
		&= \frac{2\omega_0}{\pi \sqrt{\kappa^2-\omega^2_r}} \left[ \psi\left( \frac{1}{2} + \frac{\kappa + \sqrt{\kappa^2-\omega^2_r}}{2\pi T} \right) - \psi\left( \frac{1}{2} + \frac{\kappa - \sqrt{\kappa^2-\omega^2_r}}{2\pi T} \right) \right] \, , 
		\label{eq:QFI_general_bound_dissipation}
	\end{align}
	where $\omega_r^2=\omega_0\alpha$ and $\psi$ is the digamma function.\footnote{The case $\kappa = \omega_r$ follows by continuity and gives $F_Q(\kappa=\omega_r;T) = \frac{2\omega_0}{\pi^2 T} \psi^{(1)} \left( \frac{1}{2} + \frac{\omega_r}{2\pi T} \right)$.} The bound is saturated if and only if $\delta\Pi^R(i\nu_n)=0$ for all $n$ and can be generalized to general baths by simply replacing $2\kappa\nu_n/\omega_0$ in Eq.~\eqref{eq:QFI_general_bound_dissipation} by $\delta\Pi^R_E(i\nu) = \Pi^R_E(i\nu_n) - \Pi^R_E(0) \geq 0$. For $\kappa=0$, the two digamma functions combine into $(4\omega_0/\pi\omega_r) \operatorname{Im} \psi(\frac{1}{2} + i\omega_r/2\pi T) = (2\omega_0/\omega_r)\tanh(\omega_r/2T)$, which is exactly Eq.~\eqref{eq:QFI_bound_sharp}. Equation~\eqref{eq:QFI_general_bound_dissipation} is thus the dissipative generalization of the closed-system bound, shown as solid lines in Fig.~\ref{fig:QFI_vs_dissipative_bound}.
	
	At zero temperature the global ground state of system and bath is pure, so that $F_Q = 4 \mathrm{Var}(X) = 4\langle X^2 \rangle$, since $\langle X \rangle = 0$. The bound~\eqref{eq:QFI_general_bound_dissipation} then becomes a statement about the zero-point fluctuations,
	\begin{equation}
		\langle X^2 \rangle_{T=0} \leq \frac{F_Q^{\rm bound}(\omega_r,\kappa;0)}{4} = \frac{\omega_0}{2\pi \sqrt{\kappa^2 - \omega_r^2}} \ln\left( \frac{\kappa + \sqrt{\kappa^2-\omega_r^2}}{\kappa - \sqrt{\kappa^2-\omega_r^2}} \right) \, ,
	\end{equation}
	where the expression for $\kappa < \omega_r$ is understood by analytic continuation and the critical-damping case by continuity. This is the ground-state position variance of an Ohmic Caldeira-Leggett oscillator with mass $1/\omega_0$, frequency $\omega_r$ and damping coefficient $2\kappa$~\cite{Grabert1984}. The bound therefore states that the quantum fluctuations of the quadrature can never exceed those of an Ohmically damped oscillator with the same static susceptibility. Since $F_Q^{\rm bound}$ decreases with temperature, the right-hand side bounds $F_Q/4$ at every $T$. This has a direct consequence for the marker. Once the damping exceeds the renormalized mass, $\kappa\gg\omega_r$---which, for any finite $\kappa$, is always the case sufficiently close to criticality, $\alpha\ll\kappa^2/\omega_0$---the right-hand side reduces to $F_Q^{\rm bound}/4\simeq(\omega_0/\pi\kappa)\ln(2\kappa/\omega_r)$. The ceiling therefore still diverges as the gap closes, but only logarithmically in $\alpha^{-1/2}$ and with an amplitude suppressed by $\omega_0/\kappa$, instead of $\alpha^{-1/2}$ as in Eq.~\eqref{eq:QFI_bound2}. Dissipation thus caps the metrological sensitivity that self-criticality can generate.

	Figure~\ref{fig:QFI_vs_dissipative_bound} shows the effect of dissipation on the constant-gap sweeps. Coming from high temperature, $F_Q$ follows the saturated SYK$_2$ branch, turns over close to the sign change of $\xi$, and decreases on the conformal side. The maximum stays close to the conformal/SYK$_2$ crossover for all values of $\kappa$, but it is progressively lowered and flattened, albeit not fully removed. The corresponding maps over the $(g,T)$-plane are shown in Fig.~\ref{fig:QFI_results_dissipation} of Appendix~ \ref{app:additional_results_diss}.
	
	Equation~\eqref{eq:D^R_complex_freq} also shows why the QFI reacts to weak losses earlier than the inverse static susceptibility $\alpha$. The Markovian kernel enters $-D^R(i\nu)^{-1}$ through the term $2\kappa\nu/\omega_0$, which is linear in $\kappa$ and grows with the frequency, but at $\omega=0$ only through the static shift $\kappa^2/\omega_0$. Expanding Eq.~\eqref{eq:QFI_in_complex_freq} at fixed $\alpha$ and $\delta\Pi^R$ gives
	\begin{equation}
		F_Q(\kappa) = F_Q(0) - \frac{16\kappa T}{\omega_0} \sum_{n\geq 0}\nu_n \left[ D^R(i\nu_n)|_{\kappa=0} \right]^2 + \mathcal{O}(\kappa^2) \, ,
	\end{equation}
	whose linear coefficient is strictly negative in every regime, since $D^R(i\nu)$ is real. The gap, on the other hand, has no linear dependence on $\kappa$ wherever the fermions see the boson only through static or infrared data. The QFI is therefore the first of the markers to register weak losses, whereas the static gap is the most robust against them.

	\section{Conclusion}
	\label{sec:conclusions}
	
	In this paper, we have characterized the Yukawa--SYK model using a real-time path-integral formalism. Formulating the large-$M,N$ saddle point on the Schwinger--Keldysh contour permitted us to obtain the spectral functions without any analytic continuation, so that the closed and the open system are treated within a single formalism. Mapping out the $(g,T)$-plane, we recover the conformal/non-Fermi-liquid and the SYK$_2$ regimes known from the imaginary-time analysis of Ref.~\cite{Esterlis_2019}. The real-time treatment also gives direct access to quantities that are less natural in Matsubara space, such as the interaction-induced boson linewidth $\gamma$ of Eq.~\eqref{eq:impurity_constants}. Both regimes rely on the fermions softening the boson, and we showed that this softening is bounded by a ``criticality budget" $\bigl|\Pi^R(0)\bigr| \;\le\; \frac{g^2}{4T}\frac{N}{M} \;\equiv\; \Pi_{\rm max}\,$, which follows only from the normalization of the fermionic spectral function and from the Fermi distribution, and confines self-tuned criticality to $T \leq N g^2 / (4M\omega_0)$. In addition, we identify two regimes that were not described in previous studies, in which the temperature is the largest scale but the coupling is not small: the boson remains essentially bare, while the fermion hybridizes with sidebands displaced by integer multiples of $\omega_0$, acquiring the structure of a Floquet-like problem. We showed that this driven-fermion physics is reproduced exactly by a quadratic SYK$_2$ model with time-dependent random hopping, Eqs.~\eqref{eq:effective_H}--\eqref{eq:effective_J}, and that the two regimes differ in whether the sidebands are resolved or merge into a single broad band, the crossover being located at $T \approx \omega_0^3/g^2$.
	
	Because the crossovers between these regimes are smooth, we constructed a minimal set of observables able to separate them without the need to fit any conformal exponents. These are built directly from the fermionic and bosonic two-point functions, which are experimentally accessible using, e.g., linear-response theory (see Appendix \ref{app:measurement} for a concrete example in optical cavities).
	The gap, or inverse static susceptibility, $\alpha$ of Eq.~\eqref{eq:gapdef} measures how far the boson sits from criticality and saturates to $\alpha=\omega_0$ once the boson becomes free, which is what allows the bare boson regimes to be separated from the conformal and SYK$_2$ regimes; extrapolating constant-$\alpha$ sweeps to this limit locates the boundary between them at $T=1.861\,g^2 /\omega_0$. The gap alone, however, closes in both the conformal and SYK$_2$ regime, and the curvature $\xi$ of Eq.~\eqref{eq:curvature} is needed to distinguish between them: at small gap, it is positive in the impurity case and negative when we approach the conformal solution, and its sign change locates the boundary between the two at $T \approx 0.135\, g^{-1.96}$, in close agreement with the analytical scaling $T \sim g^{-2}$. Both markers are local properties of $D^R(\omega)^{-1}$ at the origin, and we therefore complemented them with the quantum Fisher information $F_Q$, which is built from the same response function but integrated over frequency. Using sum rules obeyed by $D^R$, we derived upper bounds on $F_Q$ that are saturated in the SYK$_2$ regime, where the boson is a single sharp mode, but not in the conformal one, where the boson spreads its weight above the thermal scale. As a result, along lines of constant gap $F_Q$ has a maximum close to the conformal/SYK$_2$ crossover at a temperature set by the renormalized boson mass, $T\simeq 0.83 \, \omega_r/2$, and with a height that grows as $\alpha^{-1/2}$ on approach to criticality. Finally, since all three are constructed from the bosonic propagator, they are blind to the high-temperature part of the diagram, where the boson is bare; there, we used the fermionic spectral weight $W_{\omega_0}$ around the first sideband at $\pm \omega_0$, whose onset separates the free-fermion from the driven-fermion regimes and reproduces the boundary $T\propto g^{-2}$, while the same weight centered at $\omega_0/2$ separates the resolved from the unresolved sidebands.
	
	We then extended the analysis to an open-system extension. Linearly coupling the bosonic sector to a thermal bath only adds an environment self-energy $\Pi^R_E$ to the Schwinger--Dyson equation of the boson, as in Eq.~\eqref{eq:Schwinger_Dyson_bath}.
	In the infrared, the only two properties of an otherwise arbitrary bath that are relevant are its static shift, $\Pi^R_{\rm E}(0)$, and its low-frequency exponent, $s$. The static shift can drive the system away from self-criticality by an amount that the interaction may be unable to compensate for, since the criticality budget $\Pi_{\rm max}$ is a property of the system that carries no dependence on the environment. The exponent $s$ instead controls the relevance of the bath with respect to the conformal scaling of the bosonic self-energy, as previously discussed in Refs.~\cite{Hosseinabadi:2023mid,Cichutek_2024}. For weak dissipation, all the regimes survive and the markers introduced above continue to separate them, with modifications that we quantify explicitly. The static shift acts as a rescaling of the coupling, $g \to g_{\rm eff}$ in Eq.~\eqref{eq:effective_coupling_dissipation}, and displaces the bare-boson boundary rigidly to large coupling, $T=1.861\, g^2_{\rm eff}/\omega_0$, as confirmed numerically by the collapse of the constant-gap amplitudes (Appendix~\ref{app:additional_results_diss}). The criticality budget then becomes a lower bound on the normalized gap $\alpha/[\omega_0 + \Pi^R_E(0)]$, which is the ratio of the static responses measured with and without the Yukawa coupling and thus requires no knowledge of the bath. The dynamical part of the bath, instead, narrows the conformal frequency window and steepens the conformal/SYK$_2$ boundary beyond what the rescaling predicts, so that at large coupling the conformal regime shrinks in favor of the SYK$_2$ one. In the bare-boson regimes, it transfers the boson linewidth to the fermion, turning the gaps between the sidebands into pseudogaps: squeezed between the free-fermion and the unresolved-sideband regimes, the resolved one survives only at weak dissipation. In the strong-dissipation limit, the fermion collapses onto a sharp quasiparticle that saturates the criticality budget, while the boson is dominated by the environment and the fermions contribute only a Drude-like self-energy; there, the curvature $\xi$ is negative even though the boson is very far from criticality, so that its sign is a reliable signature of self-criticality only when read together with the normalized gap.
	
	For a fixed gap $\alpha$, the quantum Fisher information keeps its maximum close to the conformal or SYK$_2$ crossover for all the loss rates we considered, but dissipation lowers and flattens it. We derived a sharper bound valid for an arbitrary bath, which, for Markovian losses, states that the quantum fluctuations of the quadrature can never exceed those of an Ohmically damped oscillator with the same static susceptibility. Close to criticality, the ceiling of $F_Q$ therefore grows only logarithmically as the gap closes, instead of as $\alpha^{-1/2}$: dissipation caps the metrological sensitivity that self-criticality can generate. Moreover, losses enter $F_Q$ already at linear order in $\kappa$, whereas they enter the gap essentially only through the static shift $\kappa^2/\omega_0$, such that the QFI is the first of the markers to register weak losses, while the gap is the most robust against them.
	
	The criticality budget can have direct experimental relevance, since it constrains the parameters at which self-tuned criticality can be observed in terms of quantities that are accessible independently. Combining it with the self-criticality condition~\eqref{eq:self_criticality_bath}, the conformal regime requires $0<\omega_0+\Pi^R_{\rm E}(0)\le\Pi_{\rm max}$, that is
	\begin{equation}
		T \;\le\; \frac{N}{4M}\,\frac{g^2}{\omega_0+\Pi^R_{\rm E}(0)}
		\quad \overset{\rm Markov}{=} \quad
		\frac{N}{4M}\,\frac{g^2\,\omega_0}{\omega_0^2+\kappa^2}\,.
		\label{eq:design_rule}
	\end{equation}
	For a given loss rate, Eq.~\eqref{eq:design_rule} fixes the temperature below which any signature of self-tuned criticality can appear, or equivalently the Yukawa
	coupling required at a given temperature; for $\kappa\gg\omega_0$ the admissible temperature decreases quadratically with the loss rate.  
	Moreover, within the Markovian model of Eq.~\eqref{eq:diss_markov_kernel}, Eq.~\eqref{eq:design_rule} indicates that, at fixed $g$ and $\kappa$, the bare boson mass that least exhausts the criticality budget is the one that matches the dissipation strength, $\omega_0=\kappa$. 
	This finding may be relevant for experiments, as it suggests the regime that is most stable towards a given dissipation is the one where $\omega_0$ is tuned close to $\kappa$.

	Several directions follow naturally from this work. The most immediate is to relax the assumption of thermality. Throughout, we imposed the fluctuation--dissipation relations in Eq.~\eqref{eq:fluctuation_dissipation_G}--\eqref{eq:fluctuation_dissipation_D}, but the Markovian dissipation we used as a limiting case describes a zero-temperature environment, which would eventually drive the system into a genuine non-equilibrium steady state. Promoting the lesser propagators to independent variables in the saddle-point equations makes the distribution function an output rather than an input, and allows one to ask whether self-tuned criticality survives when the boson occupation is not thermal. This differs from the situation addressed by quench studies with a thermal phonon bath, where the system relaxes through a sequence of quasi-thermal states at a drifting effective temperature~\cite{Hosseinabadi:2023mid}. It is also particularly interesting to understand in this setting if the criticality budget is modified, since the non-positivity of $\Pi^R(0)$ follows from $n_{\rm F}$ being monotonically decreasing.
	
	A second direction is to consider sources of dissipation other than the boson losses studied here, and in particular those that are intrinsic to a given platform. In a cavity-QED realization~\cite{cavity_proposal,Baumgartner:2024ysk,Baumgartner:2025uue,Solis_2026,Ferrari:2026zkn}, photon leakage is only one of the two unavoidable channels,
	the other being atomic spontaneous emission, which acts directly on the fermionic sector. Both rates are controlled by the same detuning, so that the relevant
	statement is a trade-off between them rather than a bound on either one separately, and quantifying it would sharpen the parameter window identified by
	Eq.~\eqref{eq:design_rule}. Spontaneous emission in cavity realizations of SYK-type models has recently been analyzed from the point of view of dissipative quantum chaos and non-Hermitian spectral statistics~\cite{Pacchioni2026}, but not, to our knowledge, in terms of its effect on the self-critical phase.

	\section*{Acknowledgments}
	We would like to thank Rahel Baumgartner, Alberto Biella, Jean-Philippe Brantut, Léa Dubois,
	Ekaterina Fedotova, Robin Löwenberg, Francesca Orsi,
	Pietro Pelliconi, Gianluca Rastelli, David Pascual Solis, Julian Sonner, and Yi-Neng Zhou for
	helpful discussions and comments.
	Claude Opus $5$ was used for text editing, help in data analysis, and critical review of the manuscript. They also contributed to the derivation of the bound in \cref{eq:QFI_bound_sharp}. All authors have reviewed, edited, and take full responsibility for the final text.
	This project has been funded by the Caritro Foundation and by the Swiss State Secretariat for Education, Research and Innovation (SERI) under contract number UeMO19-5.1. This work was supported by the Provincia Autonoma di Trento, and Q@TN, the joint lab between the University of Trento, FBK—Fondazione Bruno Kessler, INFN—National Institute for Nuclear Physics, and CNR—National Research Council, Italy.

	\appendix
	
	\section{Schwinger--Keldysh path integral for the Yukawa-SYK model}
	\label{app:path_integral}
	
	In this appendix, we present the path integral calculation needed to reproduce the saddle-point equations of the YSYK model coupled to a bosonic environment. In the first section, we will consider a bosonic mode coupled to a generic free environment, which will then be coupled to the YSYK bosons in the second section. In the last section, we will derive the saddle-point equations for the system. 
	
	\subsection{Bosonic field coupled to an environment}
	\label{app:boson_environment}
	
	We consider the case where the environment to which the YSYK bosons $a_m$ are coupled is very large and does not generate effective interactions between different bosons. In this case the description is equivalent to having each mode coupled to its own environment. The aim of this section is to derive an effective model for the $x$ quadrature of the YSYK boson, when it is coupled in a number-conserving manner to the environment.  
	
	For this purpose, it is sufficient to consider the quadratic bosonic part of the YSYK Hamiltonian \eqref{eq:YSYK_H}
	\begin{equation}
		H_a = \omega_0 \sum_{m=1}^M a^\dag_m a_m 
	\end{equation}
	coupled to an external environment:
	\begin{equation}
		H_\mathrm{B} = H_a + \sum_{m=1}^M \left( \sum_{k} \epsilon_k b^\dag_{k,m} b_{k,m} + g_k a_m b^\dag_{k,m} + \bar{g}_k b_{k,m} a^\dag_m  \right)
		\label{eq:H_boson_with_environment}
	\end{equation}
	where $\epsilon_k$ are the internal energy levels of the environment, while $g_k$ controls the coupling between system and environment bosons. 
	
	The expectation values of the observables can be calculated from the real-time partition function:
	\begin{equation}
		Z_\mathrm{B}(t)=\Tr \left(e^{-i H_\mathrm{B} t} \rho_0 e^{i H_\mathrm{B} t} \right) \, ,
	\end{equation}
	where we take the initial state $\rho_0$ to be a thermal state at inverse temperature $\beta$ and we also assume the environment to be at thermal equilibrium at the same temperature. 
	$Z_\mathrm{B}(t)$ can be expressed using a Schwinger--Keldysh path integral. We will denote with a superscript $\mu = +(-)$ the fields on the forward (backward) contour:
	\begin{equation}
		Z_\mathrm{B}= \int \mathcal{D}a^\pm \mathcal{D}b^\pm \langle a^+(-\infty) b^+(-\infty) | \rho_0 |a^-(-\infty) b^-(-\infty) \rangle  e^{i S_\mathrm{B}[a^\pm,b^\pm]} \, ,
	\end{equation}
	where we extended the time evolution over all the real line, since we are interested in the steady-state physics. We also defined
	\begin{align}
		S_\mathrm{B}[a^\pm,b^\pm]  &= \int \dd t \sum_{m,\mu=\pm} \mu \left[ \bar{a}_{m}^\mu (-i\partial_t-\omega_0) a_{m}^\mu + \sum_k \bar{b}_{k,m}^\mu (-i\partial_t-\epsilon_k) b_{k,m}^\mu - g_k \bar{b}_{k,m}^\mu a^\mu_m- \bar{g}_k \bar{a}^\mu_m b_{k,m}^\mu \right] \nonumber \\
		&= \int \frac{\dd \omega}{2\pi} \sum_{m,\mu=\pm} \mu \left[ \bar{a}_{m}^\mu (\omega-\omega_0) a_{m}^\mu + \sum_k \bar{b}_{k,m}^\mu (\omega-\epsilon_k) b_{k,m}^\mu - g_k \bar{b}_{k,m}^\mu a^\mu_m- \bar{g}_k \bar{a}^\mu_m b_{k,m}^\mu \right] \, ,
	\end{align}
	where in the second line we transformed to frequency space using the convention
	\begin{equation}
		a(t) = \int \frac{\dd \omega}{2\pi} \alpha(\omega) e^{-i \omega t} .
	\end{equation}
	Due to the linear coupling between environment and system we can integrate out the $b$ field, obtaining an exact action for $a$ only:
	\begin{equation}
		\label{eq:bosonic_action}
		S_\mathrm{B}[a^\pm]  = \int \frac{\dd \omega}{2\pi} \sum_{m} \sum_{\mu,\nu=\pm} \bar{a}_m^{\mu} (\mu \delta^{\mu\nu}(\omega-\omega_0)- \Pi_{b}(\omega)^{\mu \nu}) a_m^\nu 
	\end{equation}
	where we introduced the environment induced self-energy:
	\begin{equation}
		\Pi_b^{\mu \nu}(\omega) =  \sum_k |g_k|^2 \mu \nu D_{b,k}^{\mu\nu} \, , \qquad D_{b,k} = \left(
		\begin{array}{cc}
			\frac{1}{\omega-\epsilon_k+i 0^+} -2\pi i n_\mathrm{B} \delta(\omega-\epsilon_k) & -2\pi i n_\mathrm{B} \delta(\omega-\epsilon_k)  \\
			-2\pi i (n_\mathrm{B}+1) \delta(\omega-\epsilon_k)  & - \frac{1}{\omega-\epsilon_k-i 0^+}-2\pi i n_\mathrm{B} \delta(\omega-\epsilon_k)  \\
		\end{array}
		\right) \, .
	\end{equation}
	One can check that the bath propagator $D_{b,k}$ already satisfies the relation $D^{++}_{b,k}+D^{--}_{b,k} = D^{+-}_{b,k}+D^{-+}_{b,k}$ and the KMS condition $D_{b,k}^{-+} (\omega)= e^{\beta \omega} D_{b,k}^{+-}(\omega)$. Notice that in the main text we adopted the notation $D^{+-} = D^{<}$, $D^{-+} = D^{>}$.
	
	We can further rewrite $\Pi_b$ by introducing the spectral density 
	\begin{equation}
		J(\omega) = \sum_{k} |g_k|^2 \delta(\omega-\epsilon_k) \,,
	\end{equation}
	which gives a convenient way of rewriting the discrete sums over environment modes as frequency integrals
	\begin{align}
		\Pi_b^{+-}(\omega) &= 2 \pi i n_\mathrm{B}(\omega) J(\omega) \, , \qquad  \Pi_b^{-+}(\omega) = 2 \pi i \left[1+n_\mathrm{B}(\omega)\right] J(\omega) \, , \\ 
		\Pi_b^{\pm \pm}(\omega) &= -\pi i \left[1+2n_\mathrm{B}(\omega)\right] J(\omega) \pm \mathcal{P} \int \frac{J(\omega')}{\omega-\omega'} \dd \omega' \, .
	\end{align}
	where $\mathcal{P}$ denotes the principal-value.
	In Sec.~\ref{sec:dissipation}, we will often adopt the Markov approximation, which is recovered in the case where the YSYK model is realized, for example, in an optical cavity the frequency scale of the relevant modes, $\omega_b$, are often much larger than $1/\beta$ \cite{Solis_2026}. In this case, the environment induced self-energies are evaluated around $\omega_b$ meaning that  $n_\mathrm{B}(\omega+\omega_b) \sim 0$ and $J(\omega+\omega_b) \sim J(\omega_b) = \kappa/\pi$ \cite{Torre_Keldysh_open_systems_2013}, so that
	\begin{equation}
		\Pi_b(\omega) = \left(
		\begin{array}{cc}
			- i \kappa & 0  \\
			2 i \kappa  & - i \kappa  \\
		\end{array}
		\right) .
	\end{equation}
	With these conventions, we reproduce the same time evolution generated by a Lindblad operator \cite{Sieberer_Keldysh_open_systeems_2016}:
	\begin{equation}
		\partial_t \rho = \mathcal{L}[\rho] = -i [H_b,\rho] + \sum_{m} (2 L_m \rho L^\dag_m-L^\dag_m L_m \rho - \rho L^\dag_m L_m ) \label{eq:markov_approx}
	\end{equation}
	with jump operators given by $L_m = \sqrt{\kappa} a_m$.
	
	The next step is to take \cref{eq:bosonic_action}, split $\sqrt{2} a_m(t) = x_m(t) + i p_m(t)$ and integrate $p$ out. Since $x$ and $p$ are real fields, in Fourier space we have
	\begin{equation}
		\bar{x}_m(\omega) = x_m(-\omega) \, , \qquad \bar{p}_m(\omega) = p_m(-\omega) \, ;
	\end{equation}
	so just the positive (negative) frequency field space is truly independent. Therefore, we can rewrite the action using auxiliary matrices $A,B$
	\begin{equation}
		S_b[x^\pm,p^\pm] = \int_0^\infty \frac{\dd \omega}{2 \pi} \sum_{m} \sum_{\mu,\nu=\pm} \left[ \bar{x}_m^\mu A^{\mu \nu} x_m^\nu+\bar{p}_m^\mu A^{\mu \nu} p_m^\nu + i \bar{x}_m^\mu B^{\mu \nu} p_m^\nu - i \bar{p}_m^\mu B^{\mu \nu} x^\nu_m  \right]
	\end{equation}
	with
	\begin{equation}
		A^{\mu \nu} = - \mu \delta^{\mu\nu} \omega_0 - \frac{\Pi_b^{\mu\nu}(\omega)+\Pi_b^{\nu\mu}(-\omega)}{2} \, , \qquad B^{\mu \nu} = \mu \delta^{\mu\nu} \omega - \frac{\Pi_b^{\mu\nu}(\omega)-\Pi_b^{\nu\mu}(-\omega)}{2} \, .
	\end{equation}
	We can now integrate $p$ out and obtain an action depending only on the $x$-quadrature:
	\begin{equation}
		S_b[x^\pm] = \int_0^\infty \frac{\dd \omega}{2 \pi} \sum_{m} \sum_{\mu,\nu=\pm} \bar{x}^\mu_m (A-B A^{-1} B)^{\mu \nu} x^\nu_m = \frac{1}{2}\int_{-\infty}^\infty \frac{\dd \omega}{2 \pi} \sum_{m} \sum_{\mu,\nu=\pm} \bar{x}^\mu_m (D_0^{-1}-\Pi_\mathrm{E})^{\mu \nu} x^\nu_m \, ,
	\end{equation}
	where, in the last step, we used the symmetry properties of $A$ and $B$ to extend the integral back over the whole frequency space, introduced the free propagator
	\begin{equation}
		(D_0^{-1}(\omega))^{\mu \nu} = \mu\frac{\omega^2-\omega_0^2}{\omega_0} \delta^{\mu\nu} \, 
	\end{equation}
	and defined the self-energy induced by the environment $\Pi_\mathrm{E}$ as the remaining contribution. The self-energy $\Pi_\mathrm{E}$ simplifies considerably in the Markovian approximation, which amounts to replacing the $i0^+$ prescription with the finite broadening $\kappa$ for the bare boson. The same broadening has been used in the numerical simulations, where $\kappa$ has been taken to be as a small but finite constant.
	The effect of the Markovian dissipation becomes more transparent after performing the Keldysh rotation:
	\begin{equation}
		U (D_0^{-1}-\Pi_\mathrm{E}) U = \left( \begin{array}{cc}
			0 & (D^A_0)^{-1}-\Pi_\mathrm{E}^A\\
			(D^{R}_0)^{-1}-\Pi_\mathrm{E}^R & (D^{K}_0)^{-1}-\Pi_\mathrm{E}^K
		\end{array}
		\right) \, , \qquad U = \frac{1}{\sqrt{2}}\left(
		\begin{array}{cc}
			1 & 1 \\
			1 & -1 \\
		\end{array}
		\right) \, ,
	\end{equation}
	with
	\begin{equation}
		D^R_0-(\Pi^R_\mathrm{E})^{-1} = \frac{\omega_0}{(\omega +i \kappa )^2-\omega_0^2} \, , \quad D^K_0-(\Pi^K_\mathrm{E})^{-1} = -\frac{2 i \kappa  \left(\kappa ^2+\omega ^2+\omega_0^2\right)}{\kappa ^4+2 \kappa ^2 \left(\omega ^2+\omega_0^2\right)+\left(\omega ^2-\omega_0^2\right)^2} \, .
	\end{equation}
	The lesser component, on the other hand, reads:
	\begin{equation}
		D_0^< - (\Pi^{-1}_\mathrm{E})^< = -\frac{i \kappa }{\kappa ^2+(\omega +\omega_0)^2} \, .
	\end{equation}
	
	Now that we have incorporated the effect of the environment in the $x$ two-point function, we can perform in the next section the Schwinger--Keldysh path integral of the full YSYK model coupled.
	
	\subsection{Schwinger--Keldysh path integral for Yukawa-SYK with dissipation} \label{app:SK_path_integral_YSYK}
	
	The Hamiltonian of the YSYK model coupled to an external environment is given by:
	\begin{equation}
		H = H_\mathrm{B} - \mu_C \sum_{j=1}^N c_j^\dag c_j + \frac{1}{\sqrt{MN}} \sum_{jkm} g^m_{jk} \, c_j^\dag c_k x_m \, ,
		\label{eq:H_YSYK_with_dissipation}
	\end{equation}
	where $H_\mathrm{B}$ is defined in~\cref{eq:H_boson_with_environment}, while the rest of the fields are defined as in Eq.~\eqref{eq:YSYK_H}.
	
	Again, we want to compute the real-time partition function with the presence of boson losses
	\begin{equation}
		Z(t)=\Tr \left(e^{-i H t} \rho_0 e^{i H t} \right) = \int \mathcal{D} a^{\pm} \mathcal{D} b^{\pm} \mathcal{D} \psi^\pm e^{iS[a^\pm,b^\pm,\psi^\pm]} = \int \mathcal{D} x^{\pm} \mathcal{D} \psi^\pm e^{iS[x^\pm,\psi^\pm]} \, ,
	\end{equation}
	where in the last step we used the fact that the interaction in the YSYK model~\eqref{eq:H_YSYK_with_dissipation} just depends on $x$ to integrate out the environment and the imaginary part of the $a$, as done in the previous section. 
	The resulting action reads:
	\begin{equation}
		S = S_\mathrm{B}[x] +\int \dd t  \sum_{\nu=\pm} \nu \left[\sum_j  \bar{\psi}_j^\nu (i \partial_t+\mu_C) \psi_j^\nu - \frac{1}{\sqrt{MN}} \sum_{jkm} g^m_{jk} \, \bar\psi_j^\nu \psi_k^\nu x_m^\nu  \right].
	\end{equation}
	
	As common in the SYK and YSYK literature \cite{Maldacena:2016hyu,Esterlis_2019}, we focus on ensemble averaged quantities:
	\begin{equation}
		\left\langle Z(t) \right\rangle = \int \mathcal{D} x^{\pm} \mathcal{D} \psi^\pm \int \prod_m \dd g^m e^{-\frac{1}{2g^2} \Tr (g^m g^m) }  e^{iS[x^\pm,\psi^\pm]}
	\end{equation}
	where $g_m$ are GUE matrices. Since the couplings appear linearly in the action, we can directly perform the Gaussian integral, which leads to the following action:
	\begin{equation}
		iS = iS_\mathrm{B}[x] + iS_\mathrm{F}[\psi] - \frac{g^2}{2} \frac{1}{MN} \sum_{jkm}\left|\int \dd t \sum_{\nu = \pm} \bar\psi_j^\nu \psi_k^\nu x_m^\nu  \right|^2,
	\end{equation}
	where $S_\mathrm{F}$ is the free fermion part of the action
	\begin{equation}
		S_\mathrm{F}[\psi] = \int \dd t \dd t' \, \sum_{j} \sum_{\mu\nu} \bar\psi^\mu_j(t) (G_0^{-1}(t,t'))^{\mu\nu} \, \psi_j^\nu(t') \, ,
		\label{eq:S_F_action_def}
	\end{equation}
	where we introduced:
	\begin{equation}
		(G_0^{-1}(t,t'))^{\mu\nu} = \mu \delta^{\mu\nu}\delta(t'-t) (i\partial_t + \mu_C) \, .
	\end{equation}

	Now, we can introduce the fermion and boson two-point functions and self-energies by inserting a resolution of unity into the path integral:
	\begin{align}
		1 &= \int \mathcal{D} G \mathcal{D} \Sigma \exp \left\{ N \int dt dt' \sum_{\mu\nu} \Sigma^{\nu \mu}(t',t) \left( G^{\mu \nu}(t,t')+ \frac{i}{N} \sum_j \psi^\mu_j(t) \bar{\psi}^\nu_j(t') \right)  \right\}\, , \label{eq:G_Sigma_fat1}\\
		1 &= \int \mathcal{D} D \mathcal{D} \Pi \exp \left\{ -\frac{M}{2} \int dt dt' \sum_{\mu \nu} \Pi^{\nu \mu}(t',t) \left( D^{\mu \nu}(t,t')+ \frac{i}{M} \sum_m x^\mu_m(t) x^\nu_m(t') \right)  \right\} \, ,
	\end{align}
	allowing to rewrite the interaction in terms of the collective fields:
	\begin{align}
		&iS = \int \dd t \dd t'\sum_{\mu\nu} \bigg[\!
		N \frac{g^2}{2i} s_{\mu \nu} G^{\mu \nu}(t,t') G^{\nu \mu}(t',t) D^{\mu \nu}(t,t') \! + \! N \Sigma^{\nu \mu}(t',t) G^{\mu \nu}(t,t') \! - \!\frac{M}{2}  \Pi^{\nu \mu}(t',t)  D^{\mu \nu}(t,t') \\
		&+\!\sum_j i \bar{\psi}^\nu_j(t')  \left((G_{0}^{-1})^{\nu \mu}(t',t)\!-\!\Sigma^{\nu \mu}(t',t) \right) \psi^\mu_j(t)\! +\! \sum_m i x^\nu_m(t')  \left(\! (D_{0}^{-1})^{\nu\mu}(t',t)\!-\!\Pi^{\nu\mu}(t',t)\!-\!\Pi^{\nu\mu}_\mathrm{E}(t',t) \! \right) x^\mu_m(t) \! \bigg] \nonumber
	\end{align}
	with $s_{\mu \nu} = \mu \cdot \nu$ and 
	\begin{equation}
		D_0^{-1}(t,t')-\Pi_\mathrm{E}(t,t') = \sum_{\omega} e^{i \omega (t'-t)} \left( D_0^{-1}(\omega)-\Pi_\mathrm{E}(\omega) \right) \, ,
	\end{equation}
	where the frequency expression for $D_0$ and $\Pi_\mathrm{E}$ were given in Sec.~\ref{app:boson_environment}.
	Since the action is now quadratic in terms of both the fermions and bosons, it is possible to perform a Gaussian integral over both fields and express the action solely in terms of the collective fields:
	\begin{equation}
		\left\langle \Tr\left(e^{T \mathcal{L}} \rho_0 \right) \right\rangle = \int \mathcal{D} G \, \mathcal{D} \Sigma \, \mathcal{D} D \, \mathcal{D} \Pi \, \, e^{i S_\text{eff}[G,\Sigma,D,\Pi]} \, ,
	\end{equation}
	with
	\begin{equation}
		\begin{split}
			i S_\text{eff} =& N \Tr \log(G_0^{-1} - \Sigma) - \frac{M}{2} \Tr \log(D_0^{-1} - \Pi-\Pi_\mathrm{E}) + N \Tr(\Sigma G) - \frac{M}{2} \Tr(\Pi D) \\
			-& i \frac{g^2 N}{2} \int \dd t \dd t'\sum_{\mu \nu} s_{\mu\nu} G^{\mu\nu}(t,t') G^{\nu\mu}(t',t) D^{\mu\nu}(t,t') \, .
		\end{split}
	\end{equation}
	In the expression above, the collective fields are treated as infinite-dimensional matrices in both the temporal and contour indices.
	
	We can now take the thermodynamic limit $N,M \to \infty$ while keeping the ratio $N/M$ fixed, which makes the saddle point equations exact:
	\begin{align}
		\frac{\delta S_\text{eff}}{\delta \Pi} = 0 \quad &\Longrightarrow \quad D(t,t') = (D_0^{-1} - \Pi-\Pi_\mathrm{E})^{-1}(t,t') \, , \\
		\frac{\delta S_\text{eff}}{\delta \Sigma} = 0 \quad &\Longrightarrow \quad G(t,t') = (G_0^{-1} - \Sigma)^{-1}(t,t') \, ,  \\
		\frac{\delta S_\text{eff}}{\delta D} = 0 \quad &\Longrightarrow \quad \Pi^{\mu\nu}(t,t') = -i g^2 \frac{N}{M} s_{\mu\nu} G^{\mu\nu}(t,t') G^{\nu\mu}(t',t) \, , \\
		\frac{\delta S_\text{eff}}{\delta G} = 0 \quad &\Longrightarrow \quad \Sigma^{\mu\nu}(t,t') = i g^2 s_{\mu\nu} G^{\mu\nu}(t,t') D^{\mu\nu}(t,t')\, .
	\end{align}
	We can restrict our analysis to the steady-state regime, as assumed in the rest of the manuscript, so that time-translation invariance holds and all collective fields depend only on time differences. We also perform the Keldysh rotation and apply the Langreth rules \cite{langreth1976linear,Haug2008}, which allow us to rewrite the Schwinger–Dyson equations in terms of only two independent components of the Green’s functions. As in the main text, we choose these to be the retarded and lesser propagators:
	\begin{align}
		\Sigma^<(t) &= i g^2 G^<(t) D^<(t) \, , \\
		\Sigma^R(t) &= i g^2 (G^<(t) D^R(t) + G^R(t) D^<(t) + G^R(t) D^R(t) ) \label{eq:SigmaR_eq_app} \\ &= i g^2 (G^<(t) D^R(t) + G^R(t) D^>(t) ) \, \nonumber ,\\
		\Pi^<(t) &= -i g^2 \frac{N}{M} G^<(t) G^>(-t) \, , \\
		\Pi^R(t) &= -i g^2 \frac{N}{M} (G^<(t) G^A(-t) + G^R(t) G^<(-t) ) \, . 
	\end{align}
	These equations must be complemented with the Schwinger--Dyson equations, which in the stationary case can be conveniently written in frequency space:
	\begin{align}
		D^R(\omega) &= \frac{1}{D_0^R(\omega)^{-1} - \Pi^R(\omega)-\Pi^R_\mathrm{E}(\omega)} \, ,\\
		D^<(\omega) &= |D^R(\omega)|^2 \left(\frac{D_0^<(\omega)}{|D_0^R(\omega)|^2} + \Pi^<(\omega) + \Pi^<_\mathrm{E}(\omega) \right) \, ,\\
		G^R(\omega) &= \frac{1}{G_0^R(\omega)^{-1} - \Sigma^R(\omega)} \, ,\\
		G^<(\omega) &= |G^R(\omega)|^2 \left(\frac{G_0^<(\omega)}{|G_0^R(\omega)|^2} + \Sigma^<(\omega) \right) \, .
	\end{align}
	This system of equations is valid for generic steady states, including situations away from thermal equilibrium. In the main text, we further simplify it by imposing the fluctuation–dissipation relations~\eqref{eq:fluctuation_dissipation_G}--\eqref{eq:fluctuation_dissipation_D}, which make the lesser and retarded propagators no longer independent.
	
	As a last step, we prove that the equations for the lesser components of the self-energy preserve the KMS conditions (equivalent to the fluctuation--dissipation relations) if we impose them on the two-point functions:
	\begin{align}
		\Sigma^>(t-i\beta) &= i g^2 G^>(t-i\beta) D^>(t-i\beta) = - i g^2 G^<(t) D^<(t) = - \Sigma^<(t) \, , \\
		\Pi^>(t-i\beta) &= -i g^2 \frac{N}{M} G^>(t-i \beta) G^<(-t+i\beta) = -i g^2 \frac{N}{M} G^<(t) G^>(-t) = \Pi^<(t) \, .
	\end{align}
	
	\section{More details about the Yukawa-SYK regimes}
	\label{app:regimes}
	
	In this appendix, we give additional details about the different phases discussed in Sec.~\ref{sec:phase_diagram}, we will start by giving extra details on the conformal regime, before moving to the SYK$_2$ one. In the last section, we will give extra details about YSYK at high temperature and the driven quadratic model introduced in Sec.~\ref{sec:high_t_regimes}.
	We set $\Pi_\mathrm{E}=0$ in this whole appendix.
	
	\subsection{Conformal regime}
	\label{app:conf_symmetry}
	
	In the low-energy limit $\omega \to 0$, the saddle-point equations simplify to
	\begin{align}
		&\sum_{\sigma} \int \dd t \, \, D^{\mu \sigma}(t_1,t) \delta\Pi^{\sigma \nu}(t,t_2) = \delta^{\mu\nu} \delta(t_1-t_2) \, , \qquad \delta\Pi^{\mu\nu}(t_1,t_2) = -i g^2 s_{\mu\nu} G^{\mu\nu}(t_1,t_2) G^{\nu\mu}(t_2,t_1) \, , \label{eq:conformal_Pi_eq} \\
		&\sum_{\sigma} \int \dd t \, \, G^{\mu \sigma}(t_1,t) \Sigma^{\sigma \nu}(t,t_2) = \delta^{\mu\nu} \delta(t_1-t_2) \, , \qquad \, \, \, \Sigma^{\mu\nu}(t_1,t_2) = i g^2 s_{\mu\nu} G^{\mu\nu}(t_1,t_2) D^{\mu\nu}(t_1,t_2) \, .
	\end{align}
	Here, we have introduced $\delta \Pi(\omega) = \Pi(\omega)-\Pi(0)$, where $\Pi(0)$ fixed by the self-critical condition in Eq.~\eqref{eq:def_self_criticality}. In this limit, $\Pi(0)$ exactly cancels the bare bosonic propagator in the Schwinger–Dyson equation and generates an ultralocal contribution in time $\delta \Pi(t_1-t_2) = \Pi(t_1-t_2)-\Pi(0) \delta(t_1-t_2)$, which can be neglected since the infrared regime corresponds to $t_1-t_2 \gg 1$. These considerations justify the form of the saddle-point equations \eqref{eq:conformal_Pi_eq}. 
	
	It is straightforward to verify that the above equations are invariant under the following reparameterization symmetry:
	\begin{align}
		&G^{\mu\nu}(t,t') \to \left[f(t)f(t')\right]^{\Delta_\mathrm{F}} G^{\mu\nu}(f(t),f(t')) \qquad D^{\mu\nu}(t,t') \to \left[f(t)f(t')\right]^{\Delta_\mathrm{B}} D^{\mu\nu}(f(t),f(t')) \\
		&\Sigma^{\mu\nu}(t,t') \to \left[f(t)f(t')\right]^{\Delta_\mathrm{\Sigma}} \Sigma^{\mu\nu}(f(t),f(t')) \qquad \delta \Pi^{\mu\nu}(t,t') \to \left[f(t)f(t')\right]^{\Delta_\Pi} \delta \Pi^{\mu\nu}(f(t),f(t')) 
	\end{align}
	with scaling dimensions
	\begin{equation}
		\Delta_\mathrm{B} = 1-2\Delta_\mathrm{F} \, , \qquad \Delta_\Sigma = 1-\Delta_\mathrm{F} \, , \qquad \Delta_\Pi = 2 \Delta_\mathrm{F} \, . \label{eq:all_conf_exp}
	\end{equation}
	At zero temperature, all collective fields are therefore determined by conformal two-point functions with anomalous dimensions fixed by \eqref{eq:all_conf_exp}. This motivates the following ansatz for the fermionic retarded propagator:
	\begin{equation}
		G^R(t) = -2ib \cos (\pi\Delta_\mathrm{F}) \frac{\theta(t)}{t^{2\Delta_\mathrm{F}}} \, ,
	\end{equation}
	and analogous conformal ansatze can be made for the remaining retarded fields.
	Starting from this expression for $G^R$, we now determine all other fields self-consistently from the infrared saddle-point equations, leading to the results reported in Eqs.~\eqref{eq:conformal_ansatz_main}–\eqref{eq:conformal_constant}.
	
	The Fourier transform of $G^R(t)$ is
	\begin{equation}
		\begin{split}
			G^R(\omega) &= 2 b\cos (\pi\Delta_\mathrm{F}) \Gamma (1-2 \Delta_\mathrm{F} ) \frac{(- i \omega + 0^+)^{2\Delta_\mathrm{F}}}{\omega} \\
			&= 2 b\cos (\pi\Delta_\mathrm{F}) \Gamma (1-2 \Delta_\mathrm{F} ) | \omega |^{2 \Delta_\mathrm{F} -1} (\cos (\pi  \Delta_\mathrm{F} ) \text{sgn}(\omega )-i \sin (\pi  \Delta_\mathrm{F} ))  \, .
		\end{split}
	\end{equation}
	Using the fluctuation–dissipation relation \eqref{eq:fluctuation_dissipation_G}, and the zero-temperature form $n_\mathrm{F}=\theta(-\omega)$, we obtain
	\begin{equation}
		G^<(\omega) = 2 i b \sin (2 \pi  \Delta_\mathrm{F} ) \Gamma (1-2 \Delta_\mathrm{F} ) \theta (-\omega ) | \omega | ^{2 \Delta_\mathrm{F} -1} \, .
	\end{equation}
	To determine $\delta\Pi^R$ from \cref{eq:PiR_t}, we need to Fourier transform the lesser propagator back to time:
	\begin{equation}
		G^<(t) =  i b |t|^{-2 \Delta_\mathrm{F}} (\cos (\pi  \Delta_\mathrm{F})+i \sin (\pi  \Delta_\mathrm{F}) \text{sgn}(t)) \, ,
	\end{equation}
	where standard Gamma-function identities have been used. We therefore have
	\begin{align}
		\delta\Pi^R(t) &= -2b^2 g^2\sin (2\pi  \Delta_\mathrm{F} ) \frac{N}{M} \frac{\theta(t)}{t^{4 \Delta_\mathrm{F}}} \, , \\
		\Longrightarrow \quad \delta\Pi^R(\omega) &= -b^2 g^2 \frac{N}{M} \Gamma (1-4 \Delta_\mathrm{F} ) | \omega | ^{4 \Delta_\mathrm{F} -1} \left(1-\cos(4\pi  \Delta_\mathrm{F})+i\text{sgn}(\omega ) \sin (4 \pi \Delta_\mathrm{F})\right) \, . \label{eq: delta_Pi^R_freq}
	\end{align}
	The Schwinger–Dyson \cref{eq:conformal_Pi_eq} in frequency space then gives the bosonic propagator
	\begin{equation}
		D^R(\omega) = -\delta\Pi^R(\omega)^{-1} \, ,
	\end{equation}
	which can in turn be used to determine $G^{<}(\omega)$.
	This time, the low-energy limit is complicated by the fact that the boson occupancy $n_\mathrm{B}$ diverges for $\omega \to 0$. At finite temperature, this behavior is regularized by the imaginary part of $D^R(\omega)$, which is odd and goes to zero linearly, so that
	\begin{equation}
		D^<(0) = 2 i T \partial_\omega\text{Im}D^R |_{\omega = 0}
	\end{equation}
	is always finite. Since we are interested in the low frequency limit of the ground state, we have to take the $\beta \to \infty$ limit first and, as long as $\omega \gg 1/\beta$, we can again approximate $n_\mathrm{B}(\omega) \sim \theta(-\omega)$, obtaining:
	\begin{align}
		D^<(\omega) =& -i\frac{M}{N} \frac{\cot (\pi  \Delta_\mathrm{F} )}{b^2 g^2  \Gamma (1-4 \Delta_\mathrm{F})}  | \omega | ^{1-4 \Delta_\mathrm{F}} \theta (-\omega ) \, , \\
		\Longrightarrow \quad D^<(t) \, =& i \frac{M}{N}  \frac{ \cot (\pi  \Delta_\mathrm{F}) \Gamma (2-4 \Delta_\mathrm{F})}{2 \pi b^2  g^2 \Gamma (1-4 \Delta_\mathrm{F})} |t| ^{4 \Delta_\mathrm{F}-2} (\cos (2 \pi  \Delta_\mathrm{F})-i \sin (2 \pi  \Delta_\mathrm{F}) \text{sgn}(t)) \, .
	\end{align}
	From the Fourier transform of the bosonic retarded propagator
	\begin{equation}
		D^R(t) = \frac{M}{N} \frac{(4 \Delta_\mathrm{F}-1) \cos (2 \pi  \Delta_\mathrm{F})}{\pi  b^2 g^2 }  t^{4 \Delta_\mathrm{F}-2} \theta(t) 
	\end{equation}
	we obtain the fermionic retarded self-energy via \eqref{eq:SigmaR_t}:
	\begin{equation}
		\Sigma^R(t) = -i \frac{M}{N}\frac{(4 \Delta_\mathrm{F}-1) \cos (2\pi  \Delta_\mathrm{F}) \csc (\pi  \Delta_\mathrm{F}) }{2 \pi  b} t^{2 \Delta_\mathrm{F}-2} \theta(t) \, .
	\end{equation}
	
	We have thus determined all fields and verified that the conformal ansatz consistently solves the infrared saddle-point equations. The remaining condition is the fermionic Dyson equation 
	\begin{equation}
		\Sigma^R(\omega) G^R(\omega) = -1 \, .
	\end{equation}
	which fixes the remaining constants and, in particular, determines $\Delta_\mathrm{F}$ as a function of $M/N$, as reported in Eq.~\eqref{eq:conformal_constant}.

	While the zero-temperature limit makes the conformal structure particularly transparent, it also introduces singularities that must be regularized in numerical simulations. These singularities are naturally resolved at finite temperature through the conformal mapping
	$f(t) = \tanh \frac{\pi t}{\beta}$ discussed in Sec.~\ref{subsec:conf_phase}. The expressions for the retarded propagators in time space are given in \eqref{eq:finite_T_conformal}, while the associated self-energies reads:
	\begin{equation}
		\Sigma^R(t) = 2i b \tilde{b} g^2 \cos(\pi \Delta_\mathrm{F}) \frac{\theta(t)}{\left[\frac{\beta}{\pi }  \sinh \left(\frac{\pi  t}{\beta }\right)\right]^{2-2 \Delta_\mathrm{F} }} \, , \qquad \delta \Pi^R(t) = -2 b^2 g^2 \sin (2 \pi  \Delta_\mathrm{F} ) \frac{\theta(t)}{\left[\frac{\beta}{\pi }  \sinh \left(\frac{\pi  t}{\beta }\right)\right]^{4 \Delta_\mathrm{F} }} \, .
	\end{equation}
	The Fourier transform of these expressions give:
	\begin{align}
		G^R(\omega) &= -i b \beta  \cos (\pi  \Delta_\mathrm{F} ) \Gamma (1-2 \Delta_\mathrm{F} ) \left(\frac{2 \pi }{\beta }\right)^{2 \Delta_\mathrm{F} } \frac{\csc \left(\pi  \Delta_\mathrm{F} -\frac{i \beta  \omega }{2}\right)}{\Gamma \left(1-\Delta_\mathrm{F} -\frac{i \beta  \omega }{2 \pi }\right) \Gamma \left(1-\Delta_\mathrm{F} +\frac{i \beta  \omega }{2 \pi }\right)} \label{eq:GR_finiteT}\\
		D^R(\omega) &= - \tilde{b} \beta  \sin (2 \pi  \Delta_\mathrm{F} ) \Gamma (4 \Delta_\mathrm{F}-1 ) \left(\frac{2 \pi }{\beta }\right)^{2-4 \Delta_\mathrm{F} } \frac{\csc \left(2\pi  \Delta_\mathrm{F} +\frac{i \beta  \omega }{2}\right)}{\Gamma \left(2\Delta_\mathrm{F} -\frac{i \beta  \omega }{2 \pi }\right) \Gamma \left(2\Delta_\mathrm{F} +\frac{i \beta  \omega }{2 \pi }\right)} \label{eq:DR_finiteT} \\
		\Sigma^R(\omega) &= i b \tilde{b} g^2 \beta  \cos (\pi  \Delta_\mathrm{F} ) \Gamma (1-2 \Delta_\mathrm{F}) \left(\frac{2 \pi }{\beta }\right)^{2-2 \Delta_\mathrm{F} } \frac{\csc \left(\pi  \Delta_\mathrm{F} +\frac{i \beta  \omega }{2}\right)}{\Gamma \left(\Delta_\mathrm{F} -\frac{i \beta  \omega }{2 \pi }\right) \Gamma \left(\Delta_\mathrm{F} +\frac{i \beta  \omega }{2 \pi }\right)} \\
		\delta \Pi^R(\omega) &= - \left(\frac{b \beta  g}{2}\right)^2 \frac{N}{M} \frac{\sec (2\pi  \Delta_\mathrm{F} ) }{\Gamma (4 \Delta_\mathrm{F} ) }\left(\frac{2 \pi }{\beta }\right)^{1+4 \Delta_\mathrm{F} } \frac{\csc \left(2\pi  \Delta_\mathrm{F} -\frac{i \beta  \omega }{2}\right)}{\Gamma \left(1-2\Delta_\mathrm{F} -\frac{i \beta  \omega }{2 \pi }\right) \Gamma \left(1-2\Delta_\mathrm{F} +\frac{i \beta  \omega }{2 \pi }\right)}  \label{eq:deltaPiR_finiteT}
	\end{align}
	while the lesser propagators can be obtained from the fluctuation-dissipation relations. analogously to what we have seen for the zero-temperature case, it is possible to show that the expressions above satisfy the infrared saddle-point equations. 
	
	\begin{figure}[htbp]
		\centering
		\includegraphics[width=0.6\textwidth]{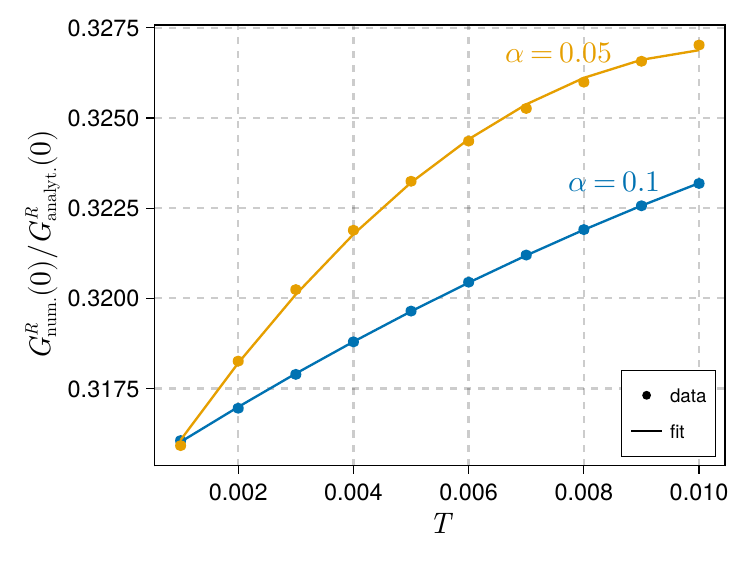}
		\caption{Ratio $G^R_{\text{num}} / G^R_{\text{analytic}}$ as a function of temperature $T$
			for fixed $\alpha = 0.1$ and $\alpha = 0.05$. Dots are numerical data, solid lines are second-order polynomial fits.}
		\label{fig:c1_fit}
	\end{figure}
	
	The only constant which is left to determine is the constant $b$, which is fixed by imposing the self-criticality condition \eqref{eq:self_fix_b}. The numerical constant $c_1$ can be determined by considering the ratio between the numerical solution over the analytical one with $c_1 = 1$. Since the numerical solutions is always at finite temperature, we considered the analytical solution \eqref{eq:GR_finiteT} and then interpolate the ratio of the two quantities evaluated at $\omega = 0$ to get the $T \to 0$ limit. The results are shown in Fig.~\ref{fig:c1_fit} and give $c_1 \approx 0.315$. This value is supported independently by the amplitude of the conformal quantum Fisher information, see Sec.~\ref{sec:QFI-impurity}

	\subsection{SYK$_2$ (impurity) regime}
	\label{app:impurity}
	
	To solve the saddle point equations in the SYK$_2$ regime discussed in Sec.~\ref{sec:impurity_phase}, we start from the ansatz that in the strong-coupling limit $g \gg \omega_0$, the boson behaves approximately as a free particle with renormalized frequency $\omega_r$:
	\begin{equation}
		D^R(\omega) = \frac{\omega_0}{(\omega+i \gamma)^2-\omega_r^2} \, ,
		\label{Eq: DR_impurity}
	\end{equation}
	where $\gamma$ is an interaction-induced broadening, assumed to be small compared to all the other scales in the problem. In particular, we assume that, on the frequency scale set by the fermionic propagator, the bosonic spectral function can effectively be approximated by a delta function. The lesser propagator therefore becomes
	\begin{equation}
		D^<(\omega) = - i \pi \frac{\omega_0}{\omega_r}(\delta(\omega-\omega_r)-\delta(\omega+\omega_r))n_b(\omega) \, .
	\end{equation}
	If we further assume, and verify self-consistently later, that $\omega_r/T \ll 1$, we can write
	\begin{equation}
		\label{eq:Dlesser_impurity}
		D^<(\omega) \sim - i \pi \frac{\omega_0}{\omega_r} \frac{T}{\omega_r}(\delta(\omega-\omega_r)+\delta(\omega+\omega_r)) \, .
	\end{equation}
	
	The equation for the fermionic self-energy, Eq.~\eqref{eq:SigmaR_eq_app}, is more conveniently analyzed in frequency space:
	\begin{equation}
		\Sigma^R(\omega) = i g^2 \int \frac{d \omega'}{2 \pi} (G^<(\omega') \left(D^R(\omega'-\omega) \right)^*+G^R(\omega') D^<(\omega'-\omega)) \sim i g^2 \int \frac{d \omega'}{2 \pi} G^R(\omega') D^<(\omega'-\omega) \, ,
	\end{equation}
	where, in the second step, we used the fact that $T/\omega_r \gg 1$, making the contribution from $D^<$ parametrically dominant. Using Eq.~\eqref{eq:Dlesser_impurity}, together $G^R(\omega \pm \omega_r) \sim G^R(\omega)$, we obtain
	\begin{equation}
		\Sigma^R(\omega) =\frac{g^2 T \omega_0}{\omega_r^2} G^R(\omega) =\frac{\Omega_0^2}{4} G^R(\omega)
	\end{equation}
	where we introduced the scale $\Omega_0$. The self-energy is therefore proportional to the propagator itself, which is the characteristic structure of the SYK$_2$ model. Substituting this expression into the Schwinger--Dyson equation for the fermion immediately gives
	\begin{equation}
		G^R(\omega) = 2 \frac{\omega - \text{sign}(\omega) \sqrt{\omega^2-\Omega^2_0}}{\Omega^2_0}
	\end{equation}
	where the branch is fixed by parity and by the requirement that $G^R(\omega)\sim 1/\omega$ at large frequencies.
	
	The scale $\Omega_0$ can be determined by imposing the self-criticality condition $\Pi^R(0) = - \omega_0$. Assuming $\Omega_0 \gg T$, consistently with the assumption that $G^R(\omega)$ varies slowly on the bosonic scale, the lesser propagator takes the form
	\begin{equation}
		G^<(\omega) = 2 \pi i \rho(\omega) \theta(-\omega) \, ,
	\end{equation}
	where $\rho$ given in  Eq.~\eqref{eq:rho_impurity}.
	Substituting this expression into the bosonic self-energy \cref{eq:PiR_t},we find
	\begin{equation}
		\Pi^R(\omega=0) = -i g^2 \frac{N}{M} \int \frac{\dd \omega}{2\pi} G^<(\omega) (G^A(\omega) + G^R(\omega)) = g^2 \frac{N}{M} \int_{-\Omega_0}^0 \dd \omega \rho(\omega) \frac{4\omega}{\Omega_0^2} = -\frac{8 g^2 N}{3 \pi M \Omega_0} \, ,
		\label{eq:impurity_self_appendix}
	\end{equation}
	and, in turn, 
	\begin{equation}
		\Omega_0 = \frac{8 g^2 N}{3 \pi M \omega_0} \, .
	\end{equation}
	For $\omega_0 = 1$, corresponding to dimensionless coupling constants, the approximations employed above are self-consistent in the regime $g^{-2} \ll T \ll g^2$. 
	
	The expression above was derived assuming  $\omega_r \ll \omega_0$.  Including the leading corrections due to a finite $\omega_r$ amounts to imposing $\Pi^R(0) = - \omega_0+ \frac{\omega_r^2}{\omega_0}$, which leads to the corrected expression
	\begin{equation}
		\Omega_0 =  \frac{8 g^2 N}{3 \pi M \omega_0} + \frac{3 \pi M T}{2  N} \, .
	\end{equation}
	
	The last step to self-consistently solve the saddle point equations is to show that the bosonic self-energy is at most linear in frequency. We begin from
	\begin{equation}
		\Pi^R(\omega) = -i g^2 \frac{N}{M} \int \frac{\dd \omega'}{2\pi} G^<(\omega') (G^A(\omega'-\omega) + G^R(\omega'+\omega)) \, .
	\end{equation}
	Since the bosonic propagators are defined on frequency scales which are much smaller then $\Omega_0$ we restrict to $\omega \ll \Omega_0$. 
	In this limit, we have
	\begin{equation}
		\text{Re} (G^A(\omega'-\omega) + G^R(\omega'+\omega)) = \frac{2}{\Omega_0^2} (\omega'-\omega + \omega'+\omega) =  \frac{4}{\Omega_0^2} \omega' \, ,
	\end{equation}
	which means that the real part of $\Pi^R(\omega)$ is $\omega$-independent and coincides with the value at $\omega=0$.
	For the imaginary part, expanding for $\omega/\Omega_0 \ll 1$, we obtain
	\begin{equation}
		\text{Im}(\Pi^R(\omega)) = -\frac{3 \omega  \omega_0}{2  \Omega_0}
	\end{equation}
	which fixes the boson broadening $\gamma$ to
	\begin{equation}
		\gamma = \frac{3 \omega_0^2}{4 \Omega_0} \, ,
	\end{equation}
	which is indeed parametrically small in the large-$g$ limit, consistently with the original ansatz.

	\subsection{Bare boson regimes}
	\label{app:high_T_model}
	
	In this appendix, we demonstrate that the effective model presented in \cref{eq:effective_H}, that describes the driven-fermion regime, can be directly derived from the YSYK Hamiltonian at high temperature and reproduces the YSYK saddle point \cref{eq:driven_SYK2}.
	
	We start by moving to the frame rotating at the bare bosonic frequency, $U(t) = \exp(i \omega_0 t \sum_m a^\dag_m a_m)$. This transformation removes the free bosonic term of \cref{eq:YSYK_H} and acts only on the bosons, so that all fermionic
	correlators are left unchanged:
	\begin{equation}
		H = - \mu_C \sum_{j=1}^N c_j^\dag c_j + \frac{1}{\sqrt{2MN}} \sum_{jkm} g^m_{jk} \, c_j^\dag c_k (a_m e^{-i \omega_0 t} + a^\dag_m e^{i \omega_0 t}) \, .
	\end{equation}
	At high-temperature the thermal fluctuations dominate over the quantum ones, and the boson becomes classical with typical magnitude
	\begin{equation}
		|a_m|_\mathrm{typ.} = \sqrt{\langle a^\dag_m a_m \rangle} = \sqrt{n_\mathrm{B}(\omega_0)} \sim \sqrt\frac{T}{\omega_0}.
	\end{equation}
	We can therefore replace the bosonic operators by static c-numbers $a_m \to \sqrt\frac{T}{\omega_0} e^{i \varphi_m}$, where $\varphi_m \in [0, 2 \pi)$ are uniformly distributed angles, and find the following effective Hamiltonian
	\begin{equation}
		H_\mathrm{eff.} = - \mu_C \sum_{j=1}^N c_j^\dag c_j + \sqrt{\frac{2T}{MN\omega_0}} \sum_{jkm} g^m_{jk} \, c_j^\dag c_k (\cos(\omega_0 t) \cos (\varphi_m)+ \sin(\omega_0 t) \sin (\varphi_m)) \, .
	\end{equation}
	Now we can define 
	\begin{align}
		J^1_{ij} &= \sqrt{\frac{2T}{M\omega_0}} \sum_{m} g^m_{ij}  \sin (\varphi_m) \, . \\
		J^2_{ij} &= \sqrt{\frac{2T}{M\omega_0}} \sum_{m} g^m_{ij}  \cos (\varphi_m) \, . 
	\end{align}
	and express the Hamiltonian exactly as in \cref{eq:effective_H}:
	\begin{equation}
		H_\mathrm{eff.} = -\mu_C \sum_{i} c_i^\dag c_i + \frac{1}{\sqrt{N}}\sum_{ij} J_{ij}(t) c_i^\dag c_j \, ,
	\end{equation}
	with
	\begin{equation}
		J_{ij}(t) = J_{ij}^1 \sin (\omega_0 t) + J_{ij}^2 \cos (\omega_0 t) \, .
	\end{equation}
	By doing an ensemble average and approximating the sum over the angle with an integral, one can also explicitly check
	\begin{equation}
		\langle |J^{1,2}_{ij}|^2 \rangle = \frac{g^2}{\beta \omega_0} \, , \qquad \langle J^1_{ij} J^2_{ij} \rangle = 0 \, ,
	\end{equation}
	as expected.
	
	Let's now move to the derivation of \cref{eq:driven_SYK2}.
	What we need to do is to perform a path-integral calculation on the Schwinger-Keldysh contour as already seen in appendix \ref{app:path_integral}. 
	The real-time partition function is given by
	\begin{equation}
		\begin{split}
			Z(t)=&\Tr \left(e^{-i H t} \rho_0 e^{i H t} \right) = \int \mathcal{D} \psi^\pm e^{iS[\psi^\pm]} \, ,\\ S[\psi^\pm] =& \int \dd t  \sum_{\nu=\pm} \nu \left[\sum_j  \bar{\psi}_j^\nu (i \partial_t+\mu_C) \psi_j^\nu - \frac{1}{\sqrt{N}} \sum_{jk} J_{jk}(t) \, \bar\psi_j^\nu \psi_k^\nu  \right]\, .
		\end{split}
	\end{equation}
	We can average over the two independent Gaussian couplings $J^1,J^2$
	\begin{equation}
		\begin{split}
			iS[\psi] =& iS_\mathrm{F}[\psi] - \frac{g^2}{2 \omega_0 \beta} \frac{1}{N} \sum_{jk} \left[ \left|\int \dd t \sum_{\nu = \pm} \bar\psi_j^\nu \psi_k^\nu \sin (\omega_0 t)  \right|^2 + \left|\int \dd t \sum_{\nu = \pm} \bar\psi_j^\nu \psi_k^\nu \cos (\omega_0 t)  \right|^2 \right] \\
			=& iS_\mathrm{F}[\psi] - \frac{g^2}{2 \omega_0 \beta} \frac{1}{N} \sum_{jk} \left[ \int \dd t \dd t' \sum_{\mu \nu} \bar\psi^\mu_j(t) \psi^\mu_k(t) \bar\psi^\nu_k(t') \psi^\nu_j(t') \cos(\omega_0(t-t')) \right],
		\end{split}
	\end{equation}
	where $S_\mathrm{F}[\psi]$ is defined in \eqref{eq:S_F_action_def}.
	We can now introduce collective fields $G,\Sigma$ as in \eqref{eq:G_Sigma_fat1} and integrate out the fermionic fields, which appear just quadratically after the redefinitions using the collective fields. The resulting action for the collective fields reads:
	\begin{equation}
		\frac{i S_\text{eff}[G,\Sigma]}{N} = \Tr \log (G_0^{-1}-\Sigma) + \Tr(\Sigma G) - \frac{g^2}{2 \omega_0 \beta} \sum_{\mu \nu} \int \dd t \dd t' s_{\mu \nu} G^{\mu \nu}(t,t') G^{\nu \mu}(t',t) \cos(\omega_0 (t-t')) \, .
	\end{equation}
	By varying the action we get the following saddle-point equations:
	\begin{align}
		\frac{\delta S_\text{eff}}{\delta \Sigma} = 0 \quad &\Longrightarrow \quad G(t,t') = (G_0^{-1} - \Sigma)^{-1}(t,t') \, , \\
		\frac{\delta S_\text{eff}}{\delta G} = 0 \quad &\Longrightarrow \quad \Sigma^{\mu\nu}(t,t') =   \frac{g^2}{\omega_0 \beta} s_{\mu\nu} G^{\mu\nu}(t,t') \cos(\omega_0 (t-t')) \, .
	\end{align}
	By performing the Keldysh rotation, we exactly get~\cref{eq:driven_SYK2}.

	\section{Extracting the two-point functions from experiment}
	\label{app:measurement}
	
	This section illustrates how the hallmarks of the rich phase diagram established in the main text can be accessed experimentally. 
	As stated in the introduction, the scenario for which we performed our analysis does not aim to quantitatively reproduce a specific experimental platform. However, we expect the main extracted features to be generic signatures of the YSYK behavior. 
	All the diagnostics introduced in Sec.~\ref{sec:observables} are functionals of retarded
	propagators. The purpose of this appendix is to make explicit that accessing them does not
	require any specialized many-body protocol: $D^R$ is, by definition, the linear susceptibility of
	the bosonic quadrature, and is therefore measured by ordinary response spectroscopy.

	Consider adding to the Hamiltonian in Eq.~\eqref{eq:YSYK_H} a weak, time-dependent field conjugate
	to the bosonic quadratures,
	\begin{equation}
		V(t) = h(t) \sum_{m=1}^{M} x_m \, ,
		\label{eq:probe}
	\end{equation}
	and let the system be in the steady state of interest before the probe is switched on. To linear
	order in $h$, the Kubo formula gives
	\begin{equation}
		\delta \langle x_m(t) \rangle
		= \sum_{n} \int_{-\infty}^{\infty} dt' \, \chi_{mn}(t-t') \, h(t') \, ,
		\qquad
		\chi_{mn}(t-t') = -i\,\theta(t-t')\,\big\langle [\, x_m(t), x_n(t') \,] \big\rangle \, .
		\label{eq:kubo}
	\end{equation}
	After disorder averaging the two-point functions are diagonal and identical in the mode index,
	such that comparing with the definitions in Eqs.~\eqref{eq:boson_<_def}--\eqref{eq:boson_R_def} gives
	$\chi_{mn} = \delta_{mn} D^R$ and hence, in frequency space,
	\begin{equation}
		\delta \langle x_m(\omega)\rangle = D^R(\omega)\, h(\omega) \, .
		\label{eq:response_master}
	\end{equation}
	If all modes are driven uniformly, one may detect the collective quadrature $\sum_m x_m$, whose
	response is $M D^R(\omega) h(\omega)$. Since $D^R$ is itself defined as an average over the $M$
	modes, one may equivalently drive and read out a single mode, in which case
	Eq.~\eqref{eq:response_master} holds up to self-averaging fluctuations that vanish as
	$M \to \infty$.
	
	We now work out an explicit example of this protocol that is relevant for cavity-QED realizations of SYK and YSYK models \cite{cavity_proposal,Baumgartner:2024ysk,Baumgartner:2025uue,Solis_2026,Ferrari:2026zkn}, where the
	bosonic modes are electromagnetic modes and the monitoring channel is the field leaking through
	an output coupler. Homodyne detection makes this example particularly convenient, since it gives
	access to the full complex response within a single apparatus. It is worth noting that the
	coupling to the outside world, which makes the measurement possible, is the very same one that
	generates the boson losses studied in Sec.~\ref{sec:dissipation}: measurability and dissipation
	are two faces of the same open-system structure.
	
	\subsection{Cavity QED example}
	\label{app:homodyne}
	
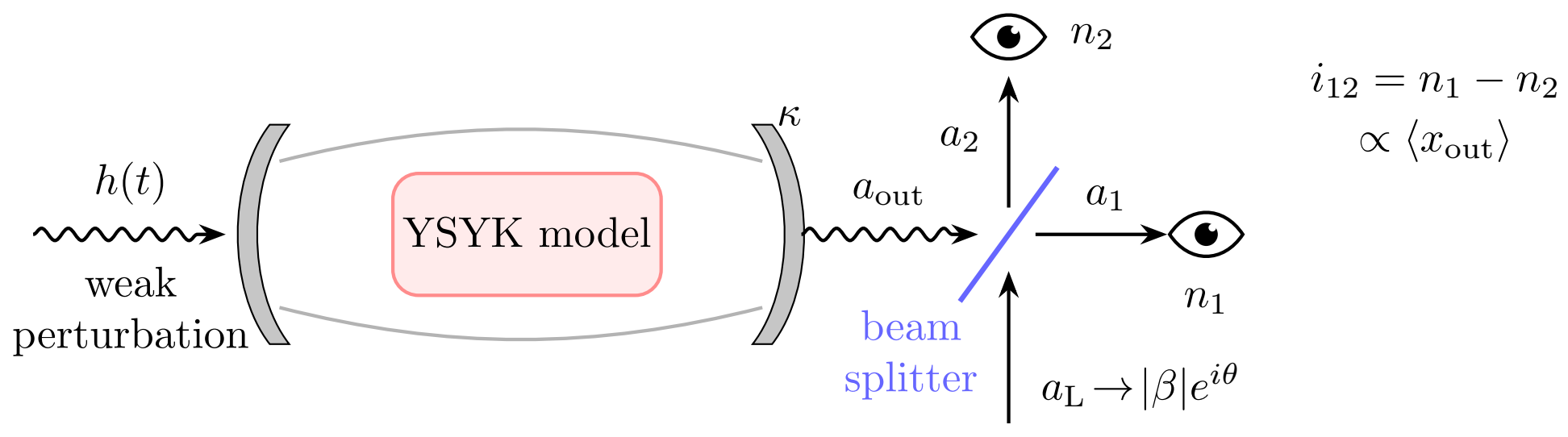
\begin{figure}[htbp]
		\centering
		\begin{tikzpicture}[
			>={Stealth[length=2.2mm]},
			photon/.style={decorate, decoration={snake, amplitude=0.5mm, segment length=2.2mm,
					post length=1.6mm}, ->, thick},
			beam/.style={thick, ->},
			mirror/.style={fill=gray!45, draw=black, thin},
			font=\small
			]
			
			\fill[mirror]
			(0.10,0.9) .. controls (-0.25,0.45) and (-0.25,-0.45) .. (0.10,-0.9)
			-- (-0.06,-0.9) .. controls (-0.41,-0.45) and (-0.41,0.45) .. (-0.06,0.9) -- cycle;
			\fill[mirror]
			(3.90,0.9) .. controls (4.25,0.45) and (4.25,-0.45) .. (3.90,-0.9)
			-- (4.06,-0.9) .. controls (4.41,-0.45) and (4.41,0.45) .. (4.06,0.9) -- cycle;
			
			\draw[gray!60, thick] (0.02,0.60) .. controls (1.4,0.95) and (2.6,0.95) .. (3.98,0.60);
			\draw[gray!60, thick] (0.02,-0.60) .. controls (1.4,-0.95) and (2.6,-0.95) .. (3.98,-0.60);
			
			\fill[red!8, draw=red!45, thick, rounded corners=6pt] (0.95,-0.5) rectangle (3.15,0.5);
			\node[align=center, font=\footnotesize] at (2.05,0.02) {YSYK model};
			
			\draw[photon] (-2.0,0) -- (-0.42,0);
			\node[font=\footnotesize, above] at (-1.2,0.10) {$h(t)$};
			\node[font=\footnotesize, below, align=center] at (-1.2,-0.12) {weak\\ perturbation};
			
			\node[font=\footnotesize, above] at (4.20,0.75) {$\kappa$};
			\draw[photon] (4.30,0) -- (5.75,0);
			\node[font=\footnotesize, above] at (5.02,0.10) {$a_{\mathrm{out}}$};
			
			\draw[very thick, blue!60] (5.60,-0.55) -- (6.40,0.55);
			\node[font=\footnotesize, blue!60, above,align=center] at (5.2,-1.5) {beam\\ splitter};
			
			\draw[beam] (6.00,-1.55) -- (6.00,-0.30);
			\node[font=\footnotesize, right, align=left] at (6.12,-1.25)
			{$a_{\mathrm{L}}\!\to\!|\beta|e^{i\theta}$};
			
			\draw[beam] (6.22,0) -- (7.30,0);
			\node[font=\footnotesize, above] at (6.80,0.05) {$a_1$};
			\draw[beam] (6.00,0.22) -- (6.00,1.30);
			\node[font=\footnotesize, left] at (5.92,0.78) {$a_2$};
			
			\pic at (7.62,0) {eye};
			\node[font=\footnotesize, below] at (7.62,-0.30) {$n_1$};
			\pic at (6.00,1.62) {eye};
			\node[font=\footnotesize, right] at (6.36,1.62) {$n_2$};
			
			\node[font=\footnotesize, align=center] at (9.5,1.0)
			{$i_{12}=n_1-n_2$\\[2pt] $\propto \langle x_{\mathrm{out}}\rangle$};
			
		\end{tikzpicture}
		\caption{Schematic of a balanced homodyne measurement of the bosonic response in a cavity-QED
			realization of the model. A weak perturbation $h(t)$, conjugate to the quadrature $x$, is applied
			to the medium; the intracavity field leaks through the cavity mirror at rate $\kappa$ and is
			combined on a balanced beam splitter with a strong classical oscillator. The
			difference photocurrent of the two detectors is proportional to the output quadrature selected by
			$\theta$, and the retarded propagator follows from Eq.~\eqref{eq:homodyne_final} as the ratio
			between the measured response and the applied perturbation.}
		\label{fig:homodyne}
	\end{figure}

	Here, the $M$ bosons are electromagnetic modes confined in a cavity, the fermions are provided by
	cold atoms placed inside it, and the random Yukawa coupling $g^m_{jk}$ is engineered as discussed
	in Ref.~\cite{Solis_2026}. The measurement chain is sketched in Fig.~\ref{fig:homodyne}.
	
	Input--output theory relates the field escaping through the output coupler, $a_\mathrm{out}$, to the intracavity
	field, $a$, and to the fields entering from outside, $a_\mathrm{in}$,\footnote{We choose the numerical factor as $\sqrt{2\kappa}$, in order to be consistent with the convention of Eq.~\eqref{eq:markov_approx}, where the jump
		operators are $L_m = \sqrt{\kappa}\,a_m$. $\kappa$ is therefore the amplitude decay rate, rather than the energy decay rate.}
	\begin{equation}
		a_{\mathrm{out}} = a_{\mathrm{in}} + \sqrt{2\kappa}\, a \, .
		\label{eq:input_output}
	\end{equation}
	In a frame rotating at a large detuning, which is precisely the setting assumed in
	Ref.~\cite{cavity_proposal}, the input channel carries no coherent amplitude,
	$\langle a_{\mathrm{in}}\rangle = 0$: the incoming field does not contribute to the mean signal,
	although it does set the shot-noise floor discussed below. Consequently, 
	\begin{equation}
		\delta \langle x_{\mathrm{out}}(t) \rangle = \sqrt{2\kappa} \; \delta \langle x(t) \rangle \, ,
		\label{eq:xout}
	\end{equation}
	and the response of the escaping field faithfully reproduces that of the mode inside.
	
	The quadratures of a propagating field are not directly accessible to a photodetector, which is
	sensitive only to intensity. Balanced homodyne detection circumvents this obstacle by interfering
	$a_{\mathrm{out}}$ with a strong coherent beam, the local oscillator $a_{\mathrm{L}}$, on a
	balanced beam splitter. Denote by $a_{1,2}$ the two output ports,
	\begin{equation}
		a_1 = \frac{a_{\text{out}} + i\, a_{\text{L}}}{\sqrt{2}} \, , \qquad
		a_2 = \frac{a_{\text{L}} + i\, a_{\text{out}}}{\sqrt{2}} \, . 
	\end{equation}
	When measuring the difference of the photon numbers of the two output ports,  terms bilinear in the same field cancel, leaving only
	the interference term
	\begin{equation}
		n_1 - n_2 = i \left( a^\dagger_{\mathrm{out}} a_{\mathrm{L}}
		- a^\dagger_{\mathrm{L}} a_{\mathrm{out}} \right) .
		\label{eq:n1_minus_n2}
	\end{equation}
	We take the local oscillator in a coherent state $\ket{\beta}$ with $\beta = |\beta| e^{i\theta}$ and $|\beta| \gg 1$. \cref{eq:n1_minus_n2} then becomes
	\begin{equation}
		i_{12} \equiv \langle n_1 - n_2 \rangle = \sqrt{2}\,|\beta| \,
		\big\langle x^{(\theta)}_{\mathrm{out}} \big\rangle \, ,
		\qquad
		x^{(\theta)} \equiv \frac{a\, e^{-i\theta} - a^\dagger e^{i\theta}}{\sqrt{2}i} \, .
		\label{eq:homodyne_current}
	\end{equation}
	The phase of the local oscillator therefore acts as a knob selecting which quadrature is recorded: $\theta = -\pi/2$ measures $x$, the field we are after, while $\theta = 0$ measures the conjugate quadrature $p$. Note also that $i_{12}$ is amplified by the macroscopic factor $|\beta|$, which is what makes a single-mode quadrature detectable in practice.
	
	Combining Eqs.~\eqref{eq:response_master},~\eqref{eq:xout}, and~\eqref{eq:homodyne_current} at
	$\theta = -\pi/2$, and Fourier transforming, the change in the difference photocurrent induced by the
	probe is
	\begin{equation}
		\delta i_{12}(\omega) = 2 |\beta| \sqrt{\kappa} \; D^R(\omega) \, h(\omega) \, .
		\label{eq:homodyne_final}
	\end{equation}
	The retarded bosonic propagator is thus obtained, as a complex function of frequency, from the
	ratio between the measured signal and the applied probe. In a realistic setup, the overall
	prefactor also contains the quantum efficiency of the detectors; since the observable $\alpha$
	introduced in Sec.~\ref{sec:observables} depends on the absolute normalization of $D^R$, this
	prefactor has to be calibrated, for example by comparison with the free theory obtained when the
	Yukawa coupling is switched off.

	\section{Numerical methods}
	\label{app:numerics}
	
	In this appendix, we describe how the saddle-point equations are solved numerically and how the crossover lines of Secs.~\ref{sec:markers} and~\ref{sec:dissipation} are extracted from the solutions. We first outline the iterative scheme on the real-frequency axis that underlies all our results. Appendix~\ref{app:stabilization} introduces the stabilization of the bosonic sector, in which the gap $\alpha$ is held fixed and the coupling $g$ is adjusted self-consistently, which makes the near-critical regimes accessible. Appendix~\ref{app:const_alpha_sweeps} specifies the resulting constant-$\alpha$ sweeps and the procedures used to locate the conformal/SYK$_2$ and bare-boson boundaries, and Appendix~\ref{app:additional_results_diss} repeats this analysis in the presence of dissipation.
	
	The saddle-point equations are solved directly on the real-frequency axis, so that
	spectral functions are obtained without any analytic continuation. We discretize
	frequency on a uniform, symmetric grid and we work with the
	conjugate time grid generated by the discrete Fourier transform.
	
	The iteration exploits the fact that the two saddle-point equations are
	local in complementary domains: the self-energies are products of propagators in the
	time domain, whereas the Dyson equations are algebraic in frequency. Each iteration
	therefore transforms the propagators to time, forms the self-energies as local products,
	and transforms back. In thermal equilibrium, the lesser
	propagators are not evolved independently but fixed by the fluctuation--dissipation
	relation~\eqref{eq:fluctuation_dissipation_G}--\eqref{eq:fluctuation_dissipation_D}. The retarded propagators are
	then updated from
	\begin{equation}
		G^R = \frac{1}{\omega + i \eta - \Sigma^R}, \qquad
		D^R = \frac{1}{\frac{(\omega + i\kappa)^2 - \omega_0^2}{\omega_0} - \Pi^R},
		\label{eq:SD_eqn_numerics}
	\end{equation}
	where $\eta$ and $\kappa$ are Markovian broadenings for the fermion and the boson, which avoids numerical instabilities while assuring causality. While $\eta$ is always small, $\kappa$ becomes large in the study of dissipation of Sec.~\ref{sec:dissipation}.
	
	We then mix the propagators with the previous iteration,
	\begin{equation}
		X_{i+1} = (1-s)\,X_i + s\,X_i^{\rm new}, \qquad 0 < s \le 1,
	\end{equation}
	which damps the oscillatory instability. The mixing parameter $s$ is adapted at run time: it is increased
	while the residual decreases and reduced whenever convergence deteriorates. Convergence is monitored through the largest pointwise change of all propagators, rescaled by the
	mixing parameter,
	$\epsilon = \max_\omega |X_{i+1}-X_i| / s$, and the loop terminates when
	$\epsilon$ falls below a threshold, typically $10^{-8}$.
	
	\subsection{Stabilization of the bosonic sector}
	\label{app:stabilization}
	
	As discussed in Sec.~\ref{sec:markers}, the self-criticality of the YSYK model
	manifests itself in the boson self-energy attempting to cancel the bare boson frequency,
	$\Pi^R(0) \to -\omega_0$, at almost every step of the iteration. For a real boson this
	is problematic: it produces a singularity of $D^R$ at $\omega = 0$ that appears well
	before the solution has converged, and which is an artifact of the convergence path
	rather than a property of the fixed point. 
	Since the intermediate iterates may cross the critical point, it is convenient to hold the
	gap fixed throughout the loop, in the same spirit as constrained schemes employed in the
	study of critical phases~\cite{Pini2019}. We therefore fix
	\begin{equation}
		\alpha \equiv -\,\mathrm{Re}\,D^R(0)^{-1} = \frac{\omega_0^2 + \kappa^2}{\omega_0} + \mathrm{Re}\,\Pi^R(0) \, ,
		\label{eq:gap_numerics}
	\end{equation}
	which coincides with the renormalized boson mass $\omega_r^2/\omega_0$. Because the solution is
	uniquely determined by $g$ and $T$, constraining $\alpha$ requires one of them to be
	adjusted; the natural choice is $g$, since $g^2$ enters the saddle-point equations
	linearly. Writing $\Pi_i^R(0) = g_i^2 c_i$, where the coefficients $c_i$ carry no explicit
	dependence on the coupling, the gap obtained at step $i+1$ with the old coupling is
	\begin{equation}
		\frac{\omega_0^2 + \kappa^2}{\omega_0} + g_i^2 c_{i+1} = \alpha + \varepsilon \, ,
	\end{equation}
	with $\varepsilon$ the deviation from the target. Demanding instead that a new coupling
	$g_{i+1}$ reproduce the target exactly,
	$(\omega_0^2 + \kappa^2)/\omega_0 + g_{i+1}^2 c_{i+1} = \alpha$, and subtracting the two conditions
	gives the update
	\begin{equation}
		g_{i+1} = \sqrt{g_i^2 - \frac{\varepsilon}{c_{i+1}}} \, .
		\label{eq:g_update}
	\end{equation}
	Since both self-energies are proportional to $g^2$, they are rescaled by
	$g_{i+1}^2/g_i^2$ once the coupling is updated, so that no additional evaluation is
	required. In practice, the relative change of $g$ per iteration is capped, and the update
	is rejected in favor of a local regularization of $D^R$ whenever
	Eq.~\eqref{eq:g_update} would drive the coupling outside a physically sensible range.
	In both the conformal and the SYK$_2$ regimes, the gap is set by the renormalized mass
	$\omega_r^2$, which is known analytically for given $g$ and $T$; this analytic estimate is
	used to target specific points of the phase diagram.
	
	Finally, we found empirically that initializing the iteration with the impurity solution
	of Sec.~\ref{sec:impurity_phase} yields convergent results in essentially all regimes,
	and we adopt it as the default starting point throughout.
	
	\subsection{Constant-$\alpha$ sweeps} \label{app:const_alpha_sweeps}
	
	All numerical results of Secs.~\ref{sec:markers},~\ref{sec:QFI} and~\ref{sec:Phase_signatures_diss} are obtained from temperature sweeps at fixed gap, computed with the scheme of Appendix~\ref{app:stabilization}, for the ten gaps $\alpha=0.0025$, $0.005$, $0.01$, $0.025$, $0.05$, $0.1$, $0.2$, $0.3$, $0.4$, $0.5$ and three values of the bosonic broadening of Eq.~\eqref{eq:SD_eqn_numerics}: $\kappa=0.1$ and $\kappa=1$, and $\kappa=10^{-10}$, a purely numerical regulator which regularizes the closed system dynamics, and which we consider as being the closed system. For the closed system, four additional sweeps at $\alpha=0.6, 0.7, 0.8$ and $0.9$ resolve the approach to the bare-boson limit (Fig.~\ref{fig:fits}). Each sweep consists of 30 temperatures, logarithmically spaced between $T=2.5\times10^{-4}$ and $T=2$. The sweeps at $\alpha=0.0025$ cover only a smaller range of temperatures, and none was computed for $\kappa=1$. The fermionic broadening is fixed to $\eta=10^{-10}$, and the frequency grid is chosen for every sweep such that it is large enough to contain the spectral weight, while still being dense enough to resolve the low-frequency structure of the Green's functions. Points at which the iteration of Appendix~\ref{app:stabilization} did not converge are discarded: one temperature in each of the closed-system sweeps at $\alpha=0.005, 0.05, 0.1$ and $0.2$, the three lowest temperatures at $\alpha=0.9$, since the required updates of $g$ in the iterative scheme fall below a minimal numerical threshold of $10^{-4}$ and are rejected, and, for $\kappa=1$, some high temperature points for smaller gaps are also not converged. From these sweeps we extract the two crossover lines of Sec.~\ref{sec:markers}: the conformal/SYK$_2$ boundary from the sign change of $\xi$, and the bare-boson boundary from the amplitudes $h(\alpha)$ of the constant-$\alpha$ curves. 
	
	\begin{figure}[t!]
		\centering
		\includegraphics[width=0.6\textwidth]{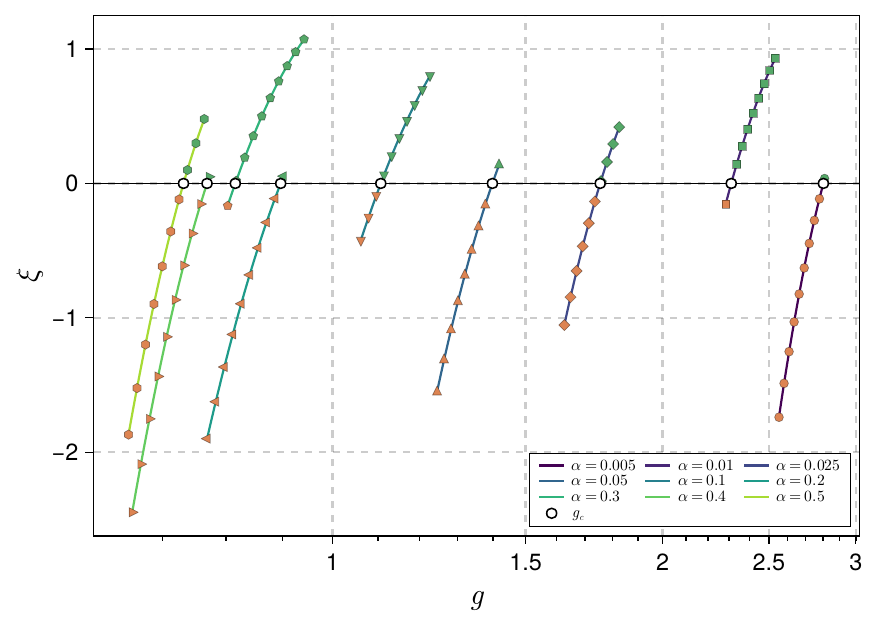}
		\caption{Curvature $\xi$ along the zoomed constant-gap sweeps ($0.005 \leq \alpha \leq 0.5$ in units of $\omega_0$) as a function of the self-consistent coupling $g$. Each sweep resolves with ten temperatures the coarse interval of Fig.~\ref{fig:Phase_diagram_plot} in which $\xi$ changes sign; along a sweep, larger $g$ means higher $T$. Marker color gives the sign of $\xi$ (orange: conformal, green: SYK$_2$). Solid lines are the fits in $\ln g$, and open circles mark their roots $g_c$, which give the critical points in the inset of Fig.~\ref{fig:Phase_diagram_plot}.}
		\label{fig:Zoom_xi_vs_g}
	\end{figure}
	
	\paragraph{Locating $\xi=0$.} To locate the critical points $(g_c, T_c)$ where the curvature vanishes, we compute $\xi$ of Eq.~\eqref{eq:curvature} at $\omega=0$. Along every sweep (with $\alpha \geq 0.005$) it changes sign exactly once. The coarse temperature grids of Figs.~\ref{fig:Phase_diagram_plot} and~\ref{fig:Phase_diagram_xi_dissipation} still leave a rather big interval on the location of the sign change. Therefore, for every gap and every $\kappa$, we computed a zoomed sweep of ten temperatures, logarithmically spaced across this interval (Fig.~\ref{fig:Zoom_xi_vs_g}). The critical point $(g_c, T_c)$ is the root of a quadratic fit of $\xi$ in $\ln g$ over the zoomed sweep, with $T_c$ read off from a quadratic fit of $\ln T$ against $\ln g$ along the same sweep. Linear interpolation between the two points that straddle the sign change, or a cubic spline instead of the quadratic fit, moves the critical points by less than $0.1\%$ in the closed system and by less than $1\%$ with dissipation which is negligible compared with the uncertainties of the fits and further corroborates our numerical results. For $\kappa=1$ and $\alpha=0.005$, three non-converged temperatures of the zoomed sweep, which produce a spurious second sign change, were removed.
	
	\paragraph{Fit of the conformal/SYK$_2$ boundary.} The critical points are fitted with $T_c = C \, g_c^{\nu}$ on windows $\alpha \leq \alpha_{\rm max}$ that start from the smallest gap. The local exponent between neighboring gaps lies between $-1.70$ and $-1.84$ for $\alpha \leq 0.1$ and steepens to $-4.7$ near $\alpha=0.5$, where the bare-boson regime takes over (cf. the inset of Fig.~\ref{fig:Phase_diagram_plot}). The window $\alpha \leq 0.3$ of Fig.~\ref{fig:Phase_diagram_plot} excludes the strongest steepening. The resulting exponent $\nu = -1.96 \pm 0.09$ is an effective one over this range, giving a more reliable fit for the pure conformal/SYK$_2$ boundary. Restricting the window to $\alpha \leq 0.1$ gives $\nu=-1.76 \pm 0.01$, while extending it to $\alpha \leq 0.5$ gives $\nu=-2.16 \pm 0.12$.
	
	\begin{figure}[t!]
		\centering
		\includegraphics[width=1\textwidth]{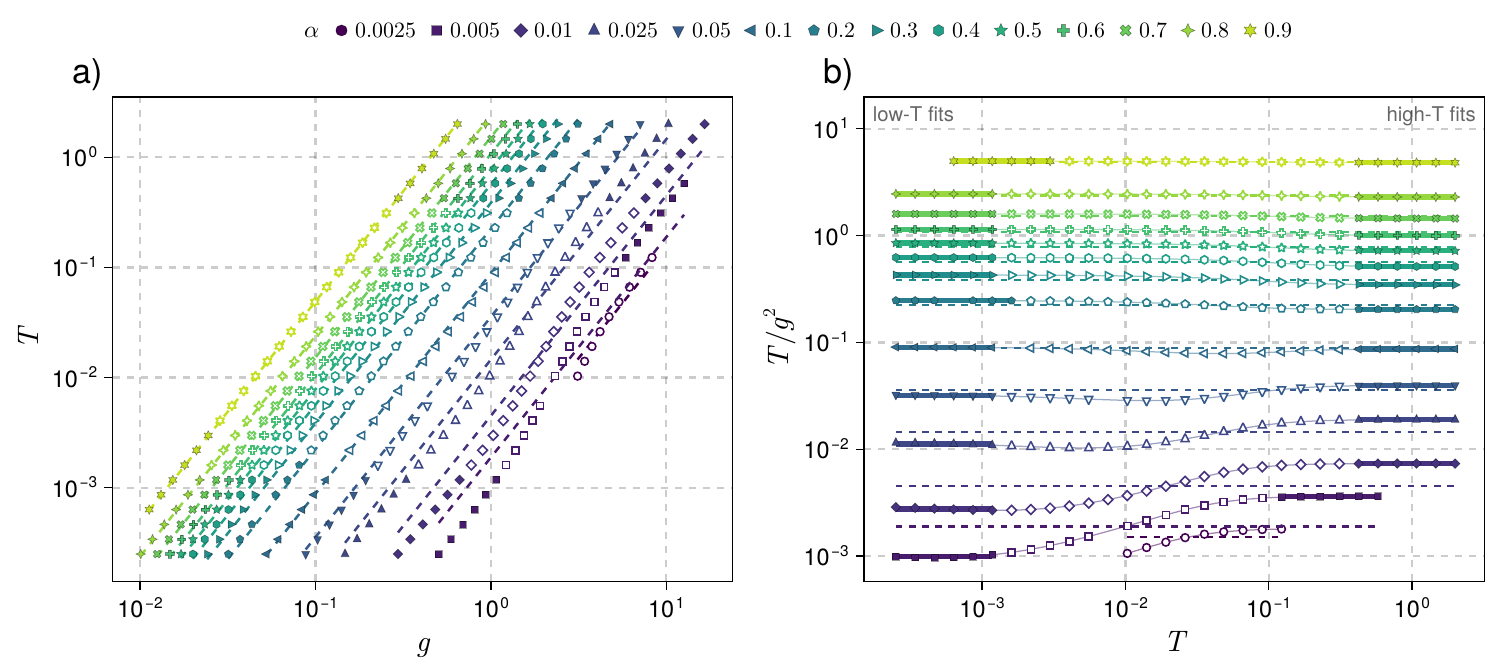}
		\caption{Two-branch fit of the closed-system constant-gap sweeps for all fourteen gaps $0.0025 \leq \alpha \leq 0.9$ (in units of $\omega_0$; marker shape and color label $\alpha$ as in the legend). {\bf a)} Sweeps in the $(g,T)$-plane. Filled markers are the six lowest- and six highest-temperature points of each sweep, which enter the fixed-exponent fits; open markers lie in the crossover regime and are not fitted. The dashed lines are $T=h(\alpha) \, g^2$ with the geometric mean, the same lines as in Fig.~\ref{fig:fits}a. {\bf b)} The same data as $T/g^2$ against $T$, where each branch appears as a plateau. Thick lines show $h_{\rm lo}$ (left) and $h_{\rm hi}$ (right) over the points they are fitted to, and dashed lines their geometric mean $h(\alpha)$. The sweep at $\alpha=0.0025$ (all markers open) starts inside the crossover, has no low-temperature branch and keeps the single fixed-exponent fit.}
		\label{fig:two_branch_fits}
	\end{figure}
	
	\paragraph{Two-branch fit of the constant-$\alpha$ curves.} Both asymptotic regimes give $T \propto g^2$ at fixed gap, Eq.~\eqref{eq:T_g2_conf_imp}, but with different amplitudes, so that a sweep does not lie on a single line $T = h \, g^2$. This is most clearly seen in the ratio $T/g^2$ along a sweep (Fig.~\ref{fig:two_branch_fits}\textbf{b}): it has one plateau at low temperature, in the conformal regime, and another one at high temperature, in the SYK$_2$ regime, joined by a smooth crossover across the sign change of $\xi$. For the smallest gaps the two plateaus are far apart, however, they move closer to each other for increasing gap and eventually cross (between $\alpha=0.05$ and $\alpha=0.1$), as at $\alpha=0.1$ and beyond the conformal plateau sits slightly higher. A single fit with fixed exponent over the whole sweep therefore returns a mixture of the two amplitudes, weighted by the number of temperatures that the grid happens to place on either side of the crossover. A fit with a free exponent, $T \propto g^\nu$, instead absorbs the crossover into $\nu$: for $\alpha \geq 0.1$ we find $\nu=1.94-1.99$, but for the smallest gaps $\nu$ drifts upwards, to $2.06, 2.18, 2.35$ and $2.56$ at $\alpha=0.05, 0.025, 0.01$ and $0.005$.
	
	Therefore, we fit the two ends of each sweep separately, each with the exponent fixed to two. We perform a least-squares fit of $T=h \, g^2$ in the $(\ln g, \ln T)$-plane to the $6$ lowest and $6$ highest temperatures and define the amplitude of the sweep as their geometric mean, 
	\begin{equation}
		h(\alpha) = \sqrt{h_{\rm lo}(\alpha) \, h_{\rm hi}(\alpha)} \, , 
		\label{eq:two_branch_amplitudes}
	\end{equation}
	where $h_{\rm lo}$ and $h_{\rm hi}$ are the amplitudes from the low- and high-temperature branch fits, corresponding to the conformal and SYK$_2$ branch, respectively. The sweep at $\alpha=0.0025$ starts inside the crossover and has no low-temperature branch; it keeps the single fixed-exponent fit and is not used in any of the regressions below. The procedure is illustrated in Fig.~\ref{fig:two_branch_fits} for all closed-system sweeps.
	
	The two branches are well defined. A free fit restricted to the six points of a branch gives $\nu_{\rm lo} = 1.91$--$2.05$ and $\nu_{\rm hi}=1.99$--$2.03$ for all closed-system gaps $0.005 \leq \alpha \leq 0.9$, and $\ln(T/g^2)$ varies within a branch by at most $0.07$ (low-temperature end of $\alpha=0.01$) and $0.023$ (high-temperature end of $\alpha=0.005$). Moreover, each branch reproduces the corresponding amplitude of Eq.~\eqref{eq:T_g2_conf_imp}. The SYK$_2$ amplitude, $h_{\text{SYK}_2} = 64\alpha/(9\pi^2)$, contains no free parameter, and the ratio $h_{\rm hi}/h_{\text{SYK}_2}$ equals $1.002, 1.018$ and $1.048$ at $\alpha=0.005, 0.01$ and $0.025$. It grows to $1.10$ at $\alpha=0.05$ and $1.20$ at $\alpha=0.1$, as the condition $\alpha \ll \omega_0$ underlying Eq.~\eqref{eq:T_g2_conf_imp} degrades. The conformal amplitude, evaluated with the numerically determined $c_1 \approx 0.315$ of Appendix~\ref{app:conf_symmetry}, agrees with $h_{\rm lo}$ to within $9\%$ for $\alpha \leq 0.1$. The $\alpha$-dependence is reproduced as well: for $\alpha \leq 0.05$ the local exponents $\dd{\ln h_{\rm lo}} / \dd{\ln \alpha} = 1.49$--$1.53$ and $\dd{\ln h_{\rm hi}} / \dd{\ln \alpha} = 1.02$--$1.07$ are close to the predicted $1/(4\Delta_F - 1)=1.467$ and $1$, respectively, and their ratio scales between $h_{\rm hi}/h_{\rm lo} \propto \alpha^{-0.44}$ and $\alpha^{-0.50}$, to be compared with $\alpha^{1-1/(4\Delta_F-1)} = \alpha^{-0.47}$. 
	
	\paragraph{Large gaps and the bare-boson boundary.} For $\alpha \geq 0.6$ the amplitude leaves the power-law behavior: the two-branch values significantly exceed the extrapolation of Eq.~\eqref{eq:halphafit} (Fig.~\ref{fig:fits}{\bf b)}. This steepening reflects the criticality budget~\eqref{eq:budget_finite_gap}: the fraction of the budget used by the fermions, $|\Pi^R(0)|/\Pi_{\max}=(4M/N)(\omega_0-\alpha)\,T/g^2$, never exceeds $1$ at every point of every sweep, and its largest value along a sweep grows from $0.85$ at $\alpha=0.5$ to $0.996$ at $\alpha=0.9$.
	
	The boundary of the bare-boson regimes, Eq.~\eqref{eq:hightempboundary}, is the line $T = h(\omega_0) \, g^2/\omega_0$ obtained from a power-law fit of the small-gap data, which corresponds to a constant-gap curve at some gap $\alpha_{\rm b} < \omega_0$. To locate $\alpha_{\rm b}$ we interpolate the large-gap amplitudes with a two-parameter form that has a simple pole at $\alpha=\omega_0$, 
	\begin{equation}
		h_{\rm tan}(\alpha)=a\,\tan\!\left[\frac{\pi}{2}\left(\frac{\alpha}{\omega_0}\right)^{b}\right] \, ,
		\label{eq:h_tan}
	\end{equation}
	fitted in $\ln h$ to the two-branch amplitudes for the gaps $\alpha \geq 0.2$, giving $a=0.884$ and $b=1.131$.
	The tangent is a particularly good choice of fitting function as it reduces to a power-law for small $\alpha$, allowing to fit a larger region than a naive single pole (the fit is indeed reasonably good over all the data points).
	Solving $h_{\rm tan} (\alpha_{\rm b}) = h(\omega_0)$ gives
	\begin{equation}
		\alpha_{\rm b}=\omega_0\left[\frac{2}{\pi}\arctan\frac{h(\omega_0)}{a}\right]^{1/b}=0.746\,\omega_0\,.
		\label{eq:alpha_b}
	\end{equation}
	The shaded band in Fig.~\ref{fig:fits}\textbf{b} is obtained by fitting Eq.~\eqref{eq:h_tan} separately to the low- and high-temperature amplitudes $h_{\rm lo}$ and $h_{\rm hi}$ of all gaps $0.005\le\alpha\le0.9$. It reaches $h(\omega_0)$ at $\alpha=0.711$ and $0.764$, which bounds the dependence of $\alpha_{\rm b}$ on the branch that is used. We therefore quote $\alpha_{\rm b}\simeq0.75\,\omega_0$.

	\begin{figure}[t!]
		\centering
		\includegraphics[width=1\textwidth]{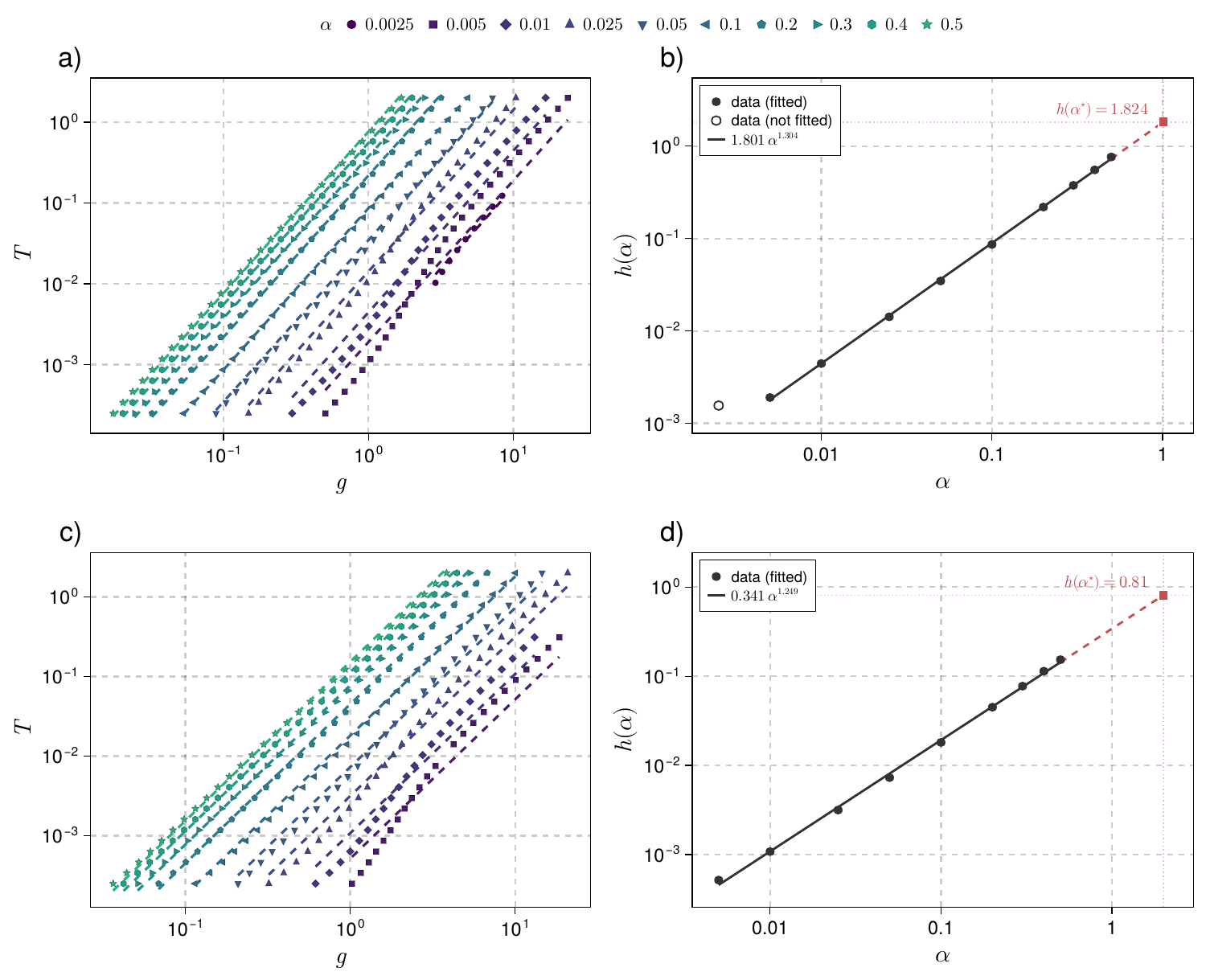}
		\caption{Bare-boson boundary with Markovian dissipation, $\kappa=0.1$ {\bf a)}, {\bf b)} and $\kappa=1$ {\bf c)}, {\bf d)}, in units of $\omega_0$ (cf.\ Fig.~\ref{fig:fits} for the closed system). Panels {\bf a)} and {\bf c)} show the constant-gap sweeps (markers, with shape and color labeling $\alpha$ as in the legend) together with the lines $T=h(\alpha)\,g^2$ (dashed), where $h(\alpha)$ is the two-branch amplitude~\eqref{eq:two_branch_amplitudes}. Panels {\bf b)} and {\bf d)} show the amplitudes $h(\alpha)$. Filled markers enter the regression $h=A \, \alpha^B$ over $0.005\le\alpha\le0.5$, as in Fig.~\ref{fig:fits}; the open marker ($\alpha=0.0025$, $\kappa=0.1$) has no low-temperature branch, shows the single fixed-exponent amplitude and is not fitted. The regression gives $A=1.801\pm0.043$, $B=1.304\pm0.008$ ($R^2=0.99975$) for $\kappa=0.1$ and $A=0.341\pm0.017$, $B=1.249\pm0.017$ ($R^2=0.99876$) for $\kappa=1$. The dashed red lines extrapolate the fit to the bare-boson limit $\alpha^\star=(\omega_0^2+\kappa^2)/\omega_0$, where the squares mark $h(\alpha^\star)=1.824\pm0.044$ and $0.811\pm0.049$ (statistical errors), to be compared with $1.861\,\omega_0^2/(\omega_0^2+\kappa^2)=1.842$ and $0.930$ from Eq.~\eqref{eq:T_g^2_diss}. The deviation at $\kappa=1$ is discussed in the text. No sweep was computed at $\alpha=0.0025$ for $\kappa=1$, and non-converged points are omitted (Appendix~\ref{app:const_alpha_sweeps}).}
		\label{fig:h_fits_dissipation}
	\end{figure}
	
	\subsection{Additional results with dissipation}
	\label{app:additional_results_diss}
	
	Here we repeat the analysis of Appendix~\ref{app:const_alpha_sweeps} for the sweeps at $\kappa=0.1$ and $\kappa=1$, on which Sec.~\ref{sec:Phase_signatures_diss} is based. We first extract the bare-boson boundary and test the static rescaling of Eq.~\eqref{eq:T_conf_imp_diss} directly on the amplitudes, then discuss the window dependence of the conformal/SYK$_2$ boundary, and finally show the quantum Fisher information over the $(g,T)$-plane.
	
	\begin{figure}[t!]
		\centering
		\includegraphics[width=0.6\textwidth]{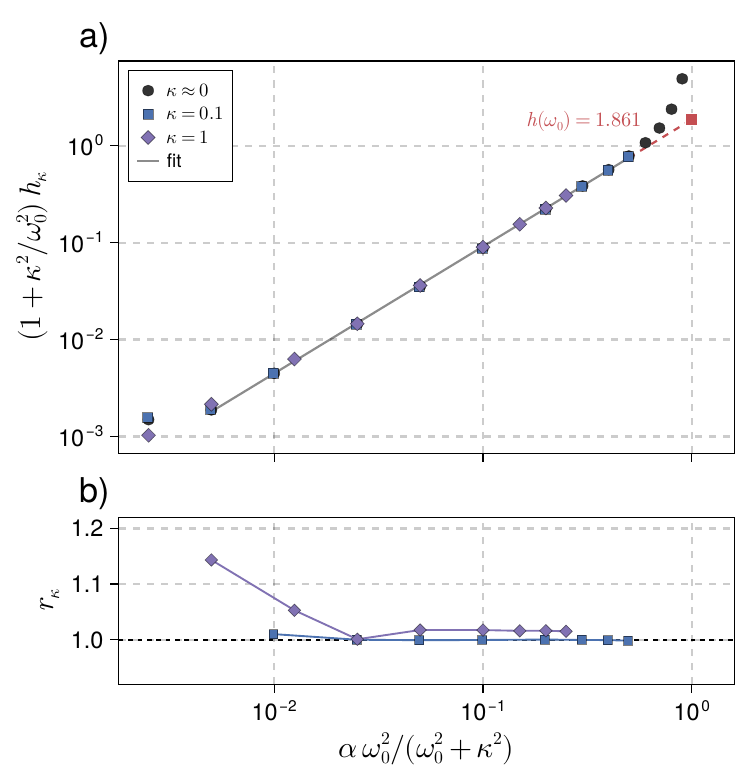}
		\caption{Test of the static rescaling, which does not rely on an extrapolation. a)~Two-branch amplitudes $h_\kappa(\alpha)$ for $\kappa\approx0$ (circles), $\kappa=0.1$ (squares) and $\kappa=1$ (diamonds), multiplied by $1+\kappa^2/\omega_0^2$ and plotted against the rescaled gap $\tilde\alpha=\alpha\,\omega_0^2/(\omega_0^2+\kappa^2)$. If the bath acted only through its static shift $\Pi^R_E(0)=\kappa^2/\omega_0$, all points would lie on the closed-system curve, Eq.~\eqref{eq:h_collapse}. The solid line is the closed-system power law~\eqref{eq:halphafit} over its fitting window, and its dashed extrapolation ends at $h_0(\omega_0)=1.861$ (square); in these variables the bare-boson boundaries~\eqref{eq:hightempboundary} and~\eqref{eq:T_g^2_diss} correspond to the same point. The closed-system amplitudes at $\alpha\ge0.6$ depart from the power law, as in Fig.~\ref{fig:fits}b. b)~Ratio $r_\kappa=(1+\kappa^2/\omega_0^2)\,h_\kappa(\alpha)/h_0(\tilde\alpha)$, with $h_0$ interpolated linearly in $(\ln\alpha,\ln h)$ between the closed-system gaps. For $\kappa=0.1$ the collapse holds to within $0.15\%$ for $\tilde\alpha\ge0.025$. For $\kappa=1$ the open-system amplitudes exceed the rescaled closed-system ones by $1.5$--$1.8\%$ for $\tilde\alpha\ge0.05$ and by up to $14\%$ at $\tilde\alpha=0.005$, which signals the dynamical part of the bath; the excess comes entirely from the low-temperature branch. Points with $\tilde\alpha<0.005$, which would require a two-branch amplitude of the closed-system sweep at $\alpha=0.0025$, are shown in {\bf a)} only.}
		\label{fig:h_collapse_dissipation}
	\end{figure}
	
	\paragraph{Bare-boson boundary.} The analysis of Fig.~\ref{fig:fits} is repeated in Fig.~\ref{fig:h_fits_dissipation} for $\kappa=0.1$ and $\kappa=1$ with the same procedure as in the closed system, i.e., two-branch amplitudes~\eqref{eq:two_branch_amplitudes} and a power law fitted over $0.005\le\alpha\le0.5$, extrapolated to the bare-boson value of the gap, $\alpha^\star\equiv\omega_0+\Pi^R_E(0)=(\omega_0^2+\kappa^2)/\omega_0$ (upper end of Eq.~\eqref{eq:rescaled_gap_diss}). Since the static shift of the bath only rescales the gap, $\alpha\to\tilde\alpha=\alpha\,\omega_0/(\omega_0+\Pi^R_E(0))$ (Eq.~\eqref{eq:T_conf_imp_diss}), it leaves the exponent $B$ unchanged and predicts $h(\alpha^\star)=1.861\,\omega_0^2/(\omega_0^2+\kappa^2)$, i.e., $1.842$ for $\kappa=0.1$ and $0.930$ for $\kappa=1$ (Eq.~\eqref{eq:T_g^2_diss}).
	
	For $\kappa=0.1$ we find $B=1.304\pm0.008$ and $h(\alpha^\star)=1.824\pm0.044$, in agreement with both predictions. For $\kappa=1$ we find $B=1.249\pm0.017$ and $h(\alpha^\star)=0.811\pm0.049$, $13\%$ below Eq.~\eqref{eq:T_g^2_diss}. The deviation comes from the smallest gaps. Since the fitting window is fixed in $\alpha$, at $\kappa=1$ it reaches rescaled gaps down to $\tilde\alpha=0.0025$, where the dynamical part of the bath raises the amplitudes (Fig.~\ref{fig:h_collapse_dissipation}) and thereby lowers $B$; the extrapolation to $\alpha^\star=2\,\omega_0$, a factor of four beyond the largest gap, amplifies this effect, and since the low-temperature end of these sweeps does not form a plateau, the extrapolated value also depends on the branch length. The rescaling itself is tested without any extrapolation below.
	
	\paragraph{Static rescaling of the amplitudes.} The rescaling can be tested more directly, without any extrapolation, on the amplitudes themselves. If the bath acted only through its static shift, Eq.~\eqref{eq:T_conf_imp_diss} would imply 
	\begin{equation}
		\left( 1 + \frac{\kappa^2}{\omega_0^2} \right) h_\kappa(\alpha) = h_0(\tilde{\alpha}) \, ,
		\label{eq:h_collapse}
	\end{equation}
	where $h_\kappa$ denotes the two-branch amplitude at dissipation $\kappa$ and $h_0$ the closed-system one. Figure~\ref{fig:h_collapse_dissipation} shows that the three data sets collapse onto a single curve. Deviations are largest at small rescaled gaps, where a growing part of each sweep lies in the conformal regime and close to the crossover. There the dynamical part of the bath, which Eq.~\eqref{eq:h_collapse} neglects, becomes noticeable. This is the same range of gaps in which the conformal/SYK$_2$ boundary steepens with respect to the closed system.

	\begin{figure}[t!]
		\centering
		\begin{minipage}[t]{0.49\textwidth}\centering
			\dissipationpanel{\bf a)}{}{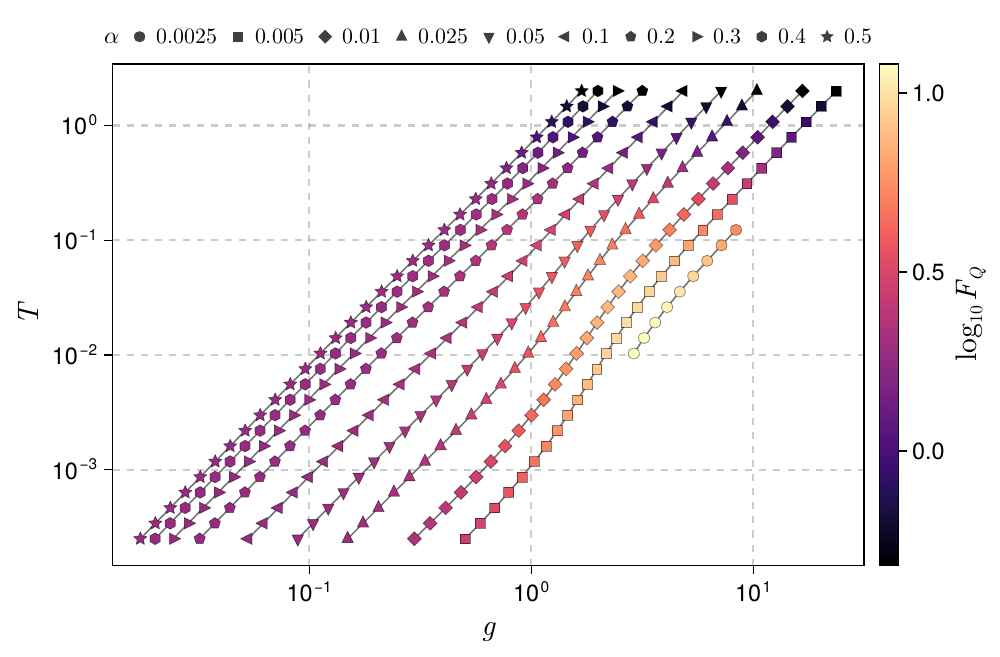}
		\end{minipage}\hfill
		\begin{minipage}[t]{0.49\textwidth}\centering
			\dissipationpanel{\bf b)}{}{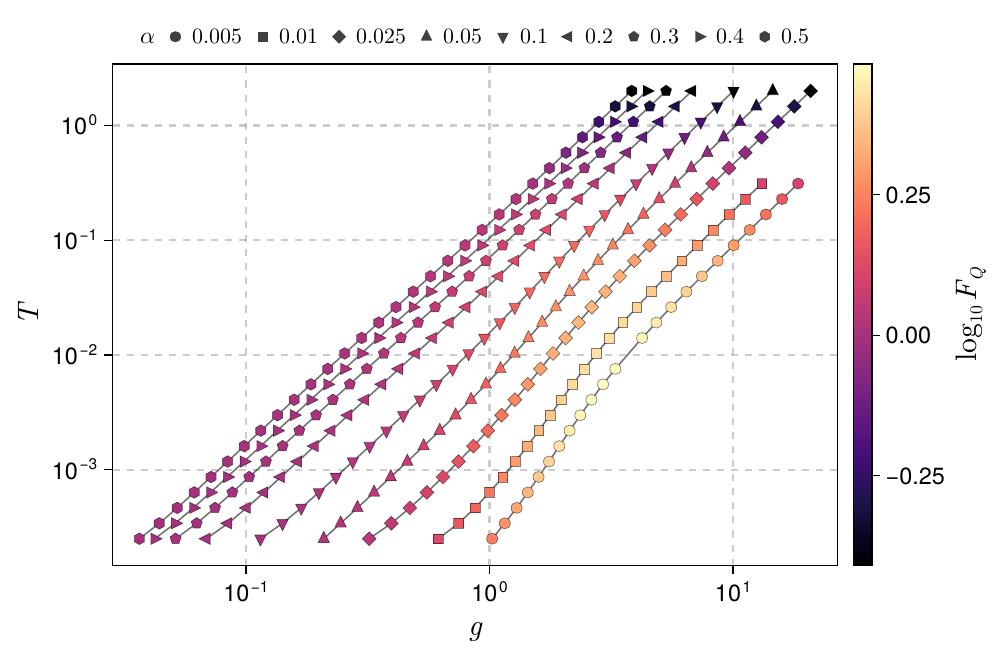}
		\end{minipage}
		
		\caption{Quantum Fisher information $F_Q$ on the constant-gap sweeps for \textbf{a)} $\kappa=0.1$ and \textbf{b)} $\kappa=1$ (in units of $\omega_0$), cf. Fig.~\ref{fig:Phase_diagram_QFI} for the closed system. Marker shape labels $\alpha$, color gives $\log_{10}F_Q$; note the different color scales between both panels. The ridge formed by the maxima along the sweeps still runs alongside the conformal/SYK$_2$ crossover, but it is lowered.}
		\label{fig:QFI_results_dissipation}
	\end{figure}
	
	\paragraph{Conformal/SYK$_2$ boundary.} For the boundary of Eq.~\eqref{eq:conf_SYK_boundary_diss} we use the window $\alpha \leq 0.3$ for $\kappa=0.1$, as in the closed system. For $\kappa=1$ the boundary is already steep at the smallest gaps, due to the broadening of the boson, and all nine gaps are fitted. For every choice of window, the exponent is lower than in the closed system, indicating that the steepening of the conformal/SYK$_2$ boundary in Fig.~\ref{fig:Phase_diagram_xi_dissipation} therefore does not depend on the fitting window. 
	
	\paragraph{Quantum Fisher information.}
	Figure~\ref{fig:QFI_results_dissipation} shows the quantum Fisher information over the $(g,T)$-plane for $\kappa=0.1$ and $\kappa=1$, complementing the individual sweeps of Fig.~\ref{fig:QFI_vs_dissipative_bound}. As in the closed system (Fig.~\ref{fig:Phase_diagram_QFI}), the maxima of $F_Q$ along the sweeps form a ridge that runs alongside the conformal/SYK$_2$ crossover, and for $\kappa=1$ lies essentially on it. The point of maximal sensitivity along a constant-$\alpha$ curve thus remains tied to the crossover in the presence of losses. What changes is the height of the ridge. Once $\alpha\ll\kappa^2/\omega_0$, the bound~\eqref{eq:QFI_general_bound_dissipation} grows only logarithmically as the gap closes. For $\kappa=0.1$ only the two smallest gaps approach this regime, and the map differs from the closed-system one mainly through a lowered ridge at large coupling. For $\kappa=1$, instead, all sampled gaps lie in it: the zero-temperature ceiling drops significantly compared with the closed system, and the whole map is compressed into less than a decade, hence the different color scales. Finally, at the high-temperature end of the sweeps $F_Q$ approaches the sum-rule value $\omega_0/T$ only for $T\gg\kappa$, which is why the upper part of the $\kappa=1$ map lies below its closed-system counterpart.

	\bibliographystyle{unsrt}
	\bibliography{references}
	
\end{document}